\documentclass[12pt,twoside]{report}
\usepackage{a4,graphicx,psfrag,amsbsy,doublespace,latexsym,graphics,calc}
\def\ra#1{{\overrightarrow#1}}
\def\la#1{{\overleftarrow#1}}
\def\gt{\tilde{g}}
\def\bt{\tilde{\beta}}
\def\dDelta{\dot{\Delta}}
\def\eqs#1#2{(\ref{#1},\ref{#2})}
\def\ins#1#2#3{\hskip #1cm \hbox{#3}\hskip #2cm}
\def\D{{\cal D}}
\def\fig#1{fig.\ \ref{#1}}
\def\F{{\cal F}}
\def\sec#1{sec.\ \ref{#1}}
\def\rtil{\tilde{r}}
\def\rt{\rtil}
\def\dephi{\!\!{\cal D}\phi\ }
\def\aka{{\it a.k.a.}\ }
\def\ie{{\it i.e.}\ }
\def\eg{{\it e.g.}\ }
\def\etc{{\it etc.}\ }
\def\etal{{\it et al.}}
\def\cf{{\it cf.}\ }
\def\viz{{\it viz.}\ }
\def\BBox{\ \includegraphics[scale=.2]{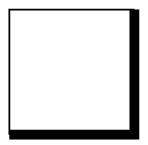}}

\def\si{\includegraphics[scale=2]{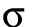}}
\def\mn{_{\mu\nu}}

\def\ma{_{\mu\alpha}}

\def\na{_{\nu\alpha}}

\def\am{_{\alpha\mu}}
\newcommand{\one}{\hbox{1\kern-2mm l}}

\newcommand{\dfrac}{\displaystyle \frac}
\newcommand{\dsum}{\displaystyle \sum}
\def\sh{\hat{S}}
\def\al{\alpha}
\def\si{\sigma}
\def\cp{c^{\prime}}
\def\lam{\lambda}
\def\Lam{\Lambda}
\newcommand{\dints}{\displaystyle \int}
\def\dirac#1{#1\llap{/}}
\def\de{\partial}

\def\be{\begin{equation}}
\def\ee{\end{equation}}
\def\ba{\begin{array}}
\def\ea{\end{array}}
\def\bea{\begin{eqnarray}}
\def\eea{\end{eqnarray}}
\def\hf{\dfrac{1}{2}}

\def\cc{{\cal C}}

\def\ct{\tilde{c}}
\def\ctp{\ct^{\prime}}

\def\CC{\mbox{C}\hspace{-1.8mm}\mbox{l}\ }

\def\ie{{\it i.e.\ }}
\def\eg{{\it e.g.\ }}
\def\etc{{\it etc.\ }}
\def\etal{{\it et al.\ }}
\def\cf{{\it c.f.\ }}

\def\k{{\bf k}}

\def\sig3{\sigma_3}
\def\Lam{\Lambda}
\def\lam{\lambda}

\def\A{{\cal A}}

\def\C{{\cal C}}

\def\J{{\cal J}}

\def\S{{\cal S}}

\def\rtil{\tilde{r}}

\def\tr{{\mathrm{tr}}}
\def\str{{\mathrm{str}}}

\def\one{\hbox{1\kern-.8mm l}}

\def\gap{\hspace{0.05in}}
\def\ph#1{\phantom{#1}}

\def\s#1#2#3{S^{(#1)}_{#2} (#3)}
\newcommand{\ldl}{\Lambda \partial_{\Lambda}}
\def\tp{\tilde{\phi}}
\newcommand{\inte}{\! \int \!\!}
\def\tS{\tilde{S}}
\def\thS{\tilde{\sh}}
\def\tg{\tilde{\gamma}}
\def\eq#1{eq.~(\ref{#1})}
\def\ceq#1{Eq.~(\ref{#1})}
\def\0{\vec{0}}
\def\ct{\tilde{c}}
\def\ctp{\ct^{\prime}}
\def\ker#1{\cdot #1 \cdot}

\begin{document}
%
%
\pagestyle{empty} 
\begin{center} 
    {\LARGE UNIVERSITY OF SOUTHAMPTON} \\
\vspace{3cm} 
    
{\Huge \bf A Gauge Invariant}\\[.6cm] {\Huge \bf Flow Equation}\\

\vspace{1cm} 
    by \\
\vspace{1cm} 
    {\LARGE Antonio Gatti} \\
\vspace{1.0cm} 
    A thesis submitted for the degree of \\
\bigskip 
    Doctor of Philosophy \\
\vspace{0.7cm} 
\bigskip 
    Department of Physics and Astronomy \\
\bigskip 
October 4, 2002
\end{center} 
\vspace{0.5cm}

%

%
\newpage 
\pagestyle{empty} 
\begin{center} 
      {\Large UNIVERSITY OF SOUTHAMPTON}  \\
\bigskip 
      \underline{\large ABSTRACT} \\
\bigskip 
      {\Large FACULTY OF SCIENCE} \\
\bigskip 
      {\Large PHYSICS} \\
\bigskip 
      \underline{\large Doctor of Philosophy} \\
\bigskip 
     {\Large  A Gauge Invariant Flow Equation}\\
\bigskip 
      {\large Antonio Gatti} \\
\end{center}

Given a Quantum Field Theory, with a particular content of fields and a
symmetry associated with them, if one wants to study the evolution of the
couplings via a Wilsonian renormalisation group, there is still a freedom on the construction of a
flow equation, allowed by scheme independence.

In the present thesis, making use of this choice, we first build up a generalisation of the Polchinski
flow equation for the massless scalar field, and,
applying it to the calculation of the beta function at one loop for the
$\lambda \phi^4$ interaction, we test its universality beyond the already
known cutoff independence. Doing so we also develop a method to perform
the calculation with this generalised flow equation for more complex cases.

In the second part of the thesis, the method is extended to $SU(N)$
 Yang-Mills gauge theory, regulated by incorporating it in a spontaneously
 broken $SU(N|N)$ supergauge group. Making use of the freedom allowed by
 scheme independence, we develop a flow
equation for a $SU(N|N)$ gauge theory, which preserves the invariance
step by step throughout the flow and demonstrate the technique with a compact calculation of the one-loop
 beta function for the $SU(N)$ Yang-Mills physical sector of $SU(N|N)$, achieving a manifestly universal result, and without
 gauge fixing, for the first time at finite $N$.

\newpage 
\pagestyle{empty} 
\begin{center} 
\vspace*{8cm} 
\hspace{-1cm}
\emph{Dedicated to my family}\\
\hspace{1cm}
\end{center}

%
%
\newpage 
\pagenumbering{roman} 
\pagestyle{plain} 
\tableofcontents 
\newpage 
\listoffigures 
\newpage 
\listoftables 
%
%
\newpage 
\chapter*{Preface} 
\addcontentsline{toc}{chapter} 
{\protect\numberline{Preface\hspace{-96pt}}} 

Original work is contained in the last section of chapter 2 and in chapters
4, 5 and 6 (in collaboration with Tim Morris and Stefano Arnone) and it can
also be found in:

\newcounter{token}
\setcounter{token}{1}
\begin{tabbing}
(\roman{token}) S.Arnone, A. Gatti and T.R. Morris, JHEP {\bf 0205} (2002) 059
\stepcounter{token} \\
(\roman{token}) S.Arnone, A. Gatti and T.R. Morris, hep-th/0205156
\stepcounter{token} \\
(\roman{token}) S.Arnone, A. Gatti and T.R. Morris, hep-th/0207153 
\stepcounter{token} \\
(\roman{token}) S.Arnone, A. Gatti and T.R. Morris, hep-th/0207154 
\stepcounter{token} \\
(\roman{token}) S.Arnone, A. Gatti and T.R. Morris, hep-th/0209162
\end{tabbing}

No claim to originality is made for the content of the rest of chapter 2 and chapter 3, which were compiled using a variety of other sources.

%
%
\newpage 
\chapter*{Acknowledgements} 
\addcontentsline{toc}{chapter}  
                {\protect\numberline{Acknowledgements\hspace{-96pt}}} 

First of all I wish to thank my supervisor Tim Morris, who has been guiding
me during these years and has always have found the time to answer all my
questions and my doubts, and my collaborator Stefano Arnone who has not
just been my colleague but also a very good friend.

I would also like to thank all the members of the SHEP group for making
such a friendly environment in which to work.

My gratitude must also go to the people I shared my house with,
here in Southampton for two years: Pier and Stefy, Mike, Giuseppe and Paolo for making me feel I am in
a family and for all the adventures we shared during this period of time,
and to all my friends without whom I might have finished earlier. Among
them I want to remember: Ajey, Scott and Tiziana, Shu, Angela and Matt, Big
Daddy J, Joao,
Fabio and Rowan, Sam, Julien, Fabrice, Sam and Sheela, Chris, Claire,
Rudi and Xana, Rui, Rosemary, Gianguido and Daniele. Last but not
least among my friends in Italy I would like to thank ``Il circolo'', which
includes Aica and Elisa, Paolo/Baolo, Fenky, Strong and Nonche'Ebano for their
always warm ``welcoming committee'' and Manu, Silvia, Fazzio and Botti. I
should add a big number of others, to whom I
apologise for not mentioning here, for a pure reason of space. 

Finally and above all I wish to thank my parents and my family for their unequivocal support without whom I could have never accomplished this result. 

\newpage

\pagenumbering{arabic}

\chapter{Introduction}\label{chap:1}
\begin{figure}[h]
\psfrag{l0}{$\Lambda_0$}
\psfrag{l}{$\Lambda$}
\psfrag{p+}{$p_>$}
\psfrag{p-}{$p_<$}
\psfrag{sl0}{$S_{\Lambda_0}$}
\psfrag{ph}{physics}
\psfrag{tt}{$\to$}
\psfrag{bb}{bare}
\psfrag{eff}{effective}
\psfrag{act}{action}
\psfrag{q}{q}
\psfrag{sl}{$S_{\Lambda}$}
\begin{center}
\includegraphics[scale=.6]{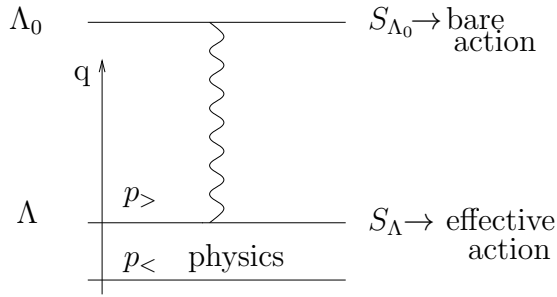}
\end{center}
\caption{Flow of $S_{\Lambda}$ as momenta are integrated out}\label{fig:1}
\end{figure}

The Exact Renormalisation Group (ERG) \cite{wilson}-\cite{mor:manerg}, is a
powerful tool to control the infinities arising in quantum field
theories. Once one has started introducing a Lagrangian which defines a
certain theory, the calculation of possible predictions from it, such as
scattering amplitudes or lifetime of particles, have in most of the cases
to pass through the process of perturbative expansion, this being often,
the only practical possibility. These quantities are evaluated at different
approximations by truncating an expansion on the coupling constant, assumed
to be small. Infinities come out already trying to calculate the second
approximation, when one faces the task to integrate over all the possible
momenta of the virtual particles taking part in the process. Rather than
integrating out all the momenta at once, by introducing a cutoff at a certain scale of momenta $\Lambda_0$ (here
imposed via a cutoff function) which regulates these integrals, making the
theory finite, one introduces another scale $\Lambda$ (much lower than
the first one), and the integral of the partition function is made from
this new scale up to the first one, we are left with an integral between
zero momenta and this new scale $\Lambda$. This integral can still be
expressed as a partition function, but the previous action (called the {\it
bare action}) which is usually chosen as simple as possible, is replaced by
a complicated {\it effective action}. This {\it transformation} of the
action in the partition function due to $integrating\ out$ momenta is a
transformation of the so called Wilsonian {\bf Renormalisation Group} (RG)
(see fig.\ref{fig:1}). Imposing the invariance of the partition function
under such transformations, one can find an equation, whose solution (with
set boundary condition at $\Lambda_0$ being the bare action of the theory),
the so called $Wilsonian\ action$, describes the RG flow of the
action. Since the equation is written non-perturbatively, the approach is
called the $Exact\ RG$. The limit that the cutoff tends to infinity of the
solution of this equation would be the action of the theory at any
scale. This will be explained in detail in the next section for the case of
a massless scalar field theory. 

When this process is applied to gauge theories, a further problem arises:
the cutoff functions usually used, break the gauge invariance. The usual
approach at this point was to recover this symmetry when the cutoff was
removed. In this way each step is done $without$ gauge invariance, which is
reinserted at the end (this approach will be briefly reviewed later in
section \ref{sec:breakGI}). The ``temporary'' non-gauge invariance limits the study of the theory in particular for what concerns non-perturbative studies. Since this is an interesting direction for a better understanding of gauge theories another way can be worth exploring, the one pioneered by Dr. Tim Morris in \cite{mor:manerg, mor:erg1, mor:erg2}, which is based on studying an ERG for a gauge theory preserving this symmetry step by step. 

This thesis is organised as follows. In the second chapter will be
considered the Polchinski approach to the Exact RG applied to scalar field
theory. A review of the concept of scheme independence will be done, 
and through the freedom derived from it, a flow equation more general than the one by Polchinski, will be introduced. In the last section
of the chapter, we will calculate making use of this generalised flow
equation, the one loop beta function for the massless scalar field theory,
developing a method to perform the calculation in this more general
framework, which will be possible to apply also to the more complicated gauge
field case.

The third chapter will be dedicated to the description of the attempts
done to regularise gauge theories. Starting from the attempts mentioned
earlier in which gauge invariance was first broken and then restored, 
and then going on to the methods involving covariant derivatives with the addition of
Pauli-Villars fields. In the last section of the chapter will be also
reviewed one of the most recent methods
pioneered by Dr. Tim Morris et. al, in which the regularisation is
performed embedding the gauge group of the theory in a bigger graded
group. The chapter will then start with some concepts related to this
peculiar group and it will explain the mechanism through which its subgroup
of physical interest comes out regulated in a gauge invariant way.

The fourth chapter will mainly be concerned with the build up of a flow equation
for the theory constructed on this bigger gauge group. Making use of the
freedom allowed by scheme independence, the equation has been constructed
in order to preserve gauge invariance through the flow. 

The fifth and sixth chapters involve a check on the flow equation introduced
in the previous chapter. Adapting the method used for the scalar field to
the gauge case, we calculate the one loop beta function for the $SU(N)$
Yang-Mills theory without fixing the gauge at any stage.

\chapter{The Scalar Field Case}

Before we start with the attempt to write a flow equation for a $SU(N|N)$ gauge theory, it might be useful to have
a look at the way a flow equation has been worked out in a much simpler
case, which is the massless scalar field theory.

Of the many possible ERG
formulations, we have chosen to describe the one by Polchinski, as in
\cite{mor:erg1,mor:erg2}, since it appeared more suitable for extracting a
generalisation of its flow equation, using scheme independence. The
concept of scheme independence, which is going to be described in the
second section of this chapter, will be central for the development of this whole work.

As it will appear clear through this
chapter, the simple massless scalar theory can give great insights on how
to proceed for more sophisticated cases such as gauge theories. As an example, the calculation of the one loop $\beta$-function performed
in the last section, with a generalised Polchinski-like flow equation for
the scalar theory, will
set up a method which will be adapted to the analogous calculation for $SU(N)$
Yang-Mills.

Let us start now with a brief introduction to the Polchinski ERG
for the massless scalar field.

\section{The Polchinski ERG equation}

The central object in the path integral formulation of quantum field
theories is the partition function $\cal Z$ from which it is possible to
extract information such as the vacuum expectation value of a product of
fields (\aka correlators). These are related to physical objects like cross sections and so
on. The partition function of a theory describing a field $\phi$ defined by
an action $S[\phi]$ is\footnote{If not otherwise specified we will always
be working in Euclidean space}:
\be\label{zed}
{\cal Z}[j]=\int \!\!{\cal D}\phi\ e^{-S[\phi]+j\cdot\phi}
\ee
Taking derivatives with respect to the current $j$ and setting it to zero,
gives expectation values of fields. When these integrals have to be
performed, depending on the form of $S$, there is sometimes the need to
make use of perturbative methods, and, as it was mentioned in the previous chapter, this gives rise to divergent integrals.

When these infinities arise from large momenta in the integrals of the particles
appearing in loops, the usual procedure is to
regularise the bare action to have finite quantum corrections, namely by substituting $S\to S_{\Lambda_0}$ (see fig.\ref{fig:1}). The partition
function regulated via $\Lambda_{_0}$ is then:
\be
{\cal Z}=\int\!\! {\cal D}\phi\ e^{-S_{\Lambda_{_0}}}
\ee
where the action of the theory is taken at $\Lam_0$ to be (bare action), in the momentum space:
\be\label{bare-action}
S_{\Lam_0}=\frac{1}{2}\int\!\! \frac{d^4p}{(2\pi)^4}p^2 c^{-1}(p^2/\Lam_0^2)\tilde{\phi}(p)\tilde{\phi}(-p)+S^{int}[\tilde{\phi},\Lam_0]
\ee
where $\tilde{\phi}(p)$ are the Fourier component of the field $\phi(x)$ and $c(p^2/\Lam^2)$ is a {\it smooth}, \ie infinitely differentiable, ultra violet  cutoff profile. The cutoff which modifies propagators $1/p^2$ to $c/p^2$, satisfies $c(0)=1$ in such a way that low energy is unaltered and $c(p^2/\Lam^2)\to 0$ as $p^2/\Lam^2\to\infty$, fast enough in order to make all the Feynman diagrams ultra-violet regulated. $S^{int}[\tilde{\phi},\Lam_0]$ is the interaction part of the bare action,
containing all the relevant (and marginal) interactions compatible with the
symmetries of the theory, considered to be the only non zero interactions
at the scale $\Lam_0$. For the case considered in section \ref{sec:1.4}, we
would normally choose the following (see also \cite{scaus}):
\be
S^{int}[\phi,\Lam_0]=\frac{\lam_0}{4!}\int\!\! d^4x\ \phi^4+\frac{m_0}{2}\int\!\! d^4x\ \phi^2
\ee

To motivate the later strategy, setting an intermediate cutoff scale
$\Lambda$, we can (at least heuristically) separate the fields
into the ones with momentum greater than $\Lambda$ ($\phi_>$) and smaller
than it ($\phi_<$) and rewrite the partition function with the new measure.
We can then perform the integral on $\phi_>$ for a certain $\Lambda$ to
get: :
\be
{\cal Z}=\int\!\! {\cal D}\phi_{>}{\cal D}\phi_{<}\ e^{-S_{\Lambda_0}}=\int\!\! {\cal D}\phi_{<}\ e^{-S_{\Lambda}}
\ee
where:
\be
e^{-S_{\Lambda}}=\int {\cal D}\phi_>\ e^{-S_{\Lambda_0}}
\ee
In principle now $S_{\Lambda}$ could contain all possible interactions
compatible with the symmetries of the theory. In our case, the RG transformation amounts to changing the cutoff from $\Lam_0$ to $\Lam << \Lam_0$ in eq.(\ref{bare-action}):
\be\label{effact}
S_{\Lambda}=\frac{1}{2}\int\!\! \frac{d^4p}{(2\pi)^4}p^2 c^{-1}(p^2/\Lam^2)\tilde{\phi}(p)\tilde{\phi}(-p)+S^{int}[\tilde{\phi},\Lam]
\ee
 where now $S^{int}[\tilde{\phi},\Lam]$ is a more complicated functional of
$\tilde{\phi}$. (The ``tilde'' for the Fourier components of $\phi(x)$ will be dropped from now on.) The choice of a flowing kinetic term which keeps the same form as the corresponding
one in the bare action, is the choice performed in \cite{pol} and it is
just for simplicities sake.
Now, changing the intermediate scale $\Lambda$, the interaction term $S^{int}_{\Lambda}$, transforms (flows) as we integrate out momenta (RG transformation). One way to get the equation describing its
flow is the following. Demanding that physics be invariant under such a
scale change, follows from asking the partition function ${\cal Z}$ to be independent on
$\Lam$. If we then require its variation under the RG transformation to vanish
\be
\delta{\cal Z}=0,
\ee
we are led to a flow equation for $S^{int}_{\Lambda}$ (Polchinski's for scalar field, see \cite{pol}):
\be\label{polflow}
\Lambda\partial_{\Lambda}
S^{int}_{\Lambda}=-\underbrace{\dfrac{1}{\Lambda^2}\dfrac{\delta
S^{int}_{\Lambda}}{\delta \phi}\cdot c^{\prime}\cdot\dfrac{\delta
S^{int}_{\Lambda}}{\delta \phi} }_{\mbox{{\small Classical
Term}}}+\underbrace{\dfrac{1}{\Lambda^2}\dfrac{\delta}{\delta\phi}\cdot
c^{\prime}\cdot\dfrac{\delta S^{int}_{\Lambda}}{\delta \phi}}_{\mbox{{\small
Quantum Correction}}}
\ee
($c^{\prime}$) is the derivative of the cutoff function with respect to its
argument ($p^2/\Lam^2$) and the following notation has been introduced:
given two functions $f(x)$ and $g(y)$ and a momentum space kernel
$W(p^2/\Lam^2)$ ($\Lam$ is the effective cutoff),
\be\label{dotnotation}
f\cdot W\cdot g=\int\int d^4x d^4 y f(x) W_{xy} g(y),
\ee
where $W_{xy}=\int\frac{d^4p}{(2\pi)^4}W(p^2/\Lam^2)e^{ip\cdot(x-y)}$.

The solution of the (exact) equation (\ref{polflow}), with
boundary condition $S^{int}_{\Lambda=\Lambda_0}=S^{int}_{\Lambda_0}$, in the continuum
limit ($\Lambda_0\to\infty$) would be the action of the theory at any scale
(as we were observing earlier in the thesis).

\section{Scheme Independence}\label{sec:1.3}

The equation derived for the interaction part of the effective action, in
the previous section was indeed a consequence of the request $\delta{\cal
Z}=0$ under a RG transformation. Following the example of J.I.Latorre and
T.R.Morris in \cite{latmor}, it is worth at this point making an
observation. This one flow equation is not necessarily the most general consequence of such a requirement.
There is a more general statement, which can be extracted from it and this
introduces a particular viewpoint on the concept of Scheme Independence (SI).

Let us first consider the effective kinetic term of the scalar field
theory, in the notation introduced
in eq.(\ref{dotnotation}):
\be\label{hat-hat}
\frac{1}{2}\de_{\mu}\phi\cdot c^{-1}\cdot\de_{\mu}\phi
\ee
This will be referred to as ``seed action'' and denoted with $\hat{S}$. The
total effective action can be written then (dropping the $\Lam$) as:
$S=\hat{S}+S^{int}$. Defining the combination $\Sigma=S-2\hat{S}$, the
Polchinski flow equation (\ref{polflow}), can be rewritten (up to a vacuum
energy term, discarded in \cite{pol}) as:
\be\label{genfloweqscalar}
\Lambda\partial_{\Lambda}S=-\dfrac{1}{\Lambda^2}\dfrac{\delta S}{\delta
\phi}\cdot c^{\prime}\cdot\dfrac{\delta\Sigma}{\delta
\phi}+\dfrac{1}{\Lambda^2}\dfrac{\delta}{\delta\phi}\cdot
c^{\prime}\cdot\dfrac{\delta\Sigma}{\delta \phi}
\ee
The invariance of the partition function is manifest from the previous
equation, since it is possible to recognise that eq.(\ref{genfloweqscalar})
can be recast as:
\be\label{newfloweqscalar}
\Lambda\partial_{\Lambda}e^{-S}=-\dfrac{1}{\Lambda^2}\dfrac{\delta}{\delta\phi}\cdot
c^{\prime}\cdot\left(\dfrac{\delta\Sigma}{\delta\phi}e^{-S}\right)
\ee
\ie the infinitesimal RG transformation on the partition function is a
change in the integrand corresponding to a total derivative. From the
previous equation we can also notice that:
\be\label{changeofvariable}
\delta{\cal Z}=\frac{\de\cal Z}{\de\Lam}\delta\Lam=\delta\Lam\int\!\! {\cal
D}\phi\left(\frac{\delta}{\delta\phi}\cdot
\psi-\psi\cdot\frac{\delta S}{\delta\phi}\right)e^{-S}
\ee
where
$\psi=-\frac{1}{\Lambda^2}c^{\prime}\frac{\delta\Sigma}{\delta\phi}$.
This establishes another result: integrating out degrees of freedom
correspond to a redefinition of the fields in the theory \cite{latmor}. In
the case we have been considering here, the change in the partition
function due to a transformation under the RG, corresponds to the variation due
to the change of variables (field redefinition):
\be
\phi\to \phi+\delta\Lam\psi
\ee
Recognising the first term in eq.(\ref{changeofvariable}) as arising from the
Jacobian and the second as arising from the variation of $S$.
$\psi$ is called the {\it kernel} of the RG transformation. Different kernels, lead to different flow
equations. If these flow equations come from different choices of kernels
connected via a field redefinition, they describe the
same physical system. This gives a great freedom on the form of the flow
equation. First of all there is a choice of the form of $\Sigma$, which could be chosen
as a polynomial in $S$.  A reason for choosing it at least linear in the effective action, as it is
done here, is to ensure a quadratic term in $S$ on
the RHS of the flow equation, which can give fixed point solutions to the
flow equation. After
this first choice is made, the freedom is on the ``seed action'' $\hat{S}$
(which will be widely used in the gauge case)
and on the choice of the cutoff function, which in principle can now
contain interactions (as will be the case for the ``wines'' which will be introduced for the
gauge case, following \cite{mor:erg1,mor:erg2}), higher functional
derivatives, and/or other more complex dependences on $S$.

Physical quantities should be independent of these choices. One of the main
purposes of this thesis is to check that the equation derived from this more general formulation of
the ERG, can give the same results that were found with previous ones. The
universal quantity examined here is the first coefficient of the beta
function for both the massless scalar field and the $SU(N)$ Yang-Mills
theory. In the former case this check leads to a proof of
universality for the beta function at one loop beyond the change of the
cutoff function and allows to develop the right procedure to deal with
calculations in this more general scheme. In the case of gauge theories,
since SI allows to write a flow equation which preserves the symmetry
for each step of the flow, the check represents also the first calculation of such a
quantity in a gauge invariant way at finite $N$.

As shown above, in the Polchinski case, the ``seed action'' coincided with
the kinetic term, as eq.(\ref{genfloweqscalar}) is equivalent to
eq.(\ref{polflow}) for this choice of $\hat{S}$. For our purposes, in
the calculation of the beta function at one loop for the massless scalar
field, the ``seed action'' will be chosen much more general. First of all we
require it to preserve the symmetry $\phi\to -\phi$, so it must
be an even functional of the fields. Second, if one wants the effective
kinetic term to flow as in eq.(\ref{effact}), the bilinear term of $\hat{S}$
must be still equal to (\ref{hat-hat}). For all the other interaction vertices
with $n>2$, we just ask them to be infinitely differentiable (Taylor
expandable to any order) to ensure they do not introduce infrared
singularities, and that they do not lead to ultraviolet divergent momentum integrals,
so that the flow described by the equation can be interpreted as
integrating out momenta. We will see from the next section that the first
coefficient of the beta function is blind to the
introduction of all these extra parameters, which can be always eliminated
in favour of the physically meaningful vertices of the effective action.

\section{One loop $\beta$-function with general $\hat{S}$: the scalar field case}\label{sec:1.4}

Before we start the discussion for the super-gauge field we will consider
the massless scalar field case, in the present formulation and show that
starting with the new form of the flow equation (\ref{genfloweqscalar}), we
can get the correct $\beta$-function at one loop, without specifying
$\hat{S}$ and without any strong constraint on it.

As we have mentioned already we expect the physical quantities to be universal, \ie independent of the
renormalization scheme. In particular, they should not be sensitive to
the particular choice of
the RG kernel, \eg on the form of the cutoff function or the expression for the
seed action.
We aim to calculate one of those, the one-loop contribution to the $\beta$
function, while keeping as general a seed action as possible.
As we will see, an elegant, clear cut way of achieving such a result is to
make use of the flow equations for the effective couplings in order
to get rid of the seed action vertices.

As usual, universal results are obtained only after the imposition of
appropriate renormalization conditions which allow us to define what we
mean by the physical (more generally renormalised) coupling and field.
(The renormalised mass must also be defined and is here set to zero
implicitly by ensuring that the only
scale that appears is $\Lambda$.)

We write the vertices of $S$ as
\be
\s{2n}{}{\vec{p};\Lam}\equiv\s{2n}{}{p_1,p_2,\cdots,p_{2n};\Lam}
\doteq (2\pi)^{8n}
{\delta^{2n} S\over\delta\phi(p_1)\delta\phi(p_2)\cdots\delta\phi(p_{2n})},
\ee
(and similarly for the vertices of $\sh$). In common with earlier works
\cite{pol,Bo}, we define the renormalised four-point coupling $\lambda$ by
the effective action's four-point vertex evaluated at zero momenta:
$\lambda(\Lambda)=\s{4}{}{\vec{0};\Lam}$. This makes sense once we
express quantities in terms of the renormalised field,
defined (as usual) to bring the kinetic term
into canonical form $\s{2}{}{p,-p;\Lam}= \s{2}{}{0,0;\Lam} + p^2
+O(p^4/\Lam^2)$.
The flow equation can then be taken to be of the form \cite{sumi,ball}:
\be \label{gammafleq}
\ldl S - {\gamma\over2}\ \phi\!\cdot\!{\delta S\over \delta \phi}
   = -{1\over\Lambda^2}
 {\delta S\over\delta\phi} \ker{c'}
{\delta \Sigma \over\delta\phi} +{1\over\Lambda^2}{\delta
\over\delta\phi} \ker{c'} {\delta \Sigma \over\delta\phi}.
\ee
We have used the short hand defined in eq.(\ref{dotnotation}), and as usual the anomalous dimension $\gamma={1\over Z}\ldl Z$,
where $Z$ is the wavefunction renormalization. As emphasised
in refs.\ \cite{mor:manerg,latmor}, although
\eq{gammafleq} is not the result of changing variables
$\phi\mapsto \phi\sqrt{Z}$ in \eq{genfloweqscalar}, it is still a perfectly
valid flow equation and a more appropriate starting point when
wavefunction renormalization has to be taken into
account. This is in fact a small example of the immense freedom we have
in defining the flow equation.
(The new term on the left hand side arises from replacing
$\partial_\Lam |_\phi$ with a partial derivative at constant renormalised
field, but in order to produce the right hand side, and in order
to reproduce the same $\sh$, we need to start with the alternative
cutoff function $cZ$ in eqs.\ (\ref{bare-action}) -- (\ref{genfloweqscalar}).
Alternatively, for the purposes of computing the $\beta$ function, we
could have simply taken account of the wavefunction renormalization
afterwards as in ref.\ \cite{Tighe}.)

We now rescale the field $\phi$ to
\be
\label{rescale}
\phi =\frac{1}{\sqrt{\lambda}}\, {\tp},
\ee
so as to put the coupling constant in
front of the action. This ensures the expansion in the coupling constant
coincides with the one in $\hbar$, the actual expansion parameter being
just $\lam \hbar$. The resulting expansion is more elegant,
being no longer tied at
the same time to the order of expansion of the field $\phi$. It is also
analogous to the treatment pursued for gauge theory in refs.\ \cite{mor:manerg,mor:erg1,mor:erg2}
(where gauge invariance introduces further simplifications in particular
forcing $\gamma=0$ for the new gauge field). The following analysis thus
furnishes a demonstration that these ideas also work within scalar
field theory.

The bare action (\ref{bare-action}) rescales as
\be
S_{\Lam_0}[\phi] =
{1\over \lam} \left[{1\over 2} \inte {d^4 p
\over (2\pi)^4} \, p^2 c^{-1}({\textstyle {p^2\over \Lambda_0^2}}) \,
{\tp}^{2} + {1
\over 4!} \inte d^4 x \, {\tp}^{4} \right] \doteq {1\over \lam}
{\tS}_{\Lam_0}[{\tp}].
\ee

Defining the ``rescaled'' effective and seed actions as $S[\phi] =
{1\over \lam} \tS[\tp],
\sh[\phi] = {1\over \lam}\thS
[\tp]$, and absorbing the change to $\partial_\Lam|_{\tp}$
in a change to $\tg$,
the flow equation (\ref{genfloweqscalar}) reads
\be
\label{tildefleq}
\ldl \left({1\over \lam} \tS\right)
- {\tg\over2\lam}\ \tp\!\cdot\!{\delta \tS\over \delta \tp}
=-\frac{1}{\lambda \Lam^2}\frac{\delta
(\tS-2\thS)}{\delta\tp}
\ker{c'} \frac{\delta \tS}{\delta
\tp}+\frac{1}{\Lam^2}\frac{\delta}{\delta
\tp}\ker{c'}\frac{\delta(\tS-2\thS)}{\delta\tp}.
\ee
Expanding the action, the beta function $\beta(\Lam) = \ldl \lam$
and anomalous dimension, in
powers of the coupling constant:
\bea \nonumber
\tS[\tp]&=&\tS_0+\lambda \tS_1+\lambda^2
\tS_2+\cdots,\\ \nonumber
\beta(\Lam)&=&\beta_1\lambda^2+\beta_2\lambda^3+\cdots,\\ \nonumber
\tg(\Lam) &=&\tg_1\lam+\tg_2\lam^2+\cdots
\eea
yields the loopwise expansion of the flow equation\footnote{In order to
simplify the notation, the tildes will be removed from now on.}
\bea
&&\Lam\de_{\Lam}S_0=-\frac{1}{\Lam^2}\frac{\delta S_0}{\delta\phi}\cdot
c'\cdot\frac{\delta (S_0-2\hat{S})}{\delta\phi}\label{scalartree},\\
&&\Lam\de_{\Lam}S_1-\beta_1 S_0-{\gamma_1\over2}\
\phi\!\cdot\!{\delta S_0\over\delta\phi}
=\nonumber \\
&&\phantom{\Lam\de_{\Lam}S_1-\beta_1 S_0}
-\frac{2}{\Lam^2}\frac{\delta S_1}{\delta\phi}\cdot
c'\cdot\frac{\delta(S_0-\hat{S})}{\delta\phi}+\frac{1}{\Lam^2}\frac{
\delta}{\delta\phi}\cdot
c'\cdot\frac{\delta(S_0-2\hat{S})}{\delta\phi}\label{scalar1loop},
\\ \nonumber
\eea
\etc
$\gamma_1$ and $\beta_1$ may now be extracted directly from
\eq{scalar1loop},  as specialised to the two-point and four-point
effective couplings, $\s{2}{}{\vec{p};\Lam}$ and $\s{4}{}{\vec{p};\Lam}$
respectively, once the renormalization conditions have been taken
into account.

We impose the wavefunction renormalization condition
in the new variables:
\be
\label{r2}
\s{2}{}{p,-p;\Lam} = \s{2}{}{0,0;\Lam}+ p^2 +O(p^4/\Lam^2).
\ee
Bearing in mind that the coupling constant has been scaled
out, we impose the condition
\be
\label{r4}
\s{4}{}{\vec{0};\Lam} = 1.
\ee
Both conditions \eq{r2} and \eq{r4} are already saturated at tree level.
(To see this it is sufficient to note that, since the theory is massless,
the only scale involved is $\Lam$.
Since $S^{(4)}_0$ is dimensionless
it must be a constant at null momenta, thus
$\s{4}{0}{\vec{0}; \Lam} = \s{4}{0}{\vec{0};
\Lam_0} = 1$. Similar arguments apply to $S^{(2)}_0$.)
Hence the renormalization condition implies that we must have
no quantum corrections to the four-point vertex at $\vec{p} =
\vec{0}$, or to the $O(p^2)$ part of the two-point vertex,
\ie
\be
\s{4}{n}{\vec{0}; \Lam} = 0 \quad{\rm and}\quad
\left.\s{2}{n}{p,-p;\Lam}\right|_{p^2} =0\qquad \forall n \geq 1,
\ee
where the notation $|_{p^2}$ means that one should take the coefficient
of $p^2$ in the series expansion in $p$.
The flow equations for these special parts of the quantum corrections
thus greatly simplify,
reducing to algebraic equations which then determine the $\beta_i$
and $\gamma_i$. In particular, from the flow of $S^{(4)}_1$
at null momenta:\footnote{Here and
later we suppress the $\Lambda$ dependence of the $S$ and $\sh$ vertices.}
\be\label{beta1}
\beta_1+2\gamma_1=\frac{8c'_0}{\Lam^2}\Big[1-\hat{S}^{(4)}(\vec{0})\Big]
\s{2}{1}{0}-\frac{1}{\Lam^2}\int_q c'({\textstyle {q^2 \over
\Lam^2}})\Big[S^{(6)}_0-2\hat{S}^{(6)}\Big](\vec{0},q,-q),
\ee
where $c'_0 = c'(0)$ and $\int_q \doteq \int {d^4 q \over (2 \pi)^4}$,
and from the flow of $S^{(2)}_1$ expanded to $O(p^2)$:
\be\label{gamma1}
\beta_1+\gamma_1= - \frac{1}{\Lam^2}\int_q c'({\textstyle {q^2 \over
\Lam^2}})
\Big[ \left.S^{(4)}_0-2 \sh^{(4)}\Big] (p,-p,q,-q)\right|_{p^2}.
\ee
Note that contrary to the standard text book derivation our one-loop
anomalous dimension is not zero, picking up a contribution from the
general field reparametrization \cite{latmor} induced by higher point terms in
$\sh$ and a contribution $-\beta_1$
due to the field rescaling \eq{rescale}.

In order to evaluate \eq{beta1}, we need to
calculate $\s{2}{1}{0}$ and $\s{6}{0}{\vec{0}, q, -q}$. We would also need
$\sh^{(4)}(\vec{0})$ and $\sh^{(6)}(\vec{0},q,-q)$, but we will see
that we can avoid using explicit expressions for them, and thus keep
$\sh$ general, by using the flow equations to express them in terms
of the effective vertices $S^{(4)}_0$ and $S^{(6)}_0$.

However, as explained in the previous section, our $\sh$ is not
completely arbitrary. Apart from some very general
requirements on the differentiability and integrability of its vertices,
for convenience we restrict $\sh$ to have only even-point vertices, as in
fact already used in eqs.\ (\ref{beta1}) and (\ref{gamma1}), and constrain
its two-point vertex so that the two-point effective coupling keeps the
same functional dependence upon $\Lambda$ as the bare one (as in
\eq{effact}). This last condition reads
\be \label{s20}
\s{2}{0}{p} = p^2 c^{-1} ({\textstyle {p^2\over \Lam^2}})
\ee
and from the two-point part of \eq{scalartree}, we immediately find
\be \label{shat20}
\sh^{(2)}(p) = p^2 c^{-1} ({\textstyle {p^2\over \Lam^2}}).
\ee

Let us start with the calculation of $\s{2}{1}{0}$. From \eq{scalar1loop},
its equation reads
\be
\ldl S_1^{(2)}(0) = \frac{1}{\Lam^2}\int_q c'({\textstyle {q^2 \over
\Lam^2}}) \Big[S^{(4)}_0 - 2 \sh^{(4)} \Big] (0,0,q,-q), \label{s12}
\ee
where eqs.\ (\ref{shat20}) and (\ref{s20}) have been already used to cancel
out the classical terms.
Pursuing our strategy, we get rid of $\sh^{(4)}$ by making use of the flow
equation for the four-point effective coupling at tree level
\be\label{sh40gen}
\ldl \s{4}{0}{\vec{p}} = {2\over \Lam^2} \sum_i
\frac{p_i^2 c'_{p_i}}{c_{p_i}} \sh^{(4)}(\vec{p}),
\ee
where $c_{p_i} \doteq c({p_i^2\over \Lam^2})$ and the invariance of
$\s{4}{0}{\vec{p}}$ under permutation of the $p_i$'s
(which it has without loss of generality) has been utilised.
Specialising the above equation to $\vec{p} = (0,0,q,-q)$, \eq{s12} becomes
\bea
\ldl \s{2}{1}{0} &=&\frac{1}{\Lam^2} \int_q
c'_q S_0^{(4)} (0,0,q,-q) -\int_q  \frac{c_q}{2 q^2} \ldl S_0^{(4)}
(0,0,q,-q)  \nonumber\\
&=& - \int_q {1\over 2 q^2} \ldl \Big(
c_q \,  S_0^{(4)}(0,0,q,-q) \Big) \nonumber\\
&=&-\ldl \int_q \frac{c_q \,  \s{4}{0}{0,0,q,-q}}{2 q^2}.\label{eqs120}
\eea
In 
the above, the derivative with respect to the cutoff may be taken after
integrating over the loop momentum since the integral is regulated both
in the ultraviolet and in the infrared as a result of the properties of
the effective couplings. \ceq{eqs120} may be now integrated
to give
\be \label{s2}
\s{2}{1}{0} = -\int_q \frac{c_q \, \s{4}{0}{0,0,q,-q}}{2 q^2},
\ee
with no integration constant since for a massless theory,
there must be no other explicit scale in the theory apart from the
effective cutoff.

Let us now move on to the tree-level six-point function. From
(\ref{scalartree}) we get
\bea
\ldl \s{6}{0}{\vec{0},q,-q} &=& \frac{4q^2}{\Lam^2}
\frac{c'_q}{c_q}\hat{S}^{(6)}(\vec{0},q,-q)\nonumber\\[3pt]
&&-\frac{8c'_0}{\Lam^2}\Big[1-\hat{S}^{(4)}(\vec{0}) \Big]
\s{4}{0}{0,0,q,-q}
+\frac{8c'_0}{\Lam^2}\hat{S}^{(4)}(0,0,q,-q)\nonumber\\[3pt]
&&-\frac{12}{\Lam^2}c'_q \, S^{(4)}_0(0,0,q,-q)
\Big[S^{(4)}_0-2\hat{S}^{(4)} \Big] (0,0,q,-q). \label{s6}
\eea
Using \eq{sh40gen}, and solving for $\hat{S}^{(6)}(\0,q,-q)$,
\bea\label{sh6}
\hat{S}^{(6)}(\0,q,-q)&=&\frac{\Lam^2}{4q^2} \frac{c_q}{c'_q} \left\{ \ldl
\s{6}{0}{\0,q,-q)} +\frac{8 c'_0}{\Lam^2}\Big [1-\hat{S}^{(4)}(\0) \Big]
\s{4}{0}{0,0,q,-q} \right.\nonumber\\[3pt]
&&-2c'_0 \frac{c_q}{q^2c'_q} \ldl \s{4}{0}{0,0,q,-q}\nonumber\\[3pt]
&&\left.-\frac{6}{q^2} \s{4}{0}{0,0,q,-q} \ldl \Big[ c_q \,
\s{4}{0}{0,0,q,-q} \Big] \right\}.
\eea
We will see that
substituting eqs.\ (\ref{s2}) and (\ref{sh6}) into \eq{beta1}
will cause almost all the non universal terms to cancel
out. The remaining ones will disappear once $\gamma_1$ is substituted
using \eq{gamma1}, leaving just the precise form of
the one-loop beta function.

Note that in \eq{sh6} and later, it appears
at first sight that we need to be able to take the inverse $1/c'_q$.
This would mean that in addition to the general restrictions on $\sh$
outlined earlier we would also require that $c'$ does not vanish at finite argument. In fact, we could arrange the
calculation more carefully so that $1/c'$ never appears, thus \eg here
we can recognize that only $c'_q \hat{S}^{(6)}(\0,q,-q)$ is needed for
\eq{beta1} and that from \eq{sh40gen}, $\ldl \s{4}{0}{0,0,q,-q}$ has
a factor of $c'_q$. For clarities sake, we will continue to write $1/c'$
in intermediate results but it is easy to check that all such inverses can be eliminated.

Returning to the calculation in detail,
the first term in (\ref{sh6}) and the $S^{(6)}_0$ term in
(\ref{beta1}) may be paired up into
\be \label{first}
\ldl \int_q \frac{c_q}{2 q^2} \, \s{6}{0}{\0,q,-q},
\ee
where again, due to the properties of the effective action vertices, the
order of the derivative and integral signs can be exchanged. Moreover, as
the integrand in \eq{first} is dimensionless, there cannot be any
dependence upon $\Lam$
after the momentum integral has been carried out, hence the result vanishes
identically!
Also, the second term in (\ref{sh6}), when substituted into
(\ref{beta1}), exactly cancels the first term of the latter once
(\ref{s2}) is used. One is then left with
\bea\label{beta1tilde1}
\beta_1+2\gamma_1&=&-c'_0 \int_q \frac{c_q^2}{q^4
c'_q} \ldl \s{4}{0}{0,0,q,-q} - 3 \int_q \frac{c_q}{q^4}
\s{4}{0}{0,0,q,-q} \ldl \Big\{ c_q \, \s{4}{0}{0,0,q,-q} \Big\}
\nonumber\\[3pt]
&=&-c'_0 \int_q \frac{c_q^2}{q^4
c'_q} \ldl \s{4}{0}{0,0,q,-q} - \frac{3}{2} \int_q \frac{1}{q^4}
\ldl \Big\{ c_q \, \s{4}{0}{0,0,q,-q} \Big\}^2.
\eea

In order to cancel out the first term in \eq{beta1tilde1}, the one-loop
contribution of the wave function renormalization coming from
\eq{gamma1} must be taken into account. Again making use of \eq{sh40gen}
to rid us of the hatted four-point coupling,
\be\label{s4hp2}
\frac{1}{\Lam^2}\,\sh^{(4)}(p,-p,q,-q)\Big|_{p^2}=
\frac{c_q}{4 q^2
c'_q} \ldl S^{(4)}_0 (p,-p,q,-q) \Big|_{p^2}\!-\, c'_0 \left(
\frac{c_q}{2q^2c'_q}\right)^2\!\! \ldl \s{4}{0}{0,0,q,-q},
\ee
and substituting back in \eq{gamma1},
\be\label{bibita}
\beta_1+\gamma_1 =\frac{1}{2} \ldl \int_q
c_q \left. \s{4}{0}{p,-p,q,-q} \right|_{p^2}
-\frac{c'_0}{2}\int_q c'_q \left(\frac{c_q}{q^2c'_q}\right)^2 \ldl
\s{4}{0}{0,0,q,-q}.
\ee
The first term on the right hand side of \eq{bibita} vanishes as it is a
dimensionless UV and IR convergent integral, and therefore $\gamma_1$
takes the form
\be\label{zed1}
\gamma_1 = -\beta_1-\frac{c'_0}{2}\int_q c'_q
\left(\frac{c_q}{q^2c'_q}\right)^2 \ldl \s{4}{0}{0,0,q,-q}.
\ee
Finally, substituting (\ref{zed1}) in (\ref{beta1tilde1}) yields
\bea
\beta_1&=&\frac{3}{2} \int_q \frac{1}{q^4}\, \ldl\, \Big\{ c_q
\s{4}{0}{0,0,q,-q} \Big\}^2\label{betaf}\\[3pt]
&=&-\frac{3}{2}{\Omega_4\over(2\pi)^4}
\int_0^{\infty}\!\!\! dq \, \de_q \left\{ c_q
\, \s{4}{0}{0,0,q,-q} \right\}^2 \nonumber\\[3pt]
&=&\frac{3}{16\pi^2},
\eea
which is the standard one-loop result \cite{PSGL}.\footnote{The term in
braces depends only on $q^2/\Lam^2$. $\Omega_4$
is the four dimensional solid angle. The last line follows from the
convergence of the integral and
normalisation conditions $c(0)=1$ and (\ref{r4}). As far as independence
with respect to the choice of cutoff function is concerned, this is
standard.}
Note that in the top equation the $\Lam$ derivative
cannot be taken outside the integral, as this latter would not then
be properly regulated in
the infrared. Moreover, had that been possible, it would have resulted in
a vanishing beta function, as the integral is actually dimensionless.

\chapter{Regularising Gauge Theories}

The first step towards an ERG approach for a quantum field theory
is to construct a regularised effective action, with a regulator
suppressing the high modes and maintaining the symmetries of the
theory. As far as a scalar field theory is concerned, the problem
to solve is quite easy and it was developed in the previous
chapter. It is well known that for gauge theories this task
represents a more complicated issue.

The notion standing at the base of the ERG is in fact the division
between small and large momenta (with respect to some effective
cutoff $\Lam$), being the high ones those that are integrated out.
This separation operated in the momentum space is at odds with the
concept of gauge invariance \cite{dattmor}. A way to notice it is
to consider a homogeneous gauge transformation $\Omega$ acting on
a field $\phi(x)$:
\be
\phi\to\Omega(x)\phi(x).
\ee
In the momentum space, $\phi(p)$ is
mapped through this transformation into a convolution with the
gauge transformation, and any division between low and high
momenta is not preserved by gauge transformations. In order to
overcome the problem, there are two options left: either one
breaks gauge invariance trying to recover it in the limit
$\Lam\to\infty$ or tries to find a generalisation of the ERG. The
former approach, which will be briefly reviewed in the next
section, has been the one mainly followed so far, as can be also
found in \cite{dattmor,bonvian,becchietal}. The second one, starts
on writing gauge invariant cutoff functions with
addition of Pauli-Villars fields, by A. Slavnov \etal, and is
continued by T.R. Morris \etal with the introduction of the
supergauge theory $SU(N|N)$ as a gauge invariant regulator for the
Yang-Mills theory. This will be reviewed in detail in the last few
sections of this chapter.

A gauge theory regulated in a gauge invariant way is then a solid
basis to build a flow equation capable to preserve this feature
while extracting information from it. This is going to be the
content of the last two chapters.

\section{Breaking the gauge symmetry and the Quantum Action Principle}\label{sec:breakGI}

If one chooses the first possibility, and introduces a scale
$\Lam$ to regularise the effective action, the result is that
whilst the classical action is invariant under the gauge
transformation, the cutoff effective action is not. The
consequence is a breaking of the effective Ward-Takahashi
identities, or Slavnov-Taylor identities, for the non-Abelian case. This
complicates the issue but it is not a problem as long as it is
possible to recover gauge symmetry when the cutoffs are removed.
Rephrasing it, it is not a problem if it is possible to identify a functional of the
effective action, representing the explicit breaking term, which
satisfies the equation $\Delta_{eff}[S_{\Lam},\Lam]=0$, in the
"physical" limit $\Lam=0$ and $\Lam_0\to\infty$.

In order to derive this symmetry breaking term, it is possible to
invoke the Quantum Action Principle. This method is used to study
the response of the action of a Quantum Field Theory under a field
transformation and it can be used to construct theories with a
given symmetry. In the case of gauge theories and for the present
purpose, one has to consider the response of the regularised
effective action under a gauge transformation and make sure that
the term arising from such a change is zero in the physical limit
described above. We will illustrate here just the procedure for
constructing an action symmetric under a simple transformation
which could then in principle be specified. Let us consider a
theory described by an action $S[\phi]$, and the corresponding
generating functional:
\be
Z[J]=\int\dephi e^{-S[\phi]+J_A\phi_A}, \ee where a source
term has been added to the action. Consider now the following
infinitesimal continuous transformation of the fields:
\be
\delta\phi_A=\epsilon P_A[\phi], \ee where $P_A$ are
(anticommuting) polynomials in the fields , which in the case of
gauge theories can correspond to a BRS transformation and
$\epsilon$ is an anticommuting parameter. Adding to the action a
source-type term for the variation of the field of the form
$-\eta_A P_A$ and performing the field transformation on the
generating functional, we get:
\begin{equation}\label{qap}
\int dx\ J_A\frac{\delta
Z}{\delta\eta_A}=\int\dephi\Delta[\phi,\eta]e^{-S[\phi]+\eta_A
P_A[\phi]}
\end{equation}
where we indicate with $\Delta$ the following:
\be
\Delta[\phi,\eta]=\int\!\! dx\ \frac{\delta^2
S}{\delta\phi_A(x)\delta\eta_A(x)}-\int\!\! dx\ \frac{\delta S
}{\delta\phi_A(x)}\frac{\delta S}{\delta\eta_A(x)}\ee As one can
notice, the first term is due to the Jacobian of the
transformation, while the second term takes into account the
change in the action due to the variation of the fields. The
response of the system is then given by the insertion of the local
operator $\Delta$. Eq.(\ref{qap}) is known as the Quantum Action
Principle.

For our purposes, as anticipated, in order to get $\Delta_{eff}$
one would have to follow a similar procedure, with a regularised
effective generating functional, performing a cutoff field
transformation. The full calculation for the present case will not
be shown here, and can be found in the literature (see \eg
\cite{simio}).

It is possible to prove that the breaking term obeys the following
equation:
\begin{equation}\label{breakingterm}
\Lam\de_{\Lam}\Delta_{eff}={\cal M}[\Delta_{eff}]
\end{equation}
where $\cal M$ is a linear operator. This implies that, if it is
possible to impose at some $\Lam_R$ zero boundary conditions for
eq.(\ref{breakingterm}), the breaking term vanishes at any $\Lam$.
The main point is then to set to zero at $\Lam_R$ those for the
relevant part of $\Delta_{eff}$. This procedure usually
overdetermines the vertices of $S_{\Lam_R}$ thus the number of
independent constraints has to be reduced making use of consistency
conditions (algebraic identities coming from anticommutativity of
the operator
$\frac{\delta}{\delta\eta_A}\frac{\delta}{\delta\phi_A}$).

In this picture it is crucial the way the relevant parts are
defined. If the boundary conditions are set at $\Lam_R\neq 0$, the
relevant parts of $\Delta_{eff}(\Lam_R)$ can be extracted by
expanding the vertices around zero momenta even in presence of
massless particles. What one gets at this point is that the
consistency conditions constrain some of the couplings in the
relevant part of $\Delta_{eff}(\Lam_R)$, which via a tuning of the
relevant couplings of the effective action must match with their
set of relations. This procedure is known as fine-tuning of the
parameters.

If one instead decides to impose the boundary conditions at the
physical point $\Lam_R=0$, if the theory includes massless
particles, one has to impose non-vanishing subtraction points.
This causes a mix of relevant and irrelevant vertices in the
consistency conditions spoiling their power.

Once all the details of this procedure have been set up, one is
left with a fine-tuning equation. If the equation is solvable, the
symmetry (the gauge symmetry in our case) is implemented at the
quantum level and no anomalies appear. A more detailed description
of the above methods can be found in the literature (see \eg
\cite{becc}). The main problem now is that the equation mentioned
above is usually difficult to solve, even at the first non trivial
order in perturbation theory.

There are successful attempts of avoiding the task to solve the
fine-tuning equation by fixing proper boundary conditions to the
RG equation (see Bonini \etal in \cite{becchietal}), but we will
not discuss them here.

Instead of doing so, since all these difficulties come from the
incompatibility between gauge invariance and the division of high
and low momenta, we try to follow the second approach mentioned in
the previous section. Following the lead of T.R.Morris we will try
here to describe first, a possible way to regularise gauge
theories without breaking their symmetry, and then how to
generalise the RG method in order to preserve the symmetry in the
flowing effective action. This will be done through the
construction of a generalised flow equation, gauge invariant
itself, capable to describe a gauge invariant flowing effective
action. Before we start, it is worth having an overview of other efforts towards a gauge invariant regulator.

\section{Higher derivatives and P-V fields}\label{sec:1.1}

The first step is to regularise the action in such a way as to
preserve its symmetry. As we have seen in the previous chapters,
in a simple case such as the scalar theory, there are many
possible choices of regularising the action, involving the
introduction of cutoff functions. These function have the r\^ole
to cut the high modes in the loop integrals in order to make them
finite. There is a wide choice for the cutoff function which can
be chosen to be either a step function (sharp cutoff) or a smooth
one as long as it preserves the symmetry of the scalar action
(for example for a single scalar field this involves the request of being
even in the fields). One possible choice, as can be found in Chapter 2, is to
introduce in the kinetic term a function in the derivatives (which
in the momentum space is a function of the momenta).

When it is the case of gauge invariance, this last requirement
causes troubles for the reason described in the previous sections.
Following the example set by the scalar case, the first attempt
towards this goal, was to introduce as a cutoff, a function in the
covariant derivative, rather than in the ordinary ones. The method
starts from the observation that a kinetic-like term (quadratic in
the fields) containing higher derivatives modifies the
propagators, conferring them a better behaviour at high momentum.
A term like this,
\begin{eqnarray}\label{hd}
\partial_{\mu}\phi\ \partial^{\mu}\phi&\to&\partial_{\mu}\phi\
c^{-1}\left(-\partial^2/\Lambda^2\right)\partial^{\mu}\phi ,
\end{eqnarray}
substituted in the Lagrangian in place of the usual kinetic term,
gives, in the momentum space, a correction to the propagator which
amounts to change the ordinary one (e.g. in the massless scalar
field) as
\begin{equation}\label{prop}
\dfrac{1}{p^2}\to\dfrac{c(p^2/\Lambda^2)}{p^2}
\end{equation}
The new propagator certainly leads to convergent momentum
integrals for a suitable choice of the function appearing in
(\ref{hd}) (for example if $c^{-1}$ is chosen to be a
polynomial for a certain choice of its degree). The idea, for a
scalar field is as simple as that, and the physical information is
restored as $\Lambda\to\infty$: at finite $\Lambda$ all loop
diagrams (responsible for divergences) are finite and the
calculations are made at this point. The physical quantities can
be calculated with a proper renormalisation condition  and sending
the scale $\Lambda$ to infinity gives a finite answer for them.

\subsection{Covariantisation}\label{sub:1.1.1}

\begin{figure}[h]
\psfrag{psi}{$\psi$}
\psfrag{phi}{$\phi$}
\begin{center}
\includegraphics[scale=.8]{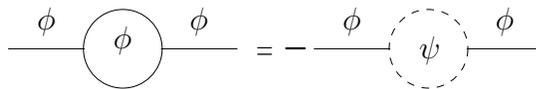}
\end{center}
\caption{Pauli-Villars field cancel out residual 1-loop
divergencies}\label{fig:1loop}
\end{figure}

This does not work for gauge theories. As we said before, a term
like the one in (\ref{hd}) would break the invariance. One way, as
we cited before, to deal with the problem is to break the
invariance and restore it when the scale is sent to infinity.
Nevertheless, since we want to follow the other path and write a
manifestly covariant Exact Renormalisation Group (ERG), our bare
action must be gauge invariantly regulated. A first attempt
embracing this philosophy was to introduce a cutoff function in
the covariant derivative, as we mentioned in the previous section,
rather than in the ordinary ones: instead of the term (\ref{hd})
we write
\begin{equation}\label{HD}
c\left(-\frac{\partial^2}{\Lambda^2}\right) \to
c\left(-\frac{\nabla^2}{\Lambda^2}\right)
\end{equation}
This is known as the higher {\it covariant derivatives}
regularisation \cite{slav,z-j}. It is known that this method
cannot work by itself since it creates a new problem: when the
higher derivatives are covariantised, divergences at one loop are
still present due to further interactions coming in with them. One way out is to introduce by hand massive (mass of order
cutoff $\Lambda$) fields with opposite spin-statistic (the so
called $Pauli-Villars$ fields) capable of cancelling these 1 loop
divergencies. Due to their statistic, they provide a "-" sign in
loops as it is shown in fig.\ref{fig:1loop}
At high momenta, in fact, when the integrals at one-loop diverge,
the propagators and the interactions of the two different fields have the same
behaviour and due to the sign difference, they cancel each other.
Once the calculation is carried out with finite integrals and the
renormalisation conditions have been applied, sending the cutoff
to infinity would eventually restore the physics, since these
fields decouple from the physical ones in this limit. Higher
covariant derivatives and Pauli-Villars (PV) fields combined
together provide a good scheme to regulate gauge theories, as is
considered in \cite{slav}. This regularisation scheme still
creates problems though. First of all when these PV fields came in
external lines there were divergences that even if discarded
assuming them non-physical, caused overlapping divergences at
higher loops containing these diagrams \cite{warrash}. Moreover,
even though this problem was solved by Bakeyev and Slavnov in
\cite{bek-slav}, the method was not straightforwardly applicable
to the RG equation approach.

A first attempt of overcoming this problem was presented in
\cite{mor:erg1,mor:erg2}, where a gauge invariant flow equation
for a free Yang-Mills (YM) theory, regulated with higher covariant
derivatives and PV fields, has already been studied and the 1-loop
$\beta$-function for $SU(N)$ YM at $N=\infty$, has been calculated
for the first time without fixing the gauge. The work was based on
insisting that the regularisation respected the flow, adding
higher order interactions for the PV fields (instead of adding
them just as mass terms), and with the aid of an auxiliary scalar
field. The regularisation was only valid for 1-loop diagrams and
at $N=\infty$ and it could not allow to perform calculations
beyond this order. On the way of doing this, it appeared clear
that all the right content of fields was contained in a bigger
group, called $SU(N|N)$, which in its bosonic sector, contains
SU(N)$\otimes$SU(N)$\otimes$U(1). Through a Higgs-type mechanism
of spontaneous symmetry breaking through an auxiliary scalar
field, one of the two $SU(N)$ sectors results in a YM theory gauge
invariantly regulated by a naturally combined action of higher
covariant derivatives and PV fields. This group and its
application for the present purposes will be better described in
the next section.

\section{Regulating via $SU(N|N)$ Gauge Theory}\label{sec:1.2}

\subsection{$SU(N|N)$ superalgebra}\label{sub:1.2.1}

Since we will have to deal with a $SU(N|N)$ gauge theory, it is
worth spending few words on $SU(N|N)$ group and related algebra.
$SU(N|N)$ is a graded Lie group, whose elements $U$ an be
represented in the exponentiated form as: \be\label{U} {\cal
U}=\exp{(i{\cal H})} \ee The set of elements $\cal H$ belong to
the corresponding Lie superalgebra $SU(N|N)$. An element of the
superalgebra, can be represented with a $2N\times 2N$ Hermitian
matrix $\cal H$: \be\label{H} {\cal H}=\left(\!\!
\begin{array}{cc}
H^1_N&\theta\\[.3cm]
\theta^{\dagger}&H^2_N
\end{array}\!\! \right)\in SU(N|N)
\ee The two $H^i_N$ are Hermitian $N\times N$ matrices whose
elements are bosonic complex numbers (commuting i.e. ordinary
$\CC$), and $\theta$ is an $N\times N$ matrix filled up with
anticommuting fermionic (Grassmann) numbers. A matrix such as the one
described in (\ref{H}) belongs to the algebra $SU(N|N)$ if it
satisfies the additional requirement of being ``supertraceless'':
\be
\mbox{str} {\cal
H}=\mbox{tr}\mbox{H}^1_N-\mbox{tr}\mbox{H}^2_N=0\label{str} \ee
The {\it Supertrace}, defined in (\ref{str}), is the natural
replacement of the trace for ordinary matrices. It is in fact
cyclically invariant because it compensates the sign picked up by
commuting the Grassmann components:
\be\label{cipeciop}
\mbox{str}XY=\mbox{str}YX \ee where $X$ and $Y$ are two general
supermatrices. In this way the supertrace of commutators vanishes,
and makes it invariant under the adjoint action of the group. Once
the matrix $\sigma_3$ is defined:
\be
\sigma_3=\left(\!\!
\begin{array}{cc}
\one_N&0\\
0&-\one_N
\end{array} \!\! \right)\label{s3}
\ee where $\one$ is the $N\times N$ identity matrix, the
supertrace of a matrix $\cal H$ can be rewritten in terms of it as
\be
\mbox{str}({\cal H})=\mathrm{tr}(\sigma_3{\cal H}). \ee The
request of being supertraceless for elements of $SU(N|N)$ is the
natural extension of the request on the elements of the ordinary
$SU(N)$ algebra: it guarantees that $\cal U$ in
eq.(\ref{U}) has unit superdeterminant. The supertraceful matrix $\sigma_3$ generates a $U(1)$
group absent from $SU(N|N)$. This $U(1)$ group though is not
orthogonal to $SU(N|N)$ because being $S_{\alpha}$ a generic
generator of $SU(N|N)$, str$(\sigma_3S_{\alpha})$ can be non zero in
the case of the identity. Moreover, even though $\sigma_3$
commutes with all the bosonic generators of $SU(N|N)$, it does not
commute with all the fermionic ones (unlike the case of
$SU(N)$ with the $U(1)$ generated by the traceful identity). This
confers to $SU(N|N)$ a different character, which will be used in
the symmetry breaking mechanism described later. The bosonic
subalgebra of $SU(N|N)$ is, as we have anticipated, $SU(N)_1\times
SU(N)_2\times U(1)$ the latter being the subgroup generated by the
unity matrix (which, since supertraceless, belongs to the
algebra).

We will consider the generators to be Hermitian matrices with
complex number entries. The superalgebra will be then defined
through a set of commutation and anticommutation rules (the
Grassmann character will be carried by the coefficients). Let us
consider an element of the Lie algebra as a linear combination of
the generators: \be\label{sunnel} H=S_{\alpha}\ H^{\alpha} \ee
where the $S_{\alpha}$'s are the generators:
\be
S_{\alpha}= \left\{\begin{array}{ll}
              \one_{2N}&\alpha=0\\
               B_a&\alpha=1,\dots, 2N^2-2\\
               F_a& \alpha=2N^2-1,\dots, 4N^2-2
            \end{array}\right.
\ee
$\one_{2N}$ is the $2N\times 2N$ identity matrix, $B_a$
are the $2N^2-2$ block diagonal traceless and supertraceless generators (along the directions of the bosonic
components) and the
$F_{\alpha}$'s are the $2N^2$ off-diagonal generators (fermionic components). The
commutation and anticommutation rules which define the algebra are:
\be\label{commrules}
\begin{array}{l}
1)\ [B_a,B_b]={\cal F}_{ab}^{\phantom{ab}c}\ B_c\\[5pt]
2)\ [B_a,F_b]={\cal G}_{ab}^{\phantom{ab}c}\ F_{c}\\[5pt]
3)\ \{F_a,F_b\}={\cal D}_{ab}^{\phantom{ab}c}\ B_c
+{\cal H}_{ab}\ \one\\[5pt]
4)\ [\one,*]=0;\\[3pt]
\phantom{4)\ } (\mbox{where\ `}*\mbox{'\ stands\ for\ any\ element})
\end{array}
\ee
All the generators are matrices with commuting numbers as entries, being the Grassmann
 character carried by the parameters. Here, ${\cal
 F}_{ab}^{\phantom{ab}c}$, ${\cal G}_{ab}^{\phantom{ab}c}$, ${\cal
 D}_{ab}^{\phantom{ab}c}$ and ${\cal H}_{ab}$ are coefficients which define
 the algebra $SU(N|N)$. Since one can get anything from first principles by using the fact that
 the generators $S_\alpha$ span the space of Hermitian matrices, it is not
 important to specify them here. However to
 be more clear, a specific choice of a basis is considered in Appendix \ref{app:bla},
 in order to write the relations of eq. (\ref{commrules}) all in terms of
 the structure constants of $SU(N)$, $f$ and $d$.
 
It is useful for future reference to list also the commutation and
aticommutation relations of the generators of $SU(N|N)$ with the generator of the $U(1)$, $\sigma_3$ defined in eq.(\ref{s3}):
\be\label{sigcom}
\begin{array}{l}
5)\ [\sigma_3,B_a]=\{\sigma_3,F_a\}=0\\[5pt]
6)\ [\sigma_3,F_a]={\cal G}_{3 a}^{\phantom{ab}b}\ F_b
\end{array}
\ee
First let us split the generators as $S_{\alpha}\equiv(\one,T_A)$\footnote{The $T_A$ are the traceless and supertraceless generators and span the
same space of matrices as $B_a$ and $F_a$}. It is
now useful to define the Killing super-metric as:
\be
h_{\alpha\beta}=2\ \mbox{str}(S_{\alpha}S_{\beta})
\ee
$h_{\alpha\beta}$ is symmetric when either index is bosonic and
antisymmetric when they are both fermionic:
\be\label{sa}
h_{\alpha\beta}=(-)^{f(\alpha)f(\beta)}h_{\beta\alpha}
\ee
where $f(\alpha)$ is 0 if the index is bosonic and 1 if it is fermionic. 
The normalisation of the generators is defined via the following form of
the metric (where all elements not indicated are zero):\\
\begin{minipage}[!h]{6in} 
\bea\label{sunmmet}
h_{\alpha\beta} = 
\left( \mbox{ {\small
\begin{tabular}{c|ccc|ccc|ccccc} 
$0$& &&& &&& &&&& \\ \hline
& $1$&&& &&& &&&& \\ 
& &$1$&& &&& &&&& \\ 
& &&$\ddots$& &&& &&&& \\ \hline
& &&& $-1$&&& &&&& \\
& &&& &$-1$&& &&&& \\ 
& &&& &&$\ddots$& &&&& \\ \hline
& &&& &&& $0$& $i$& && \\ 
& &&& &&& $-i$& $0$& && \\ 
& &&& &&& && $0$& $i$&  \\ 
& &&& &&& && $-i$& $0$&  \\ 
& &&& &&& &&&& $\ddots$ \\
\end{tabular} } }
\right) \\
\underbrace{\hspace{0.8cm}}_{U(1)}\,
\underbrace{\hspace{2cm}}_{SU_1(N)}\,
\underbrace{\hspace{2.6cm}}_{SU_2(N)}\,
\underbrace{\hspace{4cm}}_{\mathrm{Fermionic}}
\hspace{0.2cm} \nonumber \\ \nonumber
\eea \end{minipage}
This super-metric has no inverse due to the presence of $\one$ in the
generators (which gives a column and row of zeros in the matrix above). It is possible to define $SU(N|N)$
consistently with or without the identity matrix, changing the definition
of the commutators (see \cite{bars}). Here the
definition including it will be considered. However, we will see that the gauge theory constructed on this group
will decouple the component in this direction. Specialising to just the
space including the $T^A$ generators,
we can consider the Killing super-metric in this subspace, which is
invertible and defined as:
\be
g_{AB}=2\mbox{str}(T_AT_B)=h_{AB}
\ee
its inverse is defined by
\be
g_{AB}g^{BC}=g^{CB}g_{BA}=\delta^C_A
\ee
$g_{AB}$ can be used to lower or raise indices as in
\be\label{lowrise}
X^A=X_Bg^{AB}
\ee
Since the ordering of the indices of the super-metric is important (see
eq.(\ref{sa})), it is worth commenting that
in (\ref{lowrise}) the sum is on the second index and in general:
\be
X^A=X_Bg^{AB} \neq X_B g^{BA}=(-)^{f(A)f(B)}X_B g^{AB} \ee For the
generators we have the dual relation given by: \be\label{dual}
T^A=T_B g^{BA}
\ee
Another useful relation which holds for the
generators of $SU(N|N)$ is finally the completeness relation:
\be\label{complrel} (T^A)^i_{\ j}(T_A)^k_{\
l}=\frac{1}{2}\delta^i_{\ l}(\sigma_3)^k_{\
j}-\frac{1}{4N}\left[\delta^i_{\ j}(\sigma_3)^k_{\
l}+(\sigma_3)^i_{\ j}\delta^k_{\ l}\right] \ee 
(see App. \ref{app:compl} for a derivation.) This is most usefully cast in the following forms
\bea
\str(X T_A) \, \str(T^A  Y) &=& {1 \over 2} \,\str(X Y) 
- {1 \over 4N} \left[ \tr X \, \str Y + \str X \, \tr Y \right],
\label{comptree} \\[2mm]
\str(T_AXT^A Y) &=& {1 \over 2}\, \str X \, \str Y - {1 \over
4N} \tr (XY +YX), \label{comploop}
\eea
for arbitrary supermatrices $X$ and $Y$. Let us now consider
the adjoint representation of the group. An element of it can be
written as: 
\be\label{adj} 
{\cal M}={\cal M}^{\alpha} S_{\alpha}={\cal M}^0\ \one+{\cal M}_B^a\ B_a +{\cal M}_F^a\ F_a 
\ee 
where the $S_{\alpha}$'s are $4N^2 -1$ ($2N\times 2N$) matrices of $SU(N|N)$. An
element of the adjoint transforms, under an infinitesimal
transformation of the group $\omega=\omega^A T_A=\omega_B^a\
B_a+\omega_F^a\ F_a$ as follows:
\be\label{adjtr} \delta {\cal M}=-i[{\cal M},\omega]. \ee In components, given the
commutation and anticommutation rules of the group, it has the
following form: 
\be\label{adjcom} 
\left\{\begin{array}{l} 
i\delta {\cal M}_B^c={\cal M}^a_B \omega^b_B\ {\cal
F}_{ab}^{\phantom{ab}c}+{\cal M}^b_F \omega^a_F\ {\cal
D}_{ab}^{\phantom{ab}c}\\ 
i\delta {\cal M}_F^c={\cal M}^a_B \omega_F^b\ {\cal
G}_{ab}^{\phantom{ab}c}+{\cal M}^a_F\omega^b_B\ {\cal
 G}_{ba}^{\phantom{ab}c}\\
i\delta {\cal M}^0={\cal M}^a_F \omega^b_B\ {\cal H}_{ab}\\
\end{array}\right.
\ee\\
For future reference it can be useful to consider also the
$2N\otimes \overline{2N}$ representation\footnote{Unlike $SU(N)$, the
group $SU(N|N)$ is indecomposable, thus this representation is not the
Adjoint$\oplus$ the singlet}. An element of it can be represented as:
\be\label{adj+1}
{\cal C}={\cal C}^{\alpha}S_{\alpha}+{\cal C}^3\sigma_3
\ee
for it also the transformation will be of the form of eq.(\ref{adjtr}) and
in components is:
\be\label{adj+1com}
\left\{\begin{array}{l}
i\delta {\cal C}_B^c={\cal C}^a_B \omega^b_B\ {\cal
F}_{ab}^{\phantom{ab}c}+{\cal C}^b_F \omega^a_F\ {\cal
D}_{ab}^{\phantom{ab}c}\\ 
i\delta {\cal C}_F^c={\cal C}^a_B \omega_F^b\ {\cal
G}_{ab}^{\phantom{ab}c}+{\cal C}^a_F\omega^b_B\ {\cal
 G}_{ba}^{\phantom{ab}c}+ {\cal C}^3 \omega^a_F\ {\cal G}_{3 a}^{\phantom{3a}c}\\
i\delta {\cal C}^0={\cal C}^a_F \omega^b_B\ {\cal H}_{ab}\\
i\delta {\cal C}^3=0\\
\end{array}\right.
\ee 
The last line shows how the component along $\sigma_3$ does not transform
under $SU(N|N)$\footnote{However, $\sigma_3$ is not a singlet of $SU(N|N)$ though (as
$\one$ is for $SU(N)$), since it takes part in the transformations of the
other components}.

Having described here the main properties of this graded group and of its
Lie algebra, it is now possible to move onto the description of the
regularisation scheme adopted making use of it.

\subsection{Regularisation: Gauge group and Higgs-type mechanism}\label{sub:1.2.2}

 Instead of working just with the $SU(N)$ gauge field, which we write as
 $A^1_\mu(x)\equiv A^1_{a\mu}\tau^a_1$, where $\tau^a_1$ are the $SU(N)$
 generators orthonormalised to $\tr(\tau^a_1\tau^b_1)=\delta^{ab}/2$, we
 embed it in a $SU(N|N)$ supergauge field \cite{su:rome,su:pap}:
 \be
 \label{defA}
 \A_\mu = {\A}^{0}_{\mu} \one
 + \left( \!\! \begin{array}{cc}
                    A^{1}_{\mu} & B_{\mu} \\
                    \bar{B}_{\mu} & A^{2}_{\mu}
                    \end{array} \!\!
             \right).
 \ee
 Here we have written $\A$ as an element of the $SU(N|N)$ Lie
 superalgebra, using the defining representation, \ie as a supermatrix
 with bosonic block diagonal terms $A^i$ and fermionic block
 off-diagonals $B$ and $\bar{B}$, together with the central term
 $\A^0\one$. As required by $SU(N|N)$, the
 supermatrix (and thus also $\A$) is supertraceless, \ie $\tr A^1 - \tr
 A^2 =0$.  This excludes in particular $\sigma_3$, defined in eq.(\ref{s3}),
 from the Lie algebra. From now on we will write simply $\sigma=\sigma_3$. 
 The supermatrix is in addition also traceless,
 the trace having been parametrised by $\A^0$.
 Equivalently, as we have seen in the previous section, we can introduce a complete set of traceless and supertraceless
 generators $T_A$ (normalised as in \eq{sunmmet}) and thus expand $\A$ as
 \be
 \label{expandA}
 \A_\mu = {\A}^{0}_{\mu} \one + \A^A_\mu T_A.
 \ee
 The $B$ fields are wrong statistics gauge fields. They will be given a 
 mass of order the cutoff $\Lambda$. The supergroup $SU(N|N)$ has 
 $SU(N)\times SU(N)\times U(1)$ as its bosonic subgroup.
 $A^2_\mu(x)\equiv A^2_{a\mu}\tau^a_2$ is the gauge field for the second 
 $SU(N)$, and $\A^0$ is the $U(1)$ connection.
 Interactions are built via commutators, using the covariant derivative:
 \be
 \nabla_\mu = \partial_\mu -i\A_\mu,
 \ee
The coupling constant $g$ does not appear in the definition of the
 covariant derivative, as it usually does, because it is considered scaled
 out. This rescaling of the fields is proved to be useful and a more detailed discussion about this issue is presented in section \ref{sub:rescalg}
 Thus the superfield strength is given by 
 $\F_{\mu\nu}=i[\nabla_\mu,\nabla_\nu]$. The kinetic term will be 
 regularised by higher derivatives which thus take the form:
 \be
 \label{keA}
 \str\ \F_{\mu\nu} \left(\nabla\over\Lambda\right)^n\!\!\!\cdot\F_{\mu\nu},
 \ee
 (where the dot means $\nabla$ acts by commutation.
 In practice we will add the higher derivatives as a power series with
 coefficients determined by the cutoff function $c$). The supertrace, which,
 from the discussion around (\ref{cipeciop}), is 
 necessary to ensure $SU(N|N)$ invariance, forces the kinetic term for $A^2$ 
 to have wrong sign action, leading to negative norms in its Fock space 
 \cite{su:pap}.  

 As can be seen from \eq{expandA}, $\A^0$ does not appear in the kinetic 
 term. Providing the interactions can be written as $\str(\A \,\times$
 commutators), $\A^0$
 will not appear anywhere in the action. More generally we will need to
 impose its non-appearance as a constraint, since otherwise $\A^0$ has
 interactions but no kinetic term and thus acts as a 
 Lagrange multiplier resulting in a non-linear constraint on the theory,
 which does not look promising for its use as a regularisation method
 for the original $SU(N)$ Yang-Mills. 

 On the other hand, if the constraint is satisfied, $\A^0$ is then
 protected from appearing by a local ``no-$\A^0$'' shift symmetry:
 $\delta\A^0_\mu(x)=\Lambda_\mu(x)$, which implies in particular that
 $\A^0$ has no degrees of freedom.
 Together with supergauge invariance the theory is then invariant under
 \be
 \label{Agauged}
 \delta\A_\mu = \nabla_\mu\cdot\omega +\Lambda_\mu \one.
 \ee
 The effect of the no-$\A^0$ symmetry is to dynamically define the gauge
 group as the quotient $SU'(N|N)$ $=$ $SU(N|N)/U(1)$, 
 in which Lie group elements are identified modulo addition of an 
 arbitrary multiple of $\one$. 

 An alternative and equivalent formulation \cite{su:pap} is to pick coset 
 representatives, which can for example be taken to be traceless, so
 that $\A^0$ is set to zero, and thus discarded. (This is the strategy
 used in ref. \cite{Berkovits} to define a $SU'(N|N)$ sigma model.
 Incidentally this paper contains arguments for finiteness of these
 models which are similar to those given for $SU(N|N)$ gauge
 theory in \cite{su:pap}.\footnote{We thank Hugh Osborn for drawing our
 attention to this paper}) In this reduced representation, \eq{Agauged} 
 is replaced by Bars' solution \cite{bars}:
 \be
 \label{gaugeBars}
 \delta\A_\mu = [\nabla_\mu,\omega]^* \equiv [\nabla_\mu,\omega]
 -{\one\over2N}\tr[\nabla_\mu,\omega].
 \ee
 The *bracket replaces the commutator as a representation of the Lie
 product so in particular $\F_{\mu\nu}=i[\nabla_\mu,\nabla_\nu]^*$
 \cite{su:pap}. 

 The lowest dimension interaction that violates no-$\A^0$ symmetry
 contains four superfield strengths, for example:
 \be
 \label{counterX}
 \str\ \left(\F_{\mu\nu}\right)^2 \!\left(\F_{\lambda\sigma}\right)^2.
 \ee
 Such terms are not invariant under the `Bars*' \eq{gaugeBars}, either.
 Since \eq{counterX} is already irrelevant, no-$\A^0$ symmetry is 
 automatic for the conventional supergauge invariant bare action
 of ref. \cite{su:pap}. Here there is no such bare action, and
 interactions are generated by a largely unspecified exact RG, so we
 need to impose no-$\A^0$ as an extra constraint.

 We introduce a superscalar field
 \be
 \label{defC}
 \C = \left( \begin{array}{cc} C^1 & D \\
                               \bar{D} & C^2       
                \end{array}
        \right)
 \ee
 in the fundamental $\otimes$ its complex conjugate representation,
 equivalently as a matrix in the defining representation of $U(N|N)$
 \cite{su:pap}. Under supergauge transformations
 \be
 \label{Cgauged}
 \delta\C = -i\,[\C,\omega].
 \ee
 In the Bars* representation we do not replace this 
 by a *bracket, since commutators are necessary for powers of $\C$
 (appearing in its potential) to transform covariantly \cite{su:pap}. 
 However, as in ref. \cite{su:pap}, 
 since working with the full cosets seems more elegant, we will employ
 \eq{Agauged} and the full representation in this thesis.

 We will arrange for $\C$ to develop a vacuum expectation value along the
 $\sigma$ direction through an appropriate Higgs-type potential, so that classically $<\C>\ =
 \Lambda\sigma$.\footnote{Later however 
 we will use an unconventional normalisation for $\C$.}
 This spontaneously breaks $SU(N|N)$ down to its $SU(N)\times SU(N)\times U(1)$ bosonic subgroup and provides the fermionic 
 fields $B$ and $D$ with masses of order $\Lambda$. In usual unitary gauge
 interpretation, $D$  is the would be
 Goldstone mode eaten by $B$. However, since we will not gauge fix, they
 instead gauge transform into each other and propagate as a composite unit (see Appendix \ref{App:intwine}). The reason why the fermionic components of
 the $\A$ super-gauge field
 (the $B$'s) get a mass, is the Higgs mechanism, being them the fields along the
 broken gauge generators. The $D$ field (fermionic component of the
 super-Higgs field $\C$), being the component of the Higgs along the direction where the symmetry is
 broken, is the Goldstone boson and its kinetic term vanishes at $p=0$, as it is stated by the choice made in
 eq.(\ref{hatdd}).  However, since the fields $B$ and $D$ are coupled, if
 we diagonalise their kinetic terms, we can indeed notice, that the
 diagonalised mass matrix describes $B$ and $D$ as two massive particles
 with masses of order $\Lambda$. Moreover since the physical mass of a particle
 corresponds to the pole in its propagator written in the Minkowski space,
 $D$ can be regarded as a massive field, since its zero point wine (\ie
 effective propagator, see App. \ref{App:intwine}) does not have a massless pole, as one can notice from eq. (\ref{app:dd}) (the only field which does is the
 bosonic gauge field $A$). In fact, the coupled fields $B$ and $D$, have
 decoupled effective propagators or, in other words, the coupled two-point functions $BB$ and $DD$ and the cross term kinetic term $BD\sigma$, have uncoupled inverses in the transverse space. 

Finally, we arrange for the remaining `Higgs'
 fields $C^i$ also to have masses of order $\Lambda$. This is done here by the
 choice made in eq.(\ref{hatcc}). The two point $C$ vertex is chosen to be
 non vanishing at $p=0$, and the coefficient is chosen positive ($\lambda
 >0$), so that it is a mass term.

All the information that was encoded in the regularisation scheme for the
gauge invariant effective action of ref. \cite{su:pap}, will be here mainly
contained in the choice of the two-point functions, eqs.(\ref{hataa})-(\ref{hatbds})

 In ref.  \cite{su:pap}, it was proved by conventional methods that if the kinetic
 term of $\A$ is supplied with covariant higher derivatives 
 (parametrised by the cutoff function $c$) enhancing
 its high momentum behaviour by a factor $c^{-1}\sim p^{2r}/\Lambda^{2r}$, 
 and the
 kinetic term of $\C$ has its high momentum behaviour similarly enhanced
 by $\ct^{-1}\sim p^{2\rt}/\Lambda^{2\rt}$, then providing
 \be
 \label{inequalities}
 r-\rt > 1\quad{\rm and}\quad \rt>1,
 \ee
 all amplitudes are ultraviolet finite to all orders of perturbation theory.
 Since the underlying theory is renormalisable, the 
 Appelquist-Carazzone theorem implies that at energies much
 lower than the cutoff $\Lambda$, the remaining massless fields $A^1$
 and $A^2$ decouple. In this way, this framework was used as a regularisation
 of the original $SU(N)$ Yang-Mills theory carried by $A^1$.

 In brief, the reasons for the above facts are as follows. Providing
 eqs.  \eq{inequalities} hold, all divergences are superficially
 regularised by the covariant higher derivatives, except for some
 `remainders' of one-loop graphs with only $\A$ fields as external legs
 and only four or less of these legs. These remainders form a symmetric
 phase contribution, in the sense that the superficially divergent
 interactions between $\C$ and $\A$ are just those that come from $\C$'s
 covariant higher derivative kinetic term, whilst all terms containing a
 $\sigma$ from the breaking are already ultraviolet finite by power
 counting.  For three or less external $\A$ legs the remainders vanish
 by the supertrace mechanism:  the fact that in the unbroken theory, the
 resultant terms contain $\str\A=0$ or $\str\one=0$.  By manifest gauge
 invariance, the four point $\A$ remainder is then actually totally
 transverse, which implies that it is already finite by power counting.
  
 The decoupling of $A^1$ and $A^2$ follows from the unbroken local
 $SU(N)\times SU(N)$ invariance since the lowest dimension effective
 interaction
 \be
 \label{12ints}
 {1\over \Lambda^4}\,\tr\left(F^1_{\mu\nu}\right)^2 
 \tr\left(F^2_{\mu\nu}\right)^2
 \ee
 is already irrelevant \cite{su:pap,applcar}.

 Actually, there are a number of differences between the treatment we
 give here and that of ref. \cite{su:pap}.  Since ref. \cite{su:pap}
 followed a conventional treatment, gauge fixing and ghosts were
 introduced, with a corresponding higher derivative regularisation for
 them; longitudinal parts of the four point $\A$ vertex were then
 related to ghost vertices using the Lee Zinn-Justin identities, which
 were separately proved to be finite. Also, a specific form of bare
 action and covariantisation was chosen.

 Here we do not fix the gauge and the regularisation scheme is much more
 general.  As well as not specifying the covariantisation or the bare
 action (see below) there is anyway much more freedom in introducing
 interactions via the flow equation. We shall not here supply a rigorous
 proof that up to appropriate restrictions, the flow equation is finite.
 Since we never have to specify the details, we only need to {\sl
 assume} that this is true for at least one choice. However, we take
 care that the scheme as described above is qualitatively correctly
 implemented. Where we do have to explicitly compare terms we can use
 \eq{inequalities} as a guide, although it should be borne in mind that
 cutoff functions with non-power law asymptotics, for example
 exponential, could also be used.\footnote{The proof given in ref.
 \cite{su:pap} could also be easily extended to these cases.}  In
 practice, it is easy to see at one loop that the high energy
 cancellations are occurring as expected.

In this scheme, higher covariant derivative and P-V fields come out
naturally combined together and $SU(N|N)$ is proved to be a finite theory
\cite{su:pap}. This results in a $SU(N)$ gauge invariantly regulated theory
suitable for a RG flow equation approach. The purpose of the present thesis
is to check the consistency of this statement, writing a flow equation for
the theory and  calculating universal quantities such as the $\beta$
function at one loop as a check.

\chapter{$SU(N|N)$ flow equation}\label{chap:2}

 In order to be able to construct a flow equation for $SU(N|N)$, we have to
 recall a set of properties that such an equation must have, in order to
 lead the right physical interpretation. Some of these requirements are more
 general, and are related to the structure which a flow equation must describe. Others are due to the symmetries which must be preserved through
 the flow, in the present case $SU(N|N)$ gauge symmetry.

 Before we continue it is also necessary to add some more preliminary
 comments. As we have mentioned already, throughout this thesis we work in Euclidean space of dimension D.
 We could formulate everything directly in dimension $D=4$ as in \cite{us}, even though strictly
 speaking the limit $D\to 4$ is necessary to rigorously
 define the $SU(N|N)$ regularisation \cite{su:pap}. However, here we want to show, for
 the calculation of terms such as the one-loop $\beta$ function in $SU(N)$
 Yang-Mills, we do not need to pay attention to this subtlety, and we will
 then keep a general dimension $D$ until the very end.

 \section{Necessary properties of the exact RG and their interpretation}
 \label{Necessary}

 The extra fields we have added form a necessary part of the
 regularisation structure. We gain an interpretation of these fields at
 the effective level by imagining integrating out the heavy fields $B$,
 $C$ and $D$ at some scale $\Lambda$. The result is an effective action
 containing only the unbroken gauge fields $A^i$, but it is not finite.
 In particular, the one-loop determinant formed from integrating out the
 heavy fields is necessarily divergent: the divergences are there to
 cancel those left by the one-loop hole in the remaining covariant higher 
 derivative regularisation \cite{oneloophole} of the $SU(N)\times SU(N)$ 
 Yang Mills theory, in a similar way to that done
 explicitly in gauge invariant Pauli-Villars regularisation
 \cite{slav}.
   
 A gauge invariant exact RG description of gauge theory thus requires
 not only an effective action but a separate measure
 term, here provided by the above functional determinant.  The measure
 term is not itself finite, but can be represented by a finite
 addition to the effective action, after introducing auxiliary fields
 (here $B$, $C$ and $D$).

 Whilst this interpretation is reasonable, similarly to the scalar field
 case, we need to be
 sure that we are still only representing the original quantum field
 theory (here $SU(N)$ Yang-Mills). In the previous chapter this was ensured
 by asking the ``seed action'' vertices neither to lead to UV divergent
 integrals nor to have IR divergences (Taylor expandable to all orders). In the present case, this demand is especially pertinent
 in (but not restricted to) the case where there are extra regulator
 fields, particularly here $A^2$ which remains massless and in this
 effective description only decouples at momenta much less than
 $\Lambda$. More generally, even if there are only physical fields in
 the effective action, we need to be sure that locality, an important
 property of quantum field theory \cite{itz,z-j}, is properly
 incorporated.\footnote{otherwise non-physical effects or other propagating fields, could be hidden in the vertices.} Note that
 $\Lambda$ is intended to be set at the energy scales of interest, which
 is why it makes sense to use the exact RG and solve for the effective
 action directly in renormalised terms, see \eg \cite{mor:elem}.  Indeed, to
 extract the physics (\eg correlation functions \etc) we will even want to
 take $\Lambda\to0$ eventually (\cite{mor:elem}).

 These demands are fulfilled implicitly through the $\Lambda\to\infty$
 limit, providing some very general requirements on the exact RG are
 implemented, as we now explain.

 Firstly, we require that all parts of the flow equation
 can be expanded in external momenta to any order, so that the solutions
 $S$ can also be required to have an all orders derivative expansion
 \cite{mor:erg1,mor:erg2,mor:elem}.\footnote{Sharp cutoff 
 realisations \cite{weg} are more subtle \cite{mor:approx,mor:trunc,mor:momexp} and will
 not be discussed here.} This `quasilocality' requirement \cite{mor:erg1}
 is equivalent to the fundamental requirement of the Wilsonian RG
 that Kadanoff blocking take place only over a localised patch \cite{wilson},
 \ie here that each RG step, $\Lambda\mapsto \Lambda-\delta\Lambda$, be 
 free from infrared singularities.

 The flow equation is written only in terms of renormalised quantities
 at scale $\Lambda$. In fact, we require that the only explicit scale
 parameter that appears in the equations is the effective cutoff
 $\Lambda$. Again this is so that the same can be required of $S$ where
 it implements the concept of self-similar flow \cite{Shirkov}. Here
 this amounts to a non-perturbative statement of renormalisability, \ie
 existence of a continuum limit, equivalent to the requirement that $S$
 lie on a renormalised trajectory \cite{mor:elem}.  This is clearer if we
 first scale to dimensionless quantities using the appropriate powers of
 $\Lambda$. Then, $S$ is required to have no dependence on $\Lambda$ at
 all except through its dependence on the running coupling(s) $g(\Lambda)$
 \cite{mor:elem}. 

 Note that the $\Lambda\to\infty$ end of the renormalised trajectory,
 \ie the perfect action \cite{Hasenfratz} in the neighbourhood of the
 ultraviolet fixed point at $\Lambda=\infty$, amounts to our choice of
 bare action. Its precise form is not determined beforehand but as a
 result of solution of the exact RG, but it is constrained by choices in
 the flow equation. Since these choices are however  here to a large
 extent unmade, we deal with an infinite class of perfect bare actions.

 Moreover, we require that the flow of the Boltzmann measure $\exp(-S)$ is a 
 total functional derivative, as we discussed below
 \eq{newfloweqscalar}. As we have seen, importantly, this ensures that the
 partition function ${\cal Z}=\int\!\!{\cal D}\phi\, \exp(-S)$, and
 hence the physics derived from it, is invariant under the RG flow.
 Since we will solve the exact RG approximately, but by controlled
 expansion in a small quantity, this property is left undisturbed.
 Therefore we may use different scales $\Lambda$ at our convenience to
 interpret the computation.

 For example, although locality is obscured in the Wilsonian effective
 action at any finite $\Lambda$, it is important to recognise that
 invariance of ${\cal Z}$ together with the existence of a derivative
 expansion and self-similar flow (\viz that the only explicit scale be
 $\Lambda$), ensure that locality is implemented, since it is then an
 automatic property of the effective action as $\Lambda\to\infty$.

 Similarly, it is as $\Lambda\to\infty$ that we confirm from the
 Wilsonian effective action that we are
 describing $SU(N)$ Yang-Mills theory: $B$, $C$ and $D$ really are
 infinitely massive, and in spacetime dimension four or less, $A^2$ is
 guaranteed decoupled by the Appelquist-Carazzone theorem and
 \eq{12ints}. In general strong quantum corrections might alter either
 of these properties. Thus in general we would need to add appropriate
 sources to the $\Lambda\to\infty$ action; compute the partition
 function by computing the  $\Lambda\to0$ limit of $\exp(-S)$; and
 finally explicitly test these properties by computing appropriate
 correlators. (This is the most general way to extract the results for
 physical quantities from $S$.) However since $g$ is perturbative at
 high energies (indeed $g\to0$ as $\Lambda\to\infty$), we can be sure
 that the above deductions about the regulator fields, drawn at the
 perturbative level, are not destroyed by quantum corrections.

 As already mentioned, we require that an ultraviolet regularisation at
 $\Lambda$, is implemented so that the right hand
 side of the flow equation makes sense. Note that this ensures that all
 further quantum corrections to $S$ (computed by solving for the flow at
 scales less than $\Lambda$) are cutoff (smoothly) at $\Lambda$.  Since
 momentum modes $p > \Lambda$ were fully contributing to the initial
 $\Lambda\to\infty$ partition function, and since ${\cal Z}$ is
 invariant under the flow, we can be sure that their effect has been
 incorporated $S$. In other words we can be sure that our final requirement
 on the flow, namely that it corresponds to integrating out momentum
 modes, has been incorporated.

 (In refs. \cite{alg,mor:erg1}, a possible further requirement on the flow
 equation, called ``ultralocality'' was discussed, replacing the usual
 notion of locality, although it was not clear that it was necessary
 however. We have seen here that the usual concept of locality is
 recovered providing the existence of a derivative expansion, invariance
 of ${\cal Z}$, and self-similar flow, are implemented.  Furthermore the
 successful calculations of ref. \cite{scaus} and here, confirm that the 
 restriction of `ultralocality' is unnecessary since they do not assume it.)

 \section{Supergauge invariance and functional derivatives}
 \label{Supergauge}

 The requirements we have mentioned in the previous section are necessary
 for a general flow equation. However, since we are dealing with a particular theory
 we have to consider some additional ones.
 The peculiarities of $SU(N|N)$, in fact,  affect functional derivatives with
 respect to $\A$ and lead to some constraints on the form of the exact
 RG if the flow equation is to be invariant under supergauge
 transformations.  

 As in refs. \cite{mor:erg2,su:pap}, it is convenient to define the functional
 derivatives of $\C$ and $\A$ so as to extract the dual from under the
 supertrace. For an unconstrained field such as $\C$ we simply have
 \cite{mor:erg2,su:pap}:
 \be
 \label{dCdef}
 {\delta \over {\delta\C}} := {
 \left(\!{\begin{array}{cc} {\delta / {\delta C^1}} & - {\delta /
 {\delta \bar{D}}} \\ {\delta / {\delta D}} & - {\delta
 / {\delta C^2}} \end{array}} \!\!\right)},
 \ee
 or in components 
 \be
 \label{Cdumbdef}
 {\delta \over {\delta\C}}^i_{\gap j} :=
 {\delta \over {\delta\C}^k_{\gap i}}\sigma^k_{\gap j}.
 \ee
 Under supergauge transformations 
 \eq{Cgauged}, the functional derivative transforms as one would hope:
 \be
 \label{dCgauged}
 \delta \left({\delta\over\delta\C}\right) =
 -i\left[{\delta\over\delta\C},\omega\right].
 \ee
 Such a derivative\footnote{for simplicity, written with partial
 derivatives, to neglect the irrelevant spatial dependence}
 has the properties of `supersowing' \cite{mor:erg2}:
 \be
 \label{sow}
 {\partial\over\partial\C}\ \str \,\C Y = Y \quad\Longrightarrow\quad
 \str X{\partial\over\partial\C}\ \str \,\C Y = \str XY,
 \ee
 and `supersplitting' \cite{mor:erg2}:
 \be
 \label{split}
 \str {\partial\over\partial\C}X\C Y = \str X \str Y,
 \ee
 \ie of sowing two supertraces together, and splitting one supertrace into
 two, where $X$ and $Y$ are arbitrary supermatrices. These two properties
 come directly from the completeness relation for the generators of the
 group $U(N/N)$ (see \eq{complrel} and below, for the case of $SU(N|N)$
 without $\one$).
 
 ({\it N.B.} it is a helpful trick to contract in arbitrary
 supermatrices at intermediate stages of the calculation: it allows
 index-free calculations in the $SU(N|N)$ algebra and more importantly
 means that we can permute overall bosonic structures past each rather
 than have to carry intermediate minus signs from fermionic parts of
 supermatrices anticommuted through each other. Its efficacy will be
 seen in examples later.  It also leads as we will show, to efficient
 diagrammatic techniques.  The arbitrary supermatrices can always be
 stripped off at the end, if necessary.)

 Since $\A$ is constrained to be supertraceless, its dual under
 the supertrace $\str\J_\mu\A_\mu$ has without loss of generality no
 $\one$ component: only 
 \be
 \label{realJmu}
 \J_\mu -{\one\over2N}\tr \J_\mu 
 \ee
 really couples. The natural construction for the $\A$ functional derivative 
 from \eq{expandA} \cite{su:pap}:
 \be
 \label{dumbdef}
 {\delta\over\delta\A_\mu}:=
 2T_A{\delta\over\delta\A_{A\,\mu}}+{\sigma\over2N}
 {\delta\over\delta\A^0_\mu}
 \ee
 pulls out precisely this combination. However from \eq{Agauged} and
 the completeness relations for the $T_A$ (\ref{complrel}), under supergauge
 transformations
 \bea
 \label{dAgauged}
 \delta \left({\delta\over\delta\A_\mu}\right) &=&
 -i\left[{\delta\over\delta\A_\mu},\omega\right] 
 +{i\one\over2N}\tr\left[{\delta\over\delta\A_\mu},\omega\right]\\
 &=& -i\left[{\delta\over\delta\A_\mu},\omega\right]^*.\nonumber
 \eea
 The correction is to be expected since it ensures that
 $\delta/\delta\A$ remains traceless, but the fact that
 $\delta/\delta\A$ does not transform homogeneously means that
 supergauge invariance is destroyed unless $\delta/\delta\A$
 is contracted under the supertrace into something that is supertraceless
 (in which case the correction term vanishes). This is an extra constraint
 on the form of the flow equation.

 [As an alternative one might try defining $\delta/\delta\A$ as only
 the $2T_A\delta/\delta\A_A$ term in \eq{dumbdef}, however one can
 show from \eq{Agauged} that this does not transform into itself but into 
 the full functional derivative given in \eq{dumbdef}. It works however
 in the Bars* representation, where the transformation again takes the form
 \eq{dAgauged}.]

 Similarly there are corrections to \eq{sow} and \eq{split} that arise because 
 the derivative is constrained:\footnote{ignoring the spacetime index and 
 spatial dependence}
 \be
 \label{sowA}
 \str X{\partial\over\partial\A}\ \str \,\A Y = \str XY -{1\over2N}\, \str X
 \tr Y
 \ee
 as expected from \eq{realJmu}, and 
 \be
 \label{splitA}
 \str {\partial\over\partial\A}X\A Y = \str X \str Y -{1\over2N}\, \tr\, Y\!X.
 \ee
These come directly from the completeness relation for $SU(N|N)$ and are a
 way to rephrase Eqs. (\ref{comptree}) and (\ref{comploop}), from the
 previous chapter.
 Since these corrections contain $\tr Z \equiv \str\,\sigma Z$ (where $Z$
 is some supermatrix), they similarly violate $SU(N|N)$ invariance.  As
 we discuss in \sec{sow-split}, they also effectively disappear with the above
 constraint that $\delta/\delta\A$ is contracted into something
 supertraceless. (This is obvious in \eq{sowA} where thus $\str X=0$.)

 In this way the supersplitting and supersowing rules actually become
 exact for both fields, even at finite $N$ (compare \cite{mor:erg1,mor:erg2}). As
 we will see, this leads to a very efficient diagrammatic
 technique incorporated into the Feynman diagrams,
 for evaluating the gauge algebra, analogous to the 't Hooft
 double line notation \cite{tHooftdouble} and utilised earlier
 \cite{alg,mor:erg1,mor:erg2}, but here applying even at finite $N$.

\section{Covariantisation}
\label{Covariantization}

Since we want to build a flow equation which is invariant under
 supergauge transformations, we need to have covariant generalisations of the
 momentum space kernels appearing in other ERG equations' formulations \eg
 in the scalar field case described in the first chapter. In that case they
 were present in the flow equation as a result of the regularisation
 scheme, which there did not have any particular invariance-preserving
 prescriptions, being just the first derivatives of the cutoff function. In the
 present formulation, making use of the freedom allowed by scheme
 independence, the flow equation will be written incorporating
 covariantised versions of these objects, which we are going to describe in the present
 section. These covariantised momentum space kernels, will be then related
 back to the gauge invariant regularisation scheme of
 \cite{su:rome,su:pap}. Their introduction will
 involve more terms in the flow equation, but will insure it describes a
 gauge invariant flowing effective action.

\begin{figure}[h]
 \psfrag{i}{i}
 \psfrag{j}{j}
 \psfrag{k}{k}
 \psfrag{l}{l}
 \psfrag{u}{u}
 \psfrag{v}{v}
\psfrag{uwv=}{$u\{W\}v\rightarrow$}
 \begin{center}
 \includegraphics[scale=.3]{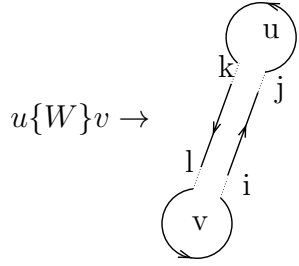}
 \end{center}
 \caption{Parallel transport between two matrix representation through the
 wine, eq.(\ref{wv}). There is no explicit representation of the $\sigma_3$
 because, it is incorporated in the closed line which defines already a supertrace.}\label{fig:cov}
 \end{figure}

 Given some momentum space kernel $W_p\equiv W(p,\Lambda)$ as the one
 defined in \eq{dotnotation} and below, we define a general covariantisation
 of any such kernel (the `wine' \cite{mor:erg1,mor:erg2}) via the supergauge invariant:
 \bea
 \label{wv}
 &&u\,\{W\}_{\!\!{}_\A} v = \\
 &&\sum_{m,n=0}^\infty\int\!\!d^D\!x\,d^D\!y\,
 d^D\!x_1\cdots d^D\!x_n\,d^D\!y_1\cdots d^D\!y_m\,
 W_{\mu_1\cdots\mu_n,\nu_1\cdots\nu_m}
 (x_1,\cdots,x_n;y_1,\cdots,y_m;x,y) \nonumber\\
 &&\phantom{\sum_{m,n=0}^\infty
     \int\!\!d^D\!x\,d^D\!y\,d^D\!x_1\cdots d^D\!x_n\,}
 \str\left[\, u(x)\, \A_{\mu_1}(x_1)\cdots \A_{\mu_n}(x_n)\, 
 v(y)\,\A_{\nu_1}(y_1)\cdots \A_{\nu_m}(y_m)\,\right],\nonumber\\
&&\ \ \ \ \ \ \ \ \ \ \ \ =\int d^Dx\ d^Dy\ [\sigma_3 u(x)]^l_i{}^{\phantom{x}i}_{{\bf x} l}\{W\}^k_{j{\bf y}}v(y)^j_k\nonumber
 \eea
 where $u$ and $v$ are any two supermatrix representations, and with the
 symbol $W$ is introduced the {\it wine} (Wilson-line) as in
\cite{mor:erg1,mor:erg2}, the Wilson line implementing the parallel transport between the two
representations (this will be seen more clearly in (\ref{wine})). A graphical representation of it is shown in
\fig{fig:cov}.
As we can notice from eq.(\ref{wv}), the wine is expandable in
 fields. Its expansion in terms of ${\cal A}$ fields is
\bea
{}^{\phantom{x}i}_{{\bf x} l}\{W\}^k_{j{\bf
y}}&\!\!\!=&(-)^h\,\,\,\!\!\!\!\!\!\!\dsum_{n,m=0}^{\infty}\int d^D\!x_1\cdots
d^D\!x_n\,d^D\!y_1\cdots d^D\!y_m\, W_{\mu_1\cdots\mu_n,\nu_1\cdots\nu_m}(x_i;y_j ;x,y)\nonumber\\[3pt]
&&\!\!\!\!\!\!\!\![{\cal A}_{\mu_1}(x_1)\cdots {\cal A}_{\mu_n}(x_n)]^i_{\ j} [{\cal A}_{\nu_1}(y_1)\cdots {\cal A}_{\nu_m}(y_m)]^k_{\ l}
\eea
where $h=f(\alpha)\sum_i f(\alpha_{i})$, $f$ is defined below \eq{sa} and
where the indices $\alpha$ and $\alpha_i$ refer to those in the expansions:
$v=v^{\alpha} S_{\alpha}$ and ${\cal A}^{\alpha_i}_{\nu_i}S_{\alpha_i}$.

\begin{figure}[h]
 \psfrag{mu1}{$\mu_1$}
 \psfrag{mu2}{$\mu_2$}
 \psfrag{nu1}{$\nu_1$}
 \psfrag{nu2}{$\nu_2$}
 \psfrag{dots}{$\cdots$}
 \psfrag{+}{$+$}
 \psfrag{=}{$=$}
 \begin{center}
 \includegraphics[scale=.5]{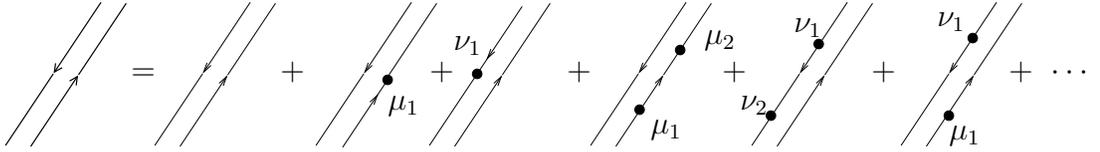}
 \end{center}
 \caption{Wine expansion. The blobs represent ${\cal A}$
 fields.}\label{fig:winexp}
 \end{figure}

\begin{figure}[h!]
\psfrag{mu1}{$p_{\mu_1}$}
\psfrag{mu2}{$p_{\mu_2}$}
\psfrag{mun}{$p_{\mu_n}$}
\psfrag{nu1}{$q_{\nu_1}$}
\psfrag{nu2}{$q_{\nu_2}$}
\psfrag{num}{$q_{\nu_m}$}
\psfrag{r}{$r$}
\psfrag{s}{$s$}
\psfrag{dot}{$\cdots$}
\psfrag{Wn1n2nnm1m2mmp1p2pnq1q2qmrs}{$W_{\mu_1\mu_2\cdots\mu_n;\nu_1\nu_2\cdots\nu_m}(p_1,p_2,\cdots
,p_n;q_1,q_2,\cdots ,q_m;r,s)$}
\begin{center}
\includegraphics[scale=.6]{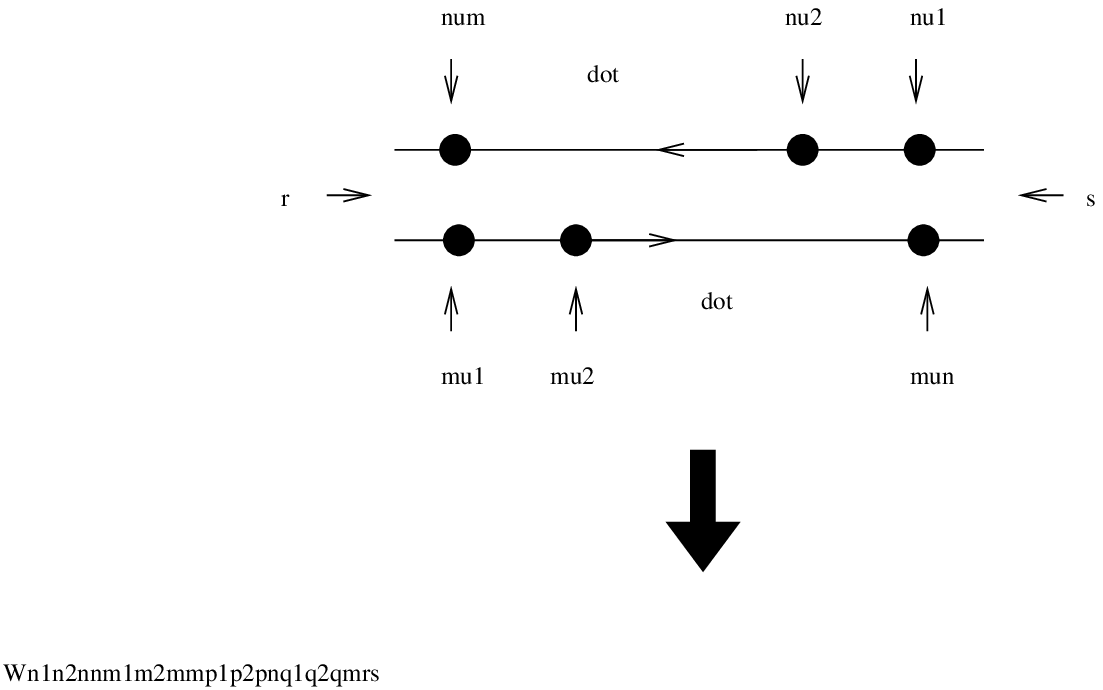}
\end{center}
\caption{Feynman rules for wine vertices}\label{fig:feynrules}
\end{figure}

A graphic representation of this expansion is shown in
fig.\ref{fig:winexp}.
In order to explain better the graphical notation, the first of the terms with one blob
is represented in formula by
\be\label{1blob}
\int d^Dx_1\ \underbrace{W_{\mu_1,}(x_1;;x,y)}_{W_{\mu_1}(x_1;x,y)} [{\cal
A}_{\mu_1}(x_1)]^i_{\ j} .
\ee
The Feynman rules in the momentum space for a general wine vertex are explained in
fig.\ref{fig:feynrules}.

\noindent Without loss of generality we may insist that 
 $\{W\}_{\!\!{}_\A}$ satisfies
 $u\,\{W\}_{\!\!{}_\A} v\equiv v\,\{W\}_{\!\!{}_\A} u$. We write the $m=0$
 vertices (where there is no second product of gauge fields), more
 compactly as
  \be
 \label{compac}
 W_{\mu_1\cdots\mu_n}(x_1,\cdots,x_n;x,y)
 \equiv W_{\mu_1\cdots\mu_n,}(x_1,\cdots,x_n;;x,y),
 \ee
 while the $m=n=0$  term is
 just the original kernel appearing in \eq{dotnotation} and below, \ie 
 \be
 \label{mno}
 W_,(;;x,y)\equiv W_{xy}.
 \ee

 We leave the covariantization general, up to
 certain restrictions. One of these is already encoded into \eq{wv},
 namely that there is just a single supertrace in \eq{wv},
 involving just two ordered products of supergauge fields.
 Another is that we require that the covariantization satisfy coincident
 line identities \cite{mor:erg1} which in particular imply that if 
 $v(y)=\one g(y)$ for all $y$, \ie is in the scalar representation of the
 gauge group, then the covariantization collapses to
 \be
 \label{Acoline}
 u\,\{W\}_{\!\!{}_\A} v = (\str\, u)\ker{W}g.
 \ee
 As shown in ref. \cite{mor:erg2}, the coincident
 line identities are equivalent to the requirement that the gauge fields
 in \eq{wv} all act by commutation. This requirement is necessary to ensure
 no-$\A^0$ remains valid and to ensure that $\delta/\delta\A$ is indeed
 contracted into something supertraceless. It is this that we need
 rather than the identities themselves, which are used occasionally,
 only to collect terms in the calculation.

 Again, although we will not use it explicitly, let us remark that
 these constraints are solved by the following general covariantization
 \cite{mor:erg1,mor:erg2}:
 \be
 \label{wine}
 u\,\{W\}_{\!\!{}_\A} v=
 \int\!\!\!\!\int\!\!d^D\!x\,d^D\!y\int\!\!\D_{\!\!{}_W}\ell_{xy}\
 \str \,u(x)\,\Phi[\ell_{xy}]\,v(y)\,\Phi^{-1}[\ell_{xy}],
 \ee
 where 
 \be
 \label{line}
 \Phi[\ell_{xy}]=P\exp-i\int_{\ell_{xy}}\!\!\!\!\!\!dz^\mu\A_\mu(z)
 \ee
 is a path ordered exponential integral, \ie a Wilson line, and the appearance
 of $\Phi^{-1}[\ell_{xy}]$ means that we traverse backwards along another
 coincident Wilson line. The covariantization is determined by the
 measure $\D_{\!\!{}_W}$ over configurations of the curves $\ell_{xy}$ and is
 so far left unspecified except for its normalisation:
 \be
 \int\!\!\D_{\!\!{}_W}\ell_{xy}\ 1 = W_{xy},
 \ee
 as follows from \eq{wv} and \eq{mno}. It is easy to see that \eq{wine}
 indeed does satisfy \eq{coline}.

 Finally, we will require that the covariantization satisfies
 \be
 \label{noATailBiting}
 {\delta\over\delta\A_\mu}\,\{W\}_{\!\!{}_\A} =0,
 \ee
(where the previous is understood contracted on a supermatrix $X$ independent
 of $\A$)
 \ie that there be no diagrams  in which the wine bites its own tail 
 \cite{alg,mor:erg1,mor:erg2}. This leads to identities for the $W$ vertices which
 again we do not need in practice: as we will confirm, such terms do not in
 any case contribute to the one-loop $\beta$ function. However 
 wine-biting-their-tail diagrams do appear in general to lead to some 
 improperly regularised terms and so some restriction
 is needed for consistency. We can use the representation \eq{wine} to see
 that sensible solutions to \eq{noATailBiting} do exist. For example we can
 simply insist that $\ell_{xy}$ is a straight Wilson line, and more
 generally that the measure  $\D_{\!\!{}_W}$ has no support on curves
 $\ell_{xy}$ that cross the points $x$ or $y$. The end points need defining
 carefully so that they only touch $x$ and $y$ after a limit has been taken \cite{alg}. However since we never specify the covariantization, we only need to assume that such a thing exists. In the calculation we just use \eq{noATailBiting} and thus just forbid all wine-biting-their-tail diagrams. 

 \subsection{Decoration with $\C$}
 \label{Decoration}

\begin{figure}[h]
 \psfrag{to}{$\to$}
 \begin{center}
 \includegraphics[scale=1]{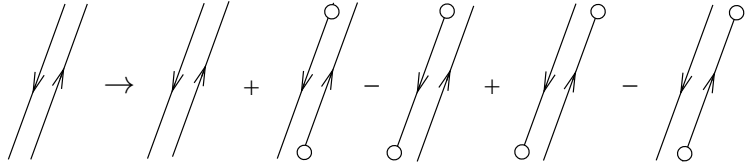}
 \end{center}
 \caption{Wines decorated with $\cal C$ fields, represented with white
 blobs. Each of the lines have an expansion in $\cal A$ fields as the one
 of \fig{fig:winexp}.}\label{fig:cwines}
 \end{figure}

 Making use of the freedom we have on the choice of various parts of the
 flow equation, given by scheme independence, and since it will prove
 convenient for later purposes, we allow having occurences of $\C$ also on the Wilson
 lines (with the obvious corresponding extension of \fig{fig:winexp})
 although we can limit their appearance to attachments at either
 end of $\ell_{xy}$.
Throughout all this thesis, as in \cite{us}, they will furthermore act only via
 commutation at both ends. Precisely, we extend the definition \eq{wv}
 so that
 \be
 \label{wev}
 u\{W\}v =  u\,\{W\}_{\!\!{}_\A}v
 -\frac{1}{4}\C\!\cdot\!u\,\{W_{m}\}_{\!\!{}_\A}\C\!\cdot\!v.
 \ee
where $W_{m}(p,\Lam)$ is some new kernel. This is represented graphically
 in \fig{fig:cwines}, where the $\cal C$ fields are drawn by a white circle.
In the expansion we now have vertices that come from both $\A$ and
 $\C$.
 Typically in this case $u$ and $v$ will actually correspond to functional
 differentials, with respect to, say, $Z_1$ and $Z_1$, and it will also be
 helpful to keep track of their flavours. by including them as labels in
 the naming convention for the kernel, \viz as $W^{Z_1 Z_2}_{(m)}$.
 The notation we will use in general is
 \bea
 \label{wcv}
 &&{\delta\over\delta Z_1^{c}}\{W^{Z_1 Z_2}\} {\delta \over\delta Z_2^{c}} =\\
 &&\sum_{m,n=0}^\infty\int\!\!d^D\!x\,d^D\!y\,
 d^D\!x_1\cdots d^D\!x_n\,d^D\!y_1\cdots d^D\!y_m\,
 W_{\ a_1\cdots\, a_n,\ b_1\cdots b_m}^{X_1\cdots X_n, Y_1\cdots Y_m, Z_1 Z_2}
 (x_1,\cdots,x_n;y_1,\cdots,y_m;x,y) \nonumber\\
 &&\phantom{\sum_{m,n=0}^\infty
     \int\!\!d^D\!x\,d^D\!y\,d^D\!x_1\cdots }
 \str\left[\, {\delta\over\delta Z_1^{c}(x)}\, X_1^{a_1}(x_1)\cdots 
 X_n^{a_n}(x_n)\, 
 {\delta\over\delta Z_2^{c}(y)}\,Y_1^{b_1}(y_1)\cdots 
 Y_m^{b_m}(y_m)\,\right],\nonumber
 \eea
 where the superfields $X_i$, $Y_i$ and $Z_i$, are $\A$ or $\C$,
 and the indices $a_i=\mu_i$, $b_i=\nu_i$ and $c=\gamma$ in the case
 that the corresponding field is $\A$ and null if the field is $\C$.
 In fact, as a consequence of the restricted structure \eq{wev}, the
 $X_2,\cdots,X_{n-1}$ and $Y_2,\cdots,Y_{m-1}$ must be $\A$s
 if they appear at all. 

 We can still insist without loss of generality
 that $u\{W\}v \equiv v\{W\}u$, and use the shorthand \eq{compac},
 where now we keep track of flavour labels as in \eq{wcv} however.
 It is still the case that with no fields on the wine, the original
 $W$ kernel is recovered as in \eq{mno}. The commutator structure in \eq{wev}
 ensures that \eq{Acoline} holds for the full wine also:
 \be
 \label{coline}
 u\,\{W\} v = (\str\, u)\ker{W}g.
 \ee
 Finally,  the $\C$s as further `decorations'
 of the covariantized kernels are required to
 partake in the restriction described below
 \eq{noATailBiting}, so this equation extends to
 \be
 \label{noTailBiting}
 {\delta\over\delta\A_\mu}\,\{W\} = {\delta\over\delta\C}\,\{W\} =0.
 \ee
 (In fact by $X=\one$ in \eq{split}, the
 contribution from differentiating the leftmost $\C$ vanishes in any
 case.)

\section{Superfield expansion}\label{Superfield}

\begin{figure}[h!]
 \begin{center}
 \psfrag{=}{$=$}
 \psfrag{dot}{$\cdots$}
 \psfrag{+}{$+$}
 \includegraphics[scale=.5]{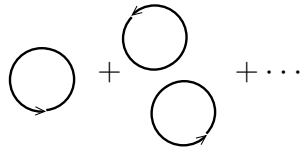}
 \end{center}
 \caption{Action's expansion in product of supertraces}\label{fig:action}
 \end{figure}

\begin{figure}[h!]
 \begin{center}
 \psfrag{=}{$=$}
 \psfrag{dot}{$\cdots$}
 \psfrag{+}{$+$}
 \includegraphics[scale=.5]{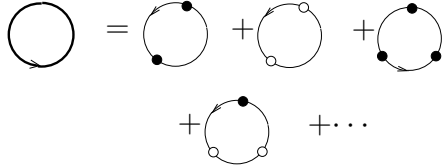}
 \end{center}
 \caption{Each supertrace in the action expansion is a sum of supertraces of
 fields}\label{fig:fieldex}
 \end{figure}

Let us consider first the effective (flowing) action $S$. We can expand it
in powers of the fields bearing in mind it must be an invariant under the
group $SU(N|N)$. The most general one, is a linear combination of product
of supertraces of fields:
 \bea
 \label{Sex}
 S &=&\sum_{n=1}^\infty{1\over s_n}\int\!\!d^D\!x_1\cdots d^D\!x_n\,
 S^{X_1\cdots X_1}_{\, a_1\cdots\,a_n}(x_1,\cdots,x_n)\ 
 \str\, X_1^{a_1}(x_1)\cdots X_n^{a_n}(x_n)\nonumber\\
 &+&{1\over2!}\sum_{m,n=1}^\infty{1\over s_ns_m}\int\!\!
 d^D\!x_1\cdots d^D\!x_n\,d^D\!y_1\cdots d^D\!y_m\,
 S_{\, a_1\cdots\, a_n,\ b_1\cdots b_m}^{X_1\cdots X_n, Y_1\cdots
 Y_m}(x_1,\cdots,x_n;y_1,\cdots,y_m)\nonumber\\
 &&\phantom{{1\over2!}\sum_{m,n=1}^\infty{1\over nm}\int\!\!
 d^D\!x_1\cdots d^D\!x_n\,}
 \str\, X_1^{a_1}(x_1)\cdots X_n^{a_n}(x_n)\
 \str\, Y_1^{b_1}(y_1)\cdots Y_m^{b_m}(y_m)\nonumber\\
 &+& \cdots,
 \eea
 where again the $X_i^{a_i}$ are $\A_{\mu_i}$ or $\C$, and $Y_j^{b_j}$
 are $\A_{\nu_j}$ or $\C$. (Note that throughout this thesis we discard
 the vacuum energy.) Only one cyclic ordering of each list 
 $X_1\cdots X_n$, $Y_1\cdots Y_m$ appears in the sum.
 Furthermore, if either list is invariant under some
 nontrivial cyclic permutations, then $s_n$ ($s_m$)
 is the order of the cyclic subgroup, otherwise $s_n=1$ ($s_m=1$).
 (For example, in the terms where every $X_i^{a_i}$ is a $\C$, $s_n=n$.)
 The expansion can be represented diagrammatically, where a thick closed line
 stands for a supertrace, as in \fig{fig:action} 
 and each blob represents a field in it (\fig{fig:fieldex}).
 In a somewhat similar way to \eq{wine} and \eq{wev}, these closed lines can
 be interpreted as decorated Wilson loops \cite{mor:erg1,mor:erg2}.

 When we spontaneously break the fermionic invariance by shifting $\C$
 in the $\sigma$ direction, it will prove to be better to work separately
 with the bosonic and fermionic parts of the superfields. Thus we write
 in the broken phase 
 \be\label{fieldsbroken}
 \A_\mu = A_\mu + B_\mu,\ins11{and} \C= C + D + \sigma.
 \ee
 where $A$ and $C$ are the block diagonals, 
 and $B$ and $D$ are the block off-diagonals in eqs. \eq{defA} and \eq{defC}
 respectively. 
 (We will see in the \sec{manifestly} that $\C$'s 
 effective vacuum expectation value is just $\sigma$.)

 Thus in the broken phase we will expand as in \eq{Sex}, but the
 flavours $X$ and $Y$ are set to $A$, $B$, $C$ or $D$. There will also
 be occurences of $\sigma$. However since $\sigma$ commutes with $A$ and
 $C$, and anticommutes with $B$ and $D$, to define the expansion we can
 take the convention that we (anti)commute all such occurences to the
 far right in the supertrace. Upon using $\sigma^2=\one$, we are then
 left with terms with either one $\sigma$ at the end or none at all.
 Since $\sigma$ has no position dependence, we put the flavour label in
 the superscript, but we omit the corresponding position label. Clearly,
 since the broken fields can still be cyclically permuted by
 (anti)commutation through $\sigma$, we also omit it from the
 determination of the symmetry factor, \ie $s_n$ is equal to the order
 of the cyclic permutation subgroup of the fields $X_i$, ignoring the
 $\sigma$ (if present). Finally note that each
 supertrace term  must separately hold only totally bosonic combinations
 since if $X_1\cdots X_n$ (or $X_1\cdots X_n\sigma$) is fermionic, it is block off-diagonal
 and has vanishing supertrace. 

 Similarly, in \eq{wcv}, in the broken phase, $X$, $Y$ and $Z$ will be
 $A$, $B$, $C$ or $D$. Note that $Z_1$ can be the opposite statistic
 partner from $Z_2$. Since it is a single supertrace, again each
 contribution in \eq{wcv} is overall bosonic however.  Single occurences
 of $\sigma$ can also appear at the ends of the Wilson lines, after
 taking into account that these can also (anti)commute through the $Z$
 functional derivatives.

 Finally, the momentum space vertices are written as
 \be
 S^{X_1\cdots X_n}_{\, a_1\cdots\,a_n}(p_1,\cdots,p_n)\
 (2\pi)^D\delta( 
 \sum_{i=1}^np_i)  
 =\int\!\!d^D\!x_1\cdots d^D\!x_n\,{\rm e}^{-i\sum_ix_i\cdot p_i}
 S^{X_1\cdots X_n}_{\, a_1\cdots\,a_n}(x_1,\cdots,x_n),
 \ee
 where all momenta are taken pointing into the vertex, and similarly for
 all the other vertices including  \eq{wcv}. 
 We use the short hand $S^{XY}_{ab}(p)\equiv S^{XY}_{ab}(p,-p)$
 and $S^{XY\sigma}_{ab}(p)\equiv S^{XY\sigma}_{ab}(p,-p)$
 for action two-point vertices.

 We will see later many examples. See also ref. \cite{mor:erg2}.

\subsection{Rescaling $g$}\label{sub:rescalg}

As in the case of the scalar field, in order to put the coupling constant
in front of the action, we want to rescale the field as:
\be\label{arescaled}
{\cal A}=\frac{1}{g}\tilde{\cal A}.
\ee
In this way, as in the previous case, the Boltzman factor in the partition function becomes:
\be\label{partresc}
e^{-S/\hbar}\to e^{-S/\hbar g^2},
\ee
and the loop ($\hbar$) expansion conicides with the coupling expansion. In
the present case, though, this rescaling give us a further nice feature. To
explain this consider just the $SU(N)$ gauge field $A_1$, with covariant
derivative $\nabla^1_{\mu}=\de_{\mu}-ig A^1_{\mu}$.
If we consider a gauge transformation on it,
\be\label{gauge-a}
\delta A^1_{\mu}=-\frac{1}{g}[\nabla^1_{\mu},\omega]=\frac{1}{g}(\de_{\mu}\omega-ig[A^1_{\mu},\omega])
\ee
If we now consider the rescaled field of \eq{arescaled}, and we perform a
gauge transformation on it:
\be\label{gaugeares}
\delta\tilde{A}^1_{\mu}=\de_{\mu}\omega-i[\tilde{A}^1_{\mu},\omega]
\ee
If we now suppose $\tilde{A}^1$ runs as:
$\tilde{A}^1_{\mu}=Z^{1/2}_{\tilde{A}^1}\tilde{A}^{1\ R}_{\mu}$, \eq{gaugeares},
would become:
\be
Z^{1/2}_{\tilde{A}^1}\tilde{A}^{1\
R}_{\mu}=\de_{\mu}\omega-iZ^{1/2}_{\tilde{A}^1}[\tilde{A}^{1\ R}_{\mu},\omega]
\ee
For gauge invariance to be preserved, $Z^{1/2}_{\tilde{A}^1}$ must be equal to
one and this ensures that the rescaled field does not renormalise. In order
to extract the one-loop beta function, it will then be enough to evaluate the
one-loop two-point equation, and not also higher points as in the
scalar field case (see sec. \ref{sec:1.4}), since here $\gamma=0$. The only
quantity that renormalises is now $g$ itself and the renormalisation
condition is set by \eq{defg}.

 \section{A manifestly $SU(N|N)$ gauge invariant ERG}
 \label{manifestly}

 Our strategy is to write down a manifestly supergauge invariant flow
 equation, obeying the rules outlined in the previous sections, and then
 spontaneously break it. Defining $\Sigma_g=g^2S-2\sh$, we simply set 
 \be
 \label{sunnfl}
 \ldl S  =
 - a_0[S,\Sigma_g]+a_1[\Sigma_g],
 \ee
 where
 \be 
 \label{a0}
  a_0[S,\Sigma_g] ={1\over2}\,\frac{\delta S}{\delta {\cal
 A}_{\mu}}\{\dDelta^{\A\A}\}\frac{\delta \Sigma_g}{\delta {\cal
 A}_{\mu}}+{1\over2}\,\frac{\delta S}{\delta {\cal C}}\{\dDelta^{\C\C}\}
 \frac{\delta \Sigma_g}{\delta {\cal C}}, 
 \ee
 and
 \be
 \label{a1}
 a_1[\Sigma_g] = {1\over2}\,\frac{\delta }{\delta {\cal
 A}_{\mu}}\{\dDelta^{\A\A}\}\frac{\delta \Sigma_g}{\delta {\cal
 A}_{\mu}} + {1\over2}\,\frac{\delta }{\delta {\cal C}}\{\dDelta^{\C\C}\}
 \frac{\delta \Sigma_g}{\delta {\cal C}}.
 \ee
where the notation for the wines is explained below eq.(\ref{wcv}). Here
 instead of indicating the kernels with $W$, we used the symbol $\dDelta$,
 because, as we will see, the integrated kernels, play here the r\^{o}le of
 effective propagators. More precisely they will be the inverse of the
 corresponding two point functions, in the transverse space (see App. \ref{App:intwine}). Eq.(\ref{sunnfl}) can be represented diagrammatically in \fig{fig:floeq}, appearing later. In the rest of this section we explain the meaning of the various
 components, at the same time developing some of the properties of this 
 exact RG. 

 The definition of $\Sigma_g$ and the form of the flow equation
 \eq{sunnfl} are the same as in refs. \cite{mor:erg1,mor:erg2}. In contrast to
 ref. \cite{mor:erg2} however, the exact RG is very simple in conception.
 The basic structure is inherited from the Wilson exact RG
 \cite{wilson,pol,alg}: the bilinear functional -$a_0$ generates the
 classical corrections, whilst the linear functional $a_1$ generates
 quantum corrections (compare with \eq{polflow}). As in refs.  \cite{mor:erg1,mor:erg2}, $a_1$ has exactly
 the same structure as $a_0$ except that the leftmost functional
 derivatives differentiate everything to their right. Consequently we
 have
 \be
 \ldl \,{\rm e}^{-S} = \sum_{X_i=\A,\C}\frac{\delta}{\delta
 X_i}\left[\{\dDelta^{X_i X_i}\}\frac{\delta\Sigma_g}{\delta X_i}{\rm e}^{-S}\right],
 \ee
(similarly to eq.(\ref{newfloweqscalar})) which shows that the condition for the Boltzman measure to be a total
 functional derivative, is fulfilled. 

 As before, $g(\Lambda)$ is the renormalised coupling of the $SU(N)$
 Yang-Mills theory carried by $A^1$. It is defined through the
 renormalization condition:
 \be
 \label{defg}
 S[\A=A^1, \C={\bar\C}] ={1\over2g^2}\,{\rm tr}\!\int\!\!d^D\!x\, 
 \left(F^1_{\mu\nu}\right)^2+\cdots,
 \ee
After the rescaling of $g$ the previous is the only condition to be set,
 $g$ being now the only quantity that runs (see sec. \ref{sub:rescalg}).
 The ellipsis in \eq{defg}, stands for higher dimension operators and the vacuum
 energy, and ${\bar\C}$ is the effective vacuum expectation value defined
 so as to minimise the effective potential $V(\C)$ in $S$:
 \be
 \label{defCbar}
 \left.{\partial V\over\partial\C}\right|_{\C={\bar\C}} =0.
 \ee
 ${\bar\C}$ is spacetime independent and generically contains terms
 proportional to $\sigma$ and $\one$ (this is explained at the end
 of this section). We will see later that for our purposes we can simply set ${\bar\C}=\sigma$.

 The strategy now to get the 1-loop $\beta$-function, will
be the same as in \cite{mor:erg1,mor:erg2} and consists in expanding the flow
equation \eq{sunnfl} in loop ($\hbar$ powers), which, at this point after
having rescaled $g$, amounts in
a coupling expansion\footnote{The redefinition of $\cal A$ described in
section \ref{sub:rescalg} led to this result}. Expanding $S$ first
\be
 \label{Sloope}
 S={1\over g^2} S_0+S_1+g^2 S_2 +\cdots,
 \ee
 where $S_0$ is the classical effective action, $S_1$ the one-loop
 correction, and so on.
 Substituting this expansion in \eq{sunnfl}, we see that
 the $\beta$ function must also take the standard form
 \be
 \label{betafn}
 \beta:=\Lambda{
 \partial g\over\partial\Lambda}=\beta_1g^3+\beta_2g^5+\cdots.
 \ee
 From \eq{Sloope} and \eq{betafn}, we obtain the 
 loopwise expansion of \eq{sunnfl}:
 \bea
 \Lambda{\partial\over\partial\Lambda}S_0 &=& -a_0[S_0,S_0-2\sh],
 \label{ergcl}\\
 \Lambda{\partial\over\partial\Lambda}S_1&=&2\beta_1S_0-2a_0[S_0-\sh,S_1]
 +a_1[S_0-2\sh], \label{ergone}\\
\Lambda{\partial\over\partial\Lambda}S_2&=&2\beta_2 S_0-2a_0[S_0-\hat{S},S_2]-a_0[S_1,S_1]+a_1[S_1]\label{2l},
\eea
\etc From the second, we will try to get $\beta_1$. Actually, we will find it convenient to add some simple quantum
 corrections to the supergauge invariant seed action $\sh$, giving it a
 $g$ dependence (as we outline below).  We also need to take account of
 the flow of $g_2$, the coupling for the second $SU(N)$ carried by
 $A_2$.  However, neither of these complications have an effect on the
 one-loop $\beta$ function computation, so will be largely ignored here.

 $\sh$ is used to determine the form of the classical effective kinetic
 terms and the kernels $\dDelta(p,\Lambda)$.  It therefore has to
 incorporate the covariant higher derivative regularisation and allow
 the spontaneous symmetry breaking we require. Unlike previously
 \cite{alg,mor:erg1,mor:erg2}, we will see that we otherwise leave it almost
 entirely unspecified. The kernels $\dDelta$ are determined 
 by the requirement that after spontaneous
 symmetry breaking, the two-point vertices of the classical effective
 action $S_0$ and $\sh$ can be set equal (see section \ref{sec:1loop}).  As previously
 \cite{alg,mor:erg1,mor:erg2}, this is imposed as a useful technical device,
 since it allows classical vertices to be immediately solved in terms of
 already known quantities.  It also means that the integral of the
 kernels defined via
 \be
 \label{defprop}
 \ldl\Delta = - \dDelta
 \ee
 will play a closely similar r\^ole to that of propagators, in particular
 being the inverse of these two-point vertices up to gauge
 transformations (see Appendix \ref{App:intwine}). 

 The $\C$ commutator terms in \eq{wev}, yield $\sigma$
 commutators on spontaneous symmetry breaking. Since $\sigma$ commutes
 with $A$ and $C$ but anticommutes with $B$ and $D$, $\Delta^{\A\A}_m$
 and $\Delta^{\C\C}_m$ allow for the addition of spontaneous mass creation
 for $B$ and $D$ whilst still keeping the two-point vertices of $\sh$ and 
 $S_0$ equal. The appearance of the $\C$ commutator on both sides allows us to
 insist that $\C\leftrightarrow-\C$ is an invariance of the symmetric phase.
 The form \eqs{a0}{a1} preserves charge conjugation symmetry $\C\mapsto \C^T$,
 $\A\mapsto-\A^T$ (using the definition of the supermatrix transpose in
 ref. \cite{su:pap}. Note that here the transformation for
 $\C$ is determined by the fact that its vacuum expectation value
 is even under charge conjugation.)

 From \eq{dCgauged} and \sec{Decoration},
 it is trivial to see that the $\delta/\delta\C$ terms are supergauge 
 invariant. Under a supergauge
 transformation we have by \eq{dAgauged} and \eq{coline},
 \be
 \delta\left(\frac{\delta S}{\delta {\cal
 A}_{\mu}}\{\dDelta^{\A\A}\}\frac{\delta \Sigma_g}{\delta {\cal
 A}_{\mu}}\right) = {i\over2N}\,
 \tr\!\left[{\delta S\over\delta\A_\mu},\omega\right]\!\cdot
 \dDelta^{\A\A}\!\cdot\str{\delta\Sigma_g\over\delta\A_\mu}
 + (S\leftrightarrow\Sigma_g),
 \ee
 where $S\leftrightarrow\Sigma_g$ stands for the same term with $S$ and
 $\Sigma_g$ interchanged. But by \eq{dumbdef} and no-$\A^0$,
 \be
 \str{\delta\Sigma_g\over\delta\A_\mu}={\delta\Sigma_g\over\delta\A^0_\mu}=0,
 \ee
 similarly for $S$, and thus the tree level terms are supergauge invariant.
 Similarly, the quantum terms are $SU(N|N)$ gauge invariant, since
 \be
 \delta\left(\frac{\delta }{\delta {\cal
 A}_{\mu}}\{\dDelta^{\A\A}\}\frac{\delta }{\delta {\cal
 A}_{\mu}}\Sigma_g\right) = {i\over N}\,\tr\!\left[{\delta \over\delta\A_\mu},
 \omega\right]\!\cdot\dDelta^{\A\A}\!\cdot
 \str{\delta\Sigma_g\over\delta\A_\mu} =0.
 \ee
 This completes the proof that the exact RG  is supergauge
 invariant!

 Note that there is no point in incorporating longitudinal terms into
 the exact RG (as was done in ref. \cite{mor:erg2}) because here the
 manifest supergauge invariance means that they can be exchanged for
 $\C$ commutators:
 \be
 \label{gaugeS}
 \nabla_\mu\!\cdot\!{\delta S\over\delta\A_\mu}\, = \,i\,\C\!\cdot\!{\delta
 S\over\delta\C}
 \ee
 (as holds for any supergauge invariant functional) and thus absorbed into the
 $\dDelta^{\C\C}_m$ term. 

It is important for the working of the $SU(N|N)$
 regularisation that the effective scale of spontaneous symmetry
 breaking is tied to the higher derivative regularisation scale, which
 thus both flow with $\Lambda$. This is not the typical situation, but
 can be arranged to happen here by constraining $\sh$ appropriately. 
 However, as we
 now show, the constraint is straightforward only if we take $\C$
 to be dimensionless in \eq{sunnfl} -- \eq{a1}.

 Contracting an arbitrary supermatrix $X$ into
 \eq{defCbar} (for convenience, \cf \sec{Supergauge}) and
 differentiating with respect to $\Lambda$, we have:
 \be
 \left[\str\,{\partial{\bar\C}\over\partial\Lambda}
 {\partial\over\partial\C}\,\str\, X{\partial V\over\partial\C}
 + \str\, X{\partial \over\partial\C}{\partial V\over\partial\Lambda}
 \right]_{\C={\bar\C}}  =0.
 \ee
 We can compute the flow ${\partial V/\partial\Lambda}$ by setting
 $\A=0$ and $\C={\bar\C}$ in \eq{sunnfl}. Taking the classical limit 
 $V\to V_0$, we find that the resulting equation simplifies dramatically.
 Using eqs. (\ref{ergcl}), (\ref{a0}), (\ref{defCbar}), (\ref{wv}), (\ref{mno}),
 the fact that vertices in the actions with only one $\A_\mu$, 
 vanish at zero momentum (by Lorentz invariance), and
 \be
 [{\bar\C},\left.{\partial{\hat V}\over\partial\C}]\
 \right|_{\C={\bar\C}}\!\!=0,
 \ee
 which follows from global $SU(N|N)$ invariance
 (where ${\hat V}$ is the potential in $\sh$), we get
 \be
 \label{minCondn}
 \str\,\left[\left(\Lambda{\partial{\bar\C}\over\partial\Lambda}
 +\dDelta^{\C\C}(0,\Lambda){\partial{\hat V}\over\partial\C}\right)
 {\partial\over\partial\C}\,\str\, X{\partial V_0\over\partial\C}
 \right]_{\C={\bar\C}}  =0.
 \ee

 With $\C$ dimensionless, we can and will insist that the classical vacuum
 expectation value ${\bar\C}=\sigma$.  eq. \eq{minCondn} is then
 satisfied if and only if\footnote{We will see that the requirement that
 $C$ has a mass in the broken phase forces
 $\dDelta^{\C\C}(0,\Lambda)\ne0$.} 
 ${\hat V}$ also has a minimum at ${\C}=\sigma$. This is
 delightful since it ensures that at the classical level at least, 
 neither action has one-point $\C$
 vertices in the broken phase.  We
 will thus impose 
 \be
 \left.{\partial {\hat V}\over\partial\C}\right|_{\C=\sigma} =0
 \ee
 as a constraint on $\sh$.

 Had we not taken $\C$ to be dimensionless, we would have had
 to require that ${\bar\C}$ depend on $\Lambda$, in order that the
 effective breaking scale flows with $\Lambda$. Since $X$ is general,
 \eq{minCondn} would then imply that ${\hat V}$ {\sl cannot} have a
 minimum also at $\C={\bar\C}$. Further analysis shows that ${\hat V}$
 is then forced to violate $\C\leftrightarrow-\C$ symmetry in the
 symmetric phase. 

 Although conventionally $\C$ would have dimension one, for these reasons
 we will take it to be dimensionless from now on. (It is intriguing that
 the conclusion that $\C$ [actually $C$] must be dimensionless was reached
 for very different reasons in refs. \cite{alg,mor:erg2} which are no longer 
 necessarily applicable, now that \eq{gaugeS} is a symmetry.)

 At the quantum level, ${\bar\C}=\sigma$ can be expected to receive loop
 corrections.  Since $SU(N)\times SU(N)$ invariance is left unbroken,
 these corrections can only be proportional to $\sigma$ or $\one$.
 Corrections proportional to the latter do not affect the breaking (but
 presumably through \eq{defg} give important contributions at higher
 loops), however corrections proportional to $\sigma$ would result,
 through \eq{gaugeS}, in broken gauge invariance identities that
 explicitly involve $g$ and thus mix different loop orders. We can avoid
 this by again using the freedom in our choice of $\sh$ to design things
 appropriately. We can constrain the appearance of ${\hat V}$ one-point 
 vertices in the broken phase  
 \be
 \label{defv}
 v^{C}\, \str\,\C\ +v^{C\sigma}\,\str\,\C\sigma
 \ee
 by imposing ${\bar\C}=\sigma$ as a renormalization condition. 
 Each $v$ is then a non-vanishing function of $g$, but from
 the analysis above, only from one-loop onwards:
 \be
 \label{expv}
 v^{C}(g) = v^C_1\,g^2+v^C_2\,g^4+\cdots\ins11{and}
 v^{C\sigma}(g) = v^{C\sigma}_1\,g^2+v^{C\sigma}_2\,g^4+\cdots.
 \ee
 We will see that these corrections in fact are already too high an order to
 affect the one-loop $\beta$ function calculation.

 \subsection{Supersowing and supersplitting in the $\A$ sector}\label{sow-split}

 The inherent supersymmetry has a remarkable effect on the gauge
 algebra: one can replace the usual manipulation of structure constants
 and reduction to Casimirs, which becomes increasingly involved 
 at higher loops, by simple steps \eq{sow} and \eq{split} which always
 either just sow together supertraces or split them open. These have an
 immediate diagrammatic interpretation.  The apparent violations present
 in \eq{sowA} and \eq{splitA} must somehow disappear since they would
 violate even global $SU(N|N)$. We first prove that this indeed the case.

 For the case where the action contains just a single supertrace, which
 will turn out to be all we need here, we could adapt the proof given in
 sec. 6.2 of ref. \cite{su:pap}. However, in preparation for future work,
 we will give a more sophisticated proof which is applicable when
 working with multiple supertrace contributions. Indeed we will see that
 there is then one special case, where the corrections in
 \eqs{sowA}{splitA} do survive, and result in a simple supergauge
 invariant correction.

 The corrections present in \eqs{sowA}{splitA} arise because $\A$ is
 constrained to be supertraceless. To compare their effect to the
 unconstrained case \eqs{sow}{split}, we momentarily `lift' $\A$ to a full
 superfield $\A^e$ by adding a $\sigma$ part:
 \be
 \A_\mu\mapsto \A^e_\mu := \A_\mu+\sigma\A^\sigma_\mu.
 \ee
 $\A^\sigma_\mu$ is taken arbitrary so the map is not at all unique. We
 similarly extend all functionals of $\A$ to the full space, simply
 by replacing $\A$ with $\A^e$, \eg
 \be
 S^e[\A^e,\C] := S[\A\mapsto\A^e,\C].
 \ee
 Again, this is a not unique procedure, as can be seen for example in
 the fact that  $\str\,\A$ vanishes, but the promoted functional
 $\str\,\A^e$ does not.  We also introduce the projection back onto the
 supertraceless space:
 \be
 \label{defpi}
 \pi \A^e_\mu = \A_\mu, \quad \pi S^e=S, \quad\etc,
 \ee
 which of course is unique. 
 Functional derivatives with respect to $\A^e$ can be written as
 \be
 \label{dAedef}
 {\delta\over\delta\A^e_\mu} = {\delta\over\delta\A_\mu} 
 + {\one\over2N} {\delta\over\delta\A^\sigma_\mu},
 \ee
 using \eq{dumbdef}, or equivalently defined as in \eq{Cdumbdef}. 
 $\delta/\delta\A^e$ thus satisfies
 the exact supersowing and supersplitting relations \eqs{sow}{split}.
 In the extended space, the constrained derivative
 \eq{dumbdef} can now be written in terms of an unconstrained 
 derivative:
 \be
 \label{dAdef}
 {\delta\over\delta\A_\mu} = {\delta\over\delta\A^e_\mu} 
 -{\one\over2N}\,\tr{\delta\over\delta\A^e_\mu}.
 \ee
 Of course $\pi$ and $\delta/\delta\A^\sigma$ do not
 commute, however
 \be
 \label{prepre}
 \frac{\delta S}{\delta {\cal
 A}_{\mu}}\{\dDelta^{\!\A\A}\}\frac{\delta \Sigma_g}{\delta {\cal
 A}_{\mu}} = \pi\left\{
 \frac{\delta S^e}{\delta {\cal
 A}_{\mu}}\{\dDelta^{\!\A\A}\}^e\frac{\delta \Sigma^e_g}{\delta {\cal
 A}_{\mu}} \right\},
 \ee
 since $\A^\sigma$ is not differentiated on the right hand side.
 Substituting \eq{dAdef} or \eq{dAedef}, and using \eq{coline} and \eq{dumbdef},

 the term in big curly braces becomes
 \be
 \label{preAAsow}
 \frac{\delta S^e}{\delta {\cal
 A}^e_{\mu}}\{\dDelta^{\!\A\A}\}^e\frac{\delta \Sigma^e_g}{\delta {\cal
 A}^e_{\mu}}
 -{1\over2N}{\delta\S^e\over\delta\A^0_\mu}\cdot\dDelta^{\!\A\A}\cdot
 {\delta\Sigma_g^e\over\delta\A^\sigma_\mu}
 -{1\over2N}{\delta\Sigma_g^e\over\delta\A^0_\mu}\cdot\dDelta^{\!\A\A}\cdot
 {\delta\S^e\over\delta\A^\sigma_\mu}.
 \ee
 Now, as we explain below, no-$\A^0$ symmetry is violated in the
 extended space. However the $\A^0$ derivatives in \eq{preAAsow} do
 vanish after the projection.  Thus \eq{prepre} becomes
 \be
 \label{sowAe}
 \frac{\delta S}{\delta {\cal
 A}_{\mu}}\{\dDelta^{\!\A\A}\}\frac{\delta \Sigma_g}{\delta {\cal
 A}_{\mu}} = \pi\left\{
 \frac{\delta S^e}{\delta {\cal
 A}^e_{\mu}}\{\dDelta^{\!\A\A}\}^e\frac{\delta \Sigma^e_g}{\delta {\cal
 A}^e_{\mu}} \right\},
 \ee
 which says precisely that the corrections in \eq{sowA} can be ignored: 
 exactly the same result is obtained if exact supersowing is used.

 However, performing the same analysis on the corresponding quantum term
 in \eq{a1}, we get a correction to exact supersplitting, consisting
 of an attachment of the (zero-point) kernel $\dDelta^{\A\A}(p,\Lambda)$
 to two $\A$ points in $\Sigma_g$:
 \be
 \label{splitAe}
 \frac{\delta }{\delta {\cal
 A}_{\mu}}\{\dDelta^{\!\A\A}\}\frac{\delta \Sigma_g}{\delta {\cal
 A}_{\mu}} = \pi\left\{
 \frac{\delta }{\delta {\cal
 A}^e_{\mu}}\{\dDelta^{\!\A\A}\}^e\frac{\delta \Sigma^e_g}{\delta {\cal
 A}^e_{\mu}} \right\} -{1\over N} \pi {\delta\over\delta\A^\sigma_\mu}\cdot
 \dDelta^{\!\A\A}\cdot{\delta\Sigma_g\over\delta\A^0_\mu}.
 \ee
 To understand when this correction is non-vanishing, we need briefly to
 analyse the consequences of no-$\A^0$ symmetry in more detail.
 Considering the transformation\footnote{there are higher order
 constraints from separating out higher powers of $\A^0$ but from
 \eq{splitAe} we only need the first order} $\delta\A_\mu=
 \lambda_\mu\one$ in \eq{Sex}, we see that the result must vanish either
 via the supergroup algebra because the corresponding vertex contains a
 factor $\str\A\A$, thus generating $\str\A=0$ (but $\str\A^e\ne0$ in
 the extended space), or because a non-trivial constraint exists on the
 corresponding vertex function.  (This is simply that the sum over all
 possible valid placings of $\A^0$s associated position and Lorentz
 argument inside a vertex function leaving other arguments alone, yields zero.) 
 This non-trivial constraint then causes the
 coefficient to vanish whether or not the remaining supergauge fields
 are extended by $\A^\sigma\sigma$.  Thus the correction in \eq{splitAe}
 vanishes in all cases except where the zero-point $\dDelta^{\A\A}$
 kernel attaches each end to a $\str\A\A$ factor. Comparing the result to the
 computation assuming exact supersplitting, \ie the first term in
 \eq{splitAe}, we see that instead of getting a supergroup factor
 $(\str\one)^2=0$ we get $-{1\over N}\str\sigma$ \ie a supergroup factor of
 $-2$.

 (Note that in deriving this rule we have assumed that vertices in $\Sigma_g$ 
 with factors $\str\A$ have been set to zero from the beginning [as would
 follow immediately from the $SU(N|N)$ group theory]. If for some reason
 this was not done then the first term in \eq{splitAe} can get a non-zero
 computation from the kernel attaching to this $\str\A=2N\A^\sigma$ point. 
 However it then also appears in the correction with precisely equal
 and opposite coefficient.) 

 This supergroup factor should have been expected since the algebra
 part of the attachment of a zero-point kernel to a two-point vertex simply
 counts the number of bosonic degrees of freedom in the algebra
 minus the number of fermionic degrees of freedom. There are $N^2$ fermionic
 such terms in $B$, but only $N^2-2$ in $A$, since both $\A^\sigma$ and,
 by no-$\A^0$ symmetry, $\A^0$, are missing. 

 Since the correction in \eq{splitAe} is non-vanishing only when using up
 a separate $\str\A\A$ factor, it is clear that the result is still supergauge
 invariant in the remaining external superfields. Furthermore in the
 present case where we will be able to work with actions with only a single 
 supertrace, the entire effect of the correction is a just vacuum energy 
 contribution, which from now on we ignore.

 \subsection{Diagrammatic interpretation}
 \label{Diagrammatic}

\begin{figure}[hh]
 \psfrag{=}{$=$}
 \psfrag{-}{$-$}
 \psfrag{+}{$+$}
 \psfrag{ldl}{$\ldl$}
 \psfrag{S}{$S$}
 \psfrag{si}{$\Sigma_g$}
 \psfrag{1/2}{$\displaystyle\frac{1}{2}$}
 \psfrag{xi}{\tiny $f$}
 \psfrag{sumi}{$\displaystyle\sum_{f=\A,\C}$}
 \begin{center}
 \includegraphics[scale=.5]{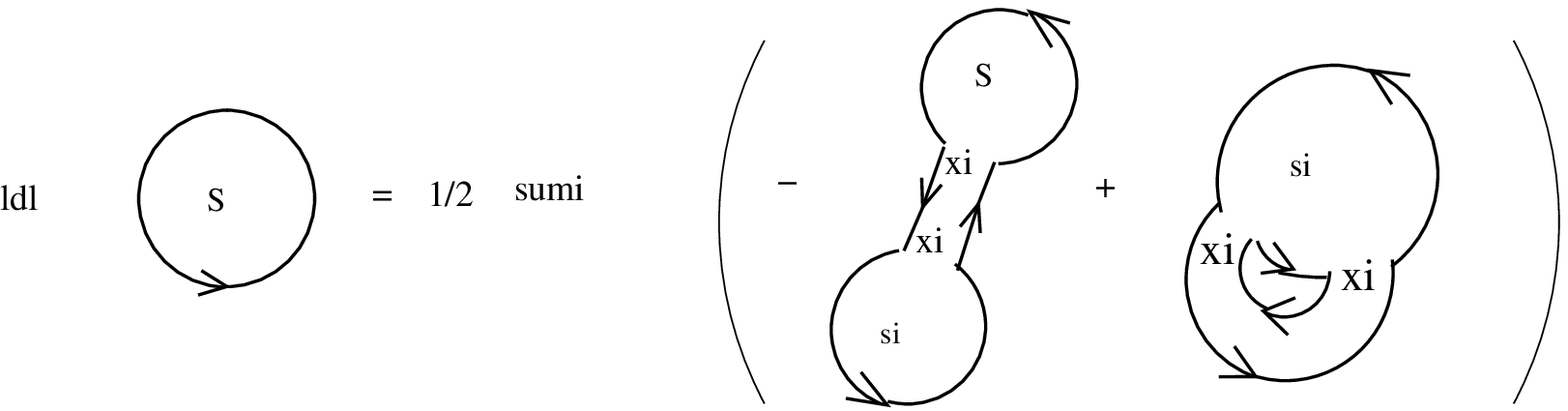}
 \end{center}
 \caption{Graphical representation of the exact RG, when $S$ and $\sh$ 
 contain only single supertraces.}\label{fig:floeq}
 \end{figure}

 $\A$ thus also effectively satisfies the exact supersowing and
 supersplitting relations \eq{sow} and \eq{split}. By using these
 equations when the covariantized kernels \eq{wcv} act on
 the actions \eq{Sex}, and comparing the result to
 the diagrammatic interpretation of the covariantized kernels 
 and actions, \fig{fig:winexp} and figs. \ref{fig:action},\ref{fig:fieldex}, 
 it is clear that the exact RG is given diagrammatically as in \fig{fig:floeq}.
 Here we have specialized to the case of our interest, where 
 $S$ and $\sh$ can be assumed to have only a single supertrace. (The
 extension to the more general contributions \fig{fig:action} is obvious.)
 Expanding the thick lines (representing any number of fields) into a
 power series in the fields, we translate the figure into individual
 Feynman diagrams, whose Feynman rules are given by the momentum space
 versions of the vertices in \eq{wcv} and \eq{Sex} (without the symmetry
 factors).\footnote{This part of the analysis is the same as in
 ref. \cite{mor:erg2}, except that here we make explicit the factor of
 $1/2$ from \eq{a0} and \eq{a1}, in \fig{fig:floeq} and the Feynman diagrams,
 and the factor of $1/\Lambda^2$ is now incorporated in the definition of
 the kernels $\dDelta$.} The points representing individual fields 
 and their associated momenta and Lorentz indices,
 appear in all places on a composite
 loop with equal weight, whilst respecting the cyclic order.
 Of course if one of the corresponding vertices
 does not appear in the expansions \eq{wcv} and \eq{Sex}, the corresponding
 Feynman rule is zero. 

 It can be seen from \fig{fig:floeq} that the tree level corrections preserve
 the assumption that there is only a single supertrace in $S$, but that
 each quantum correction results in
 an extra supertrace factor. Thus in general $S$ has terms with any number
 of supertraces, and already a minimum of a product of two supertraces at 
 one-loop. However for the computation of the $\beta$ function, we need only
 look at contributions to the $AA$ two-point vertex (see \eq{defg} and later,
 or refs. \cite{alg,mor:erg1,mor:erg2}). Since $A$ is both traceless and supertraceless,
 to get a non-vanishing answer both $A$s must lie in the same
 supertrace, leaving the other one empty of fields. In this way, $S$ 
 effectively contains only a single supertrace to the order in which
 we are working.

 \subsection{After spontaneous breaking}

 We substitute $\C\mapsto\C+\sigma$, and from now on work in the
 spontaneously broken phase. Working with fields appropriate for the
 remaining $SU(N)\times SU(N)$ symmetry, we break $\A$ and $\C$ down
 to their bosonic and fermionic parts $A$, $B$, $C$ and $D$ as in
 \eq{fieldsbroken}. 

\begin{figure}[h]
\psfrag{1/2}{$1\over 2$}
\psfrag{+}{$\pm$}
\psfrag{=}{$=$}
\psfrag{d}{\small $\frac{\delta}{\delta Y}$}
 \begin{center}
 \includegraphics[scale=.8]{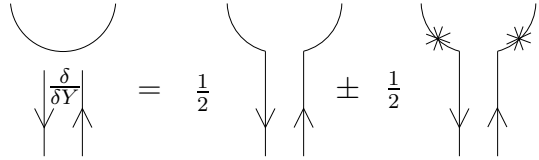}
 \end{center}
 \caption{Feynman diagram representation of attachment via a
 partial supermatrix.}\label{partialAttach}
 \end{figure}

 The diagrammatic interpretation is still the same, except that
 we now have the four flavours to scatter around the composite loops,
 and appearances of $\sigma$, which can be
 simplified as explained in \sec{Superfield}. In addition, we
 must recall the corrections to supersplitting and supersowing arising
 from differentiating only partial supermatrices \cite{mor:erg2}. These 
 lead to further appearances of $\sigma$ which are
 easily computed by expressing the partial supermatrices in terms
 of full supermatrices via the projectors ${\rm d}_\pm$ onto the block
 (off)diagonal components
 \be
 {\rm d}_\pm X = {1\over2}(X\pm\sigma X\sigma),
 \ee
 (hence $C={\rm d}_+\C$, $D={\rm d}_-\C$, \etc).
 Diagrammatically this simply amounts to
 corrections involving a pair of $\sigma$s inserted either side of the 
 attachment as in \fig{partialAttach} \cite{mor:erg2}.
 
 For tree-level type attachments as in \eq{sow}, the corrections merely
 ensure that the coefficient supermatrices ($X$ and $Y$) have the appropriate
 statistics to make each supertrace term totally bosonic (\cf 
 \sec{Superfield}), but this has already been taken into account in the 
 Feynman rules. Thus these corrections have no effect at tree level 
 \cite{mor:erg2}. 

 Since the classical action $S_0$ (similarly $\sh$) has only a single
 supertrace and respects $\C\leftrightarrow-\C$ invariance in the symmetric
 phase (\cf \sec{manifestly}), upon spontaneous breaking
 we have the `theory space' symmetry 
 \bea
 \label{z2}
 C &\leftrightarrow& -C,\nonumber\\ 
 D &\leftrightarrow& -D,\nonumber\\
 \sigma &\leftrightarrow& -\sigma.
 \eea

 The single supertrace part of the one-loop effective action $S_1$ 
 has a single supertrace because it also has
 a supertrace void of fields (\cf \sec{Diagrammatic}). 
 In order for this not to vanish it must `trap' a $\sigma$ (so that we
 get $\str\,\sigma=2N$ rather than $\str\,\one=0$). Therefore, the non-trivial
 supertrace has one less $\sigma$ (mod two) and is thus odd under the symmetry
 \eq{z2}. 

 These observations, which can be easily extended to multiple loops
 and supertraces, are useful in limiting the possible vertices.

\chapter{One-loop $\beta$-function}\label{sec:2.1}

\section{The one-loop equation}\label{sec:1loop}

To start with the calculation of the first coefficient of the
$\beta$-function, we can now consider \eq{ergone} which we rewrite here in
its extended form:
\bea
\Lam \frac{\de}{\de\Lam}S_1&=&2\beta_1 S_0-\sum_{X_i}\left[\frac{\delta (S_0-\sh)}{\delta X_i}\{\dDelta^{X_iX_i}\}\frac{S_1}{\delta X_i}+{1\over2}\,\frac{\delta }{\delta X_i}\{\dDelta^{X_iX_i}\}\frac{\delta \Sigma_0}{\delta X_i}\right]\label{oneeq}
\eea
Where $X_i$ can be either $\A_{\mu}$ or $\C$. Before we continue, this is
the right time to point out that, unlike the formulation
described in \cite{us}, for the rest of the calculation
here\footnote{Unless stated differently}, the wines
with $\C$ will not be incorporated with the ``undecorated'' ones. This was
only done in section \ref{Decoration} and \ref{manifestly} in order to
write the properties of the wines and the flow equation more
compactly. To get to the new formulation, recall first that from \eq{wev},
we can recast $a_0$ and $a_1$ appearing in the flow equation \eq{sunnfl} as follows:
\be\label{newa0}
 a_0[S,\Sigma_g] ={1\over2}\,\sum_{X_i}\left[\frac{\delta S}{\delta X_i}\{\dDelta^{X_iX_i}\}\frac{\delta \Sigma_g}{\delta X_i}+{1\over4}\,\frac{\delta S}{\delta X_i}\!\cdot\C\{\dDelta^{X_iX_i}_m\}
 \frac{\delta \Sigma_g}{\delta X_i}\!\cdot\C\right], 
 \ee
and
\be
 a_1[\Sigma_g] = {1\over2}\,\sum_{X_i}\left[\frac{\delta }{\delta X_i}\{\dDelta^{X_iX_i}\}\frac{\delta \Sigma_g}{\delta X_i} + {1\over4}\,[\frac{\delta }{\delta X_i}\!\cdot\C\{\dDelta^{X_iX_i}_m\}
 \frac{\delta \Sigma_g}{\delta X_i}\!\cdot\C\right].
 \ee
The wine vertices will then be considered from now on in a different
 basis. For the kernels: 
\bea\label{kernels}
\dDelta^{AA}=\dDelta^{\A\A} & \dDelta^{BB}=\dDelta^{\A\A}+\dDelta^{\A\A}_m\\
\dDelta^{CC}=\dDelta^{\C\C} & \dDelta^{DD}=\dDelta^{\C\C}+\dDelta^{\C\C}_m
\eea
We are ready now to proceed. In order to extract $\beta_1$ from \eq{oneeq}, let us first consider the
renormalisation condition for the coupling $g$. After the redefinition of
the gauge field $\cal A_{\mu}$ (see section \ref{sec:1.3}) we have already
mentioned that the only quantity which renormalises is now the gauge
coupling itself. We can then set the
renormalisation condition asking for $g$ to be, in the expansion of $S(\Lam)$, the coefficient of the
quadratic term in the two-point function for the bosonic gauge field $A$ at
order $p^2$ (following \cite{mor:erg2}), in order to have the physical
$\beta$-function for $SU(N)$ Yang-Mills. In formulae it is expressed in
\eq{defg}. That expression is non-perturbative. At any order, this means;
\be\label{pininfarina}
S^{AA}_{\,\mu\,\nu}+S^{AA\sigma}_{\,\mu\,\nu}=\frac{2}{g^2(\Lam)}\BBox_{\,\mu\,\nu}(p)+{\cal O}(p^3)
\ee
where we define $\BBox_{\mu\nu}(p)=p^2\delta_{\mu\nu}-p_{\mu}p_{\nu}$, and
\bea
S^{AA}_{\,\mu\,\nu}&=&\frac{1}{g^2}S^{(0)AA}_{\ \ \ \,\mu\,\nu}+S^{(1)AA}_{\ \ \ \,\mu\,\nu}+\cdots\\
S^{AA\sigma}_{\,\mu\,\nu}&=&\frac{1}{g^2}S^{(0)AA\sigma}_{\ \ \ \,\mu\,\nu}+S^{(1)AA\sigma}_{\ \ \ \,\mu\,\nu}+\cdots\nonumber\\
&=&S^{(1)AA\sigma}_{\ \ \ \,\mu\,\nu}+\cdots
\eea
but at tree level we will see we already have:
\be\label{trillo}
S^{(0)AA}_{\ \ \ \,\mu\,\nu}=2\BBox_{\,\mu\,\nu}(p)+{\cal O}(p^3)
\ee
The condition is already fulfilled, so:
\be
\frac{1}{g^2}S^{(0)AA}_{\ \ \ \,\mu\,\nu}+\sum_{n=1}
g^{2n-2}(S^{(n)AA}_{\ \ \ \,\mu\,\nu}+S^{(n)AA\sigma}_{\ \ \ \,\mu\,\nu})=\frac{1}{g^2}\left.S^{(0)AA}_{\ \ \ \,\mu\,\nu}\right|_{p^2}+{\cal
O}(p^3)
\ee
This means that as for the scalar field case, considering the one-loop
equation \eq{oneeq} if we take the combination
$S^{AA}_{\,\mu\,\nu}+S^{AA\sigma}_{\,\mu\,\nu}$ in the expansion of $S$ and its order $p^2$,
we have a great simplification due to the term on the LHS which now
vanishes. Moreover, we can simplify even further \eq{oneeq} if we take into
account the freedom we still have on the choice of $\sh$. As we have
mentioned in the previous chapter, if we choose it
so that all its two point functions are equal to the two point effective
action at tree level, the classical term in \eq{oneeq} is zero (for the
component in the expansion we are looking for). In formulae:
\be\label{s-shat}
\hat{S}^{X_1X_2}_{\,a_1\,a_2}=S^{X_1X_2}_{\,a_1\,a_2}
\ee
where $X_i$ can be any field and $a_i$ is a Lorentz index if $X_i$ is a
gauge field and nothing otherwise. This is not a big restriction and it is
the same request we set in the scalar field case. It just amounts, as we
pointed out in section \ref{sec:1.4} for the scalar field, to asking the two-point effective tree
level vertices, to flow keeping the same functional dependence upon
$\Lam$. It is an arbitrary choice which is worth taking since it greatly
simplify the calculation. These two-point vertices can be almost uniquely
determined via dimensional analysis, gauge invariance (in the next section
the Ward identities relating them are considered) and recalling they must
be derived from $SU(N|N)$ theory used as a regulator. We will briefly
discuss their form here. 

The first one, the $AA$ vertex, turns out to be the most general transverse function of
$p^2$ of dimension $\Lam^2$. The third, the $CC$ vertex, must be a general function of $p^2$, of
dimension $4$, which is not zero at $p=0$. Since we want to interpret this
term as a mass for $C$ (of order the
cutoff $\Lambda$, in order for the regularisation to work), we have also
to choose its coefficient positive. The $BB$ vertex, does not have to be transverse, and it is written as a combination
of a transverse term and a non transverse one. The former is chosen equal to the $AA$ one, in order for the
regularisation to work (so to have the right cancellations of the
propagators at high momenta). The latter has a form, constrained in order to
produce through Ward identities the last two in the list ($BD\sigma$ first,
and from it $DD$). The $DD$ one is required to have the momentum dependence of $C$'s for the
regularisation to work, and it must vanish at $p=0$, since it's the kinetic
term of a Goldstone (massless) field. Finally $BD\sigma$ is obtained by dimensional analysis and
through Ward identities from the $BB$ vertex.
A list of two-point vertices with such requirements, follows below:
\bea
\hat{S}^{AA}_{\,\mu\,\nu}(p)&=&2\frac{\BBox_{\mu\nu}}{c_p}\equiv \hat{S}_{\mu\nu}\label{hataa}\\
\hat{S}^{BB}_{\,\mu\,\nu}(p)&=&\hat{S}_{\mu\nu}+4\frac{\Lambda^2}{\tilde{c}_p}\delta_{\mu\nu}\label{hatbb}\\
\hat{S}^{CC}(p)&=&\frac{\Lambda^2 p^2}{\ct_p}+2\lambda\Lambda^4\label{hatcc}\\
\hat{S}^{DD}(p)&=&\frac{\Lambda^2 p^2}{\ct_p}\label{hatdd}\\
\hat{S}^{BD\sigma}_{\,\mu}(p)&=&-2\frac{\Lambda^2 p_{\mu}}{\ct_p}\label{hatbds}
\eea
where $\BBox_{\mu\nu}$ was defined below \eq{pininfarina}, $c$ and
$\tilde{c}$ are general functions of $x=p^2/\Lam^2$ and $\lambda$ is
a constant. The only necessary requirement on $c$ is $c_0=1$, from the
normalisation condition and \eq{trillo}, as for the scalar field. We will
also require $\tilde{c}_0=1$ and, as we have mentioned earlier, in order for
$C$ to have a mass, we choose $\lambda >0$.
In principle we could have
chosen two pairs of different functions, $c$ and $\hat{c}$ for the $A$ and the
$B$ and $\tilde{c}$ and $\hat{\tilde{c}}$ for the $C$
and the $D$ vertices. In fact, even though they come from the same supermultiplets, when
one considers the broken phase, the two pairs of two-point function, can
pick different contributions due to their different statistics. Although
the most general requirements would then be that the functions chosen must
have the same behaviour as $p/\Lam\to \infty$, we decide to choose them
equal. 

The form of $\sh$ is then still left quite general. Apart from having to
preserve all the symmetries of the theory, and the request
expressed by \eq{s-shat} on its two-point vertices, we just add as for the
scalar field case, the restriction on its higher vertices to be Taylor
expandable to any order and to keep UV finite all the integral in which they appear.

Considering again \eq{oneeq} we are then left with:
\be\label{1-loop}
a_1[\Sigma^{(0)}]^{AA}_{\,\mu\,\nu}+a_1[\Sigma^{(0)}]^{AA\sigma}_{\,\mu\,\nu}=-4\beta_1\BBox_{\,\mu\,\nu}+{\cal
O}(p^3)
\ee
where $\Sigma^{(0)X_1X_2}_{\ \ \ \,a_1\,a_2}=S^{(0)X_1X_2}_{\ \ \ \,a_1\,a_2}-2\hat{S}^{X_1X_2}_{\,a_1\,a_2}$. Because of (\ref{z2}) 
though, the 1-loop vertices must contain a $\sigma$, so \eq{1-loop}
becomes finally
\be\label{ooo}
a_1[\Sigma^{(0)}]^{AA\sigma}_{\,\mu\,\nu}=-4\beta_1\BBox_{\,\mu\,\nu}+{\cal
O}(p^3)
\ee
Computing all the diagram contributing to the vertex $S^{AA\sigma}_{\,\mu\,\nu}$
at one loop, and considering the form of the flow equation, we can find the
following expression for \eq{ooo}:
\bea
-4\beta_1\BBox_{\,\mu\,\nu}(p^2)&=&\nonumber\\[3pt]
&\!\!\!\!\!\!\!\!\!\!\!\!\!\!\!\!2N\displaystyle\int_k&\!\!\!\!\!\left\{\dDelta^{AA}_k\Sigma^{AAAA}_{\,\alpha\,\alpha\,\mu\,\nu}(-k,k,-p,p)+\dDelta^{A,AA}_{\,\mu}(p;k-p,-k)\Sigma^{AAA}_{\,\alpha\,\alpha\,\nu}(p-k,k,-p)\right.\nonumber\\[3pt]
&&\!\!\!\!\!\!\!\!\!\!\!\!\!+\dDelta^{AA,AA}_{\,\mu\,\nu}(p,-p;k,-k)\Sigma^{AA}_{\,\alpha\,\alpha}(k)\nonumber\\[3pt]
&&\!\!\!\!\!\!\!\!\!\!\!\!\!-\dDelta^{BB}_k\Sigma^{BBAA}_{\,\alpha\,\alpha\,\mu\,\nu}(-k,k,p,-p)-\dDelta^{A,BB}_{\,\mu}(p;k-p,-k)\Sigma^{BBA}_{\,\alpha\,\alpha\,\nu}(p-k,k,-p)\nonumber\\[3pt]
&&\!\!\!\!\!\!\!\!\!\!\!\!\!-\dDelta^{AA,BB}_{\,\mu\,\nu}(p,-p;k,-k)\Sigma^{BB}_{\,\alpha\,\alpha}(k)\nonumber\\[3pt]
&&\!\!\!\!\!\!\!\!\!\!\!\!\!+\dDelta^{CC}_k\Sigma^{CCAA}_{\ \ \ \,\,\mu\,\nu}(-k,k,p,-p)+\dDelta^{A,CC}_{\,\mu}(p;k-p,-k)\Sigma^{CCA}_{\ \ \ \,\,\nu}(p-k,k,-p)\nonumber\\[3pt]
&&\!\!\!\!\!\!\!\!\!\!\!\!\!+\dDelta^{AA,CC}_{\,\mu\,\nu}(p,-p;k,-k)\Sigma^{CC}(k)\nonumber\\[3pt]
&&\!\!\!\!\!\!\!\!\!\!\!\!\!-\dDelta^{DD}_k\Sigma^{DDAA}_{\ \ \
\,\,\mu\,\nu}(-k,k,p,-p)-\dDelta^{A,DD}_{\,\mu}(p;k-p,-k)\Sigma^{DDA}_{\ \ \ \,\,\nu}(p-k,k,-p)\nonumber\\[3pt]
&&\!\!\!\!\!\!\!\!\!\!\!\!\!\left.\left.-\dDelta^{AA,DD}_{\,\mu\,\nu}(p,-p;k,-k)\Sigma^{DD}(k)\right\}\right|_{p^2}\label{5.19}
\eea
where $\int_k=\int\frac{d^Dk}{(2\pi)^D}$ and all the $\Sigma$'s are at tree level but the subscript ``0'' has been
omitted. What we have to find at this stage are the vertices at tree level
shown in Tab.\ref{tab:vertices}. 
\begin{table}[h!]
\begin{center}
\begin{tabular}{|c|c|c|}
\hline
\ \ {\bf 4-point}&{\bf 3-point}&{\bf 2-point}\ \ \\
\hline
\hline
& &\\[-0.3cm]
\ \
$S^{(0)AAAA}_{\ \ \ \,\mu\,\nu\,\rho\,\sigma}(p,q,r,s)$&$S^{(0)AAA}_{\ \ \ \,\mu\,\nu\,\rho}(p,q,r)$&$S^{(0)AA}_{\,\mu\,\nu}(p)$\
\ \\
& &\\[-0.3cm]
\ \ $S^{(0)BBAA}_{\ \ \ \,\mu\,\nu\,\rho\,\sigma}(p,q,r,s)$&$S^{(0)BBA}_{\ \ \ \,\mu\,\nu\,\rho}(p,q,r)$&$S^{(0)BB}_{\,\mu\,\nu}(p)$\
\ \\
& &\\[-0.3cm]
\ \ $S^{(0)CCAA}_{\ \ \ \ \ \ \,\,\mu\,\nu}(p,q,r,s)$&$S^{(0)CCA}_{\ \ \ \ \ \ \,\,\mu}(p,q,r)$&$S^{(0)CC}(p)$\ \ \\
&& \\[-0.3cm]
\ \ $S^{(0)DDAA}_{\ \ \ \ \ \ \,\,\mu\,\nu}(p,q,r,s)$&$S^{(0)DDA}_{\ \ \ \ \ \ \,\,\mu}(p,q,r)$&$S^{(0)DD}(p)$\ \ \\
&& \\[-0.3cm]
\hline
\hline
&& \\[-0.3cm]
&$S^{(0)ABD\sigma}_{\ \ \ \,\mu\,\nu}(p,q,r)$&$S^{(0)BD\sigma}_{\,\mu}(p)$\\
&& \\[-0.3cm]
&$S^{B(0)AD\sigma}_{\ \ \ \,\mu\,\nu}(p,q,r)$&\\[.3cm]
\hline
\end{tabular}
\end{center}
\caption{Tree level vertices needed to calculate $\beta$ at one loop}\label{tab:vertices}
\end{table}
The last two lines list the vertices needed in order to calculate the others. The two three-point vertices in the second part of Table \ref{tab:vertices}
are related via a symmetry of the theory, which will be discussed in the
next section, called ``Charge Conjugation''. In particular: $S^{ABD\sigma}_{\,\mu\,\nu}(p,q,r)=-S^{BAD\sigma}_{\,\nu\,\mu}(q,p,r)$.

\section{Symmetries}\label{simsim}

Before we continue studying the equations of the vertices needed to perform
our calculation, it is worth spending some words about the symmetries of
the theory we are considering. 

\begin{figure}[h]
\psfrag{p}{$p$}
\psfrag{qnt}{$q_{\nu}$}
\psfrag{qnb}{$q^{\nu}$}
\psfrag{r}{$r$}
\psfrag{q}{$q$}
\psfrag{=}{$=$}
\psfrag{+}{$+$}
\psfrag{-}{$-$}
\begin{center}
\includegraphics[scale=.6]{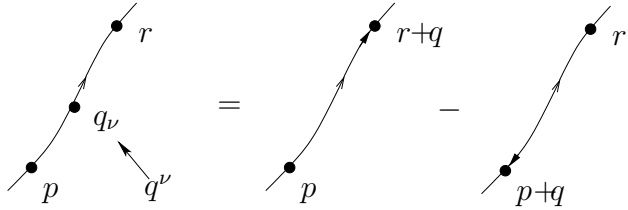}
\end{center}
\caption{Graphical representation of gauge invariance identities.}\label{fig:gi}
\end{figure}

The main symmetry involved here is, of course, supergauge invariance under
$SU(N|N)$ transformation. This invariance must hold for the action we want
to describe, and it is preserved along its flow due to the properties of the
flow equation and the way it has been constructed. Expanding the action in
fields and imposing it is invariant under applying to them the
transformations expressed by \eqs{Agauged}{Cgauged}, it is possible to get
constraint on the vertices, \ie Ward identities. These identities are
expressed by eqs. (\ref{bosonwi}), (\ref{fermionwi}) and (\ref{winewi}), while a  pictorial representation
is given in \fig{fig:gi}.
This symmetry is of great importance and preserving it in the
flowing action was the purpose of the whole thesis. It has been used
through the various steps of the calculation of $\beta$ at one-loop as a
check, helping us to be sure it was correctly performed. Since the
supergauge theory will have to be spontaneously broken, in order to the
regularisation to work, it is convenient to find an expansion for the
action in terms of a different basis, \ie the shifted fields. Imposing the corresponding
transformations to hold, it is possible then to find the Ward identities
for the vertices in the broken phase, which is the basis that will be used
in the calculation. This issue will be addressed in section \ref{sub:bwi}. 

\begin{figure}[h]
\psfrag{mu}{$\mu$}
\psfrag{nu}{$\nu$}
\psfrag{rho}{$\rho$}
\begin{center}
\includegraphics[scale=.3]{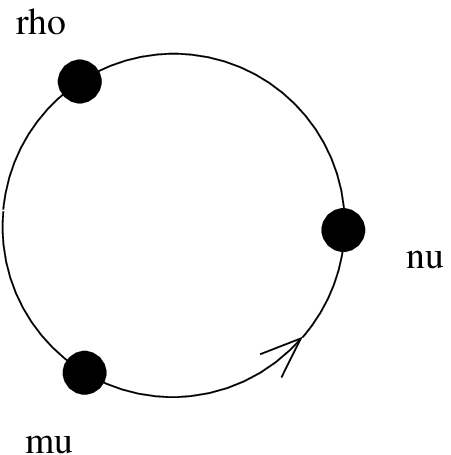}
\end{center}
\caption{Diagram representing the three vertices of \eq{trea}}\label{fig:trea}
\end{figure}

Another relevant symmetry involved here, and which follows directly from
the construction of the action expanded in supertraces, is the
cyclicity. This property ensures that vertices can be set equal if their
arguments (\ie momenta and indices), are related by cyclic permutations. An example of this property is expressed
below for the three-point vertex of pure $A$:
\be\label{trea}
S^{AAA}_{\,\mu\,\nu\,\rho}(p,q,r)=S^{AAA}_{\,\nu\,\rho\,\mu}(q,r,p)=S^{AAA}_{\,\rho\,\mu\,\nu}(r,p,q)
\ee
Cyclicity is automatically incorporated in the diagrammatic representation
and it is one of the reasons that makes it so powerful. As an example, all
the three vertices of \eq{trea} are represented by the same diagram in
\fig{fig:trea}

\begin{figure}[h]
\psfrag{a1}{$a_1$}
\psfrag{a2}{$a_2$}
\psfrag{an}{$a_n$}
\psfrag{(-)^r}{$(-)^r$}
\begin{center}
\includegraphics[scale=.7]{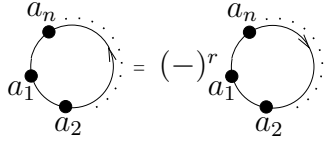}
\end{center}
\caption{Diagrammatic representation of CC invariance}\label{fig:cctr}
\end{figure}

A further symmetry, which we have already mentioned in the previous section
is ``Charge Conjugation'' (CC). This invariance has to hold since the final
goal is to describe a $SU(N)$ Yang-Mills theory, which has this symmetry, and it is imposed by asking the
action to be invariant under the transformation:
\bea\label{CCtran}
\A_{\mu}&\to&-\A_{\mu}^T\\
\C&\to&\C^T
\eea
This requirement sets as well relations on the vertices, and in
particular sets equal those with inverted set of arguments, up to a minus
sign to the power of the number of gauge fields. Namely:
\be
S^{X_1\cdots X_n}_{\,a_1\cdots a_n}(p_1,\dots,p_n)=(-)^r S^{X_n\cdots
X_1}_{\,a_n\cdots a_1}(p_n,\dots,p_1)
\ee
where $r$ is the number of gauge fields in $X_1,\dots,X_n$. This property is easily
expressed in diagrammatic form, as can be seen in \fig{fig:cctr}.
Through CC symmetry, it has been possible to rule out some vertices, which
might have appeared in the action otherwise, and might have caused
trouble. One of them is the two-point vertex $AC\sigma$, which, assuming the
action invariant for charge conjugation, is identically zero (it also
vanishes for other reasons, such as gauge invariance and Lorentz invariance). This fact
will lead the two equations for $AA$ and $CC$ to be decoupled, while those
for $BB$ and $DD$ will be coupled through and with the equation for the non-zero vertex $BD\sigma$.
Other important properties can be extracted imposing this symmetry , also
on wine vertices. One of them, widely used throughout the whole thesis,
relates one-point wine vertices and reads (in the case of a pure gauge wine vertex):
\be
\dDelta^{A,AA}_{\,\mu}(p;q,r)=-\dDelta^{A,AA}_{\,\mu}(p;r,q)
\ee

\begin{figure}[h]
\psfrag{si}{$\sigma$}
\psfrag{mu}{$\mu$}
\psfrag{nu}{$\nu$}
\psfrag{rho}{$\rho$}
\psfrag{+}{$+$}
\psfrag{0}{$0$}
\psfrag{=}{$=$}
\begin{center}
\includegraphics[scale=.85]{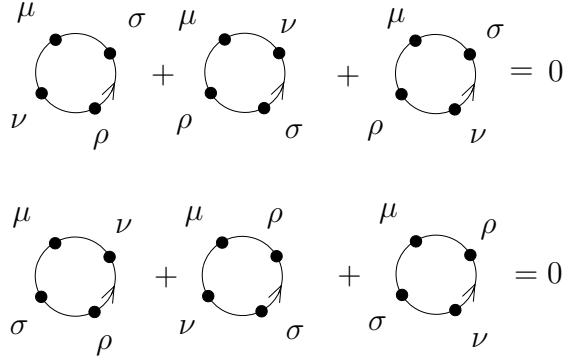}
\end{center}
\caption{Diagrammatic representation of no-$\A^0$ symmetry for the
four-point pure $A$ vertex.}\label{fig:noa0}
\end{figure}

The last invariance we will describe in this section and which was already
discussed in the previous chapter, is the no-$\A^0$ symmetry. As it was
already discussed there, this symmetry has to be imposed for the action as
we start considering the supergauge field $\A_{\mu}$ containing the
identity matrix in its expansion on the generators of $SU(N|N)$, if we do
not want its presence to create a non linear constraint on the
theory. Its requirement reflects on the vertices through a set of
constraints, obtained imposing the transformation
\be\label{noa0}
\delta\A^0_{\mu}=\Lam_{\mu}(x),
\ee
to be an invariance for the expanded action.
Given for example the four-point pure $A$ vertex, we can imagine for
example that one of them were an $\A^0_{\mu}$. In this case invariance for
\eq{noa0} would give the following relation:
\be\label{noa01}
S^{AAAA}_{\,\mu\,\nu\,\rho\,\sigma}(p,q,r,s)+S^{AAAA}_{\,\mu\,\rho\,\sigma\,\nu}(p,r,s,q)+S^{AAAA}_{\,\mu\,\rho\,\nu\,\sigma}(p,r,q,s)=0
\ee
If two of them were $\A^0_{\mu}$, the symmetry would give us a different
constraint:
\bea
&S^{AAAA}_{\,\mu\,\nu\,\rho\,\sigma}(p,q,r,s)+S^{AAAA}_{\,\mu\,\rho\,\sigma\,\nu}(p,r,s,q)+S^{AAAA}_{\,\mu\,\rho\,\nu\,\sigma}(p,r,q,s)+\nonumber\\
&S^{AAAA}_{\,\mu\,\sigma\,\nu\,\rho}(p,s,q,r)+S^{AAAA}_{\,\mu\,\sigma\,\rho\,\nu}(p,s,r,q)+S^{AAAA}_{\,\mu\,\nu\,\sigma\,\rho}(p,q,s,r)=0
\eea
This second relation, since must hold together with the first one give us
in particular:
\be\label{noa02}
S^{AAAA}_{\,\mu\,\sigma\,\rho\,\nu}(p,s,r,q)+S^{AAAA}_{\,\mu\,\nu\,\sigma\,\rho}(p,q,s,r)+S^{AAAA}_{\,\mu\,\sigma\,\nu\,\rho}(p,s,q,r)=0
\ee
\eqs{noa01}{noa02} must hold for the four-point pure $A$ vertices, for $S$
to be invariant under no-$\A^0$ symmetry. Their diagrammatic form is
expressed in \fig{fig:noa0}. As can be seen more clearly from
\fig{fig:noa0}, the two previous equations are the same
constraint through Charge Conjugation invariance.
If three or four of them were $\A^0$'s we would not get any constraint
since, we would be left respecively with str$(\A_{\mu})$ and str$(\one)$ which
are already zero.

\noindent Relations like these can be found for all possible vertices. This
symmetry will be particularly useful in the calculation for the second
coefficient of the $\beta$-function, as can be found in
\cite{sequel}. As far as the present work is concerned, this symmetry played
a crucial r\^{o}le together with CC invariance, in order to rule out the
vertex $AAC\sigma$. As we will see in the last chapter, this allowed us to
treat the three different sectors $A$, $C$ and the sector $B$ and
$D$\footnote{$B$ and $D$ will be considered together, for reasons which will become clear later}, in a similar way. 

Other important consequences of this invariance, relevant to
the present work were already described in section \ref{sub:1.2.2} and will
not be repeated here.

\subsection{Ward identities in the broken phase}\label{sub:bwi}

As we have already mentioned earlier in this chapter, supergauge invariance
is the most relevant symmetry throughout the present work, and since we will be
working in the broken phase of the theory, it is worth exploring the Ward
identities in this regime.
According to the split
fields notation which has just been introduced in \eq{fieldsbroken}, they read:
\bea
q^{\nu}U^{\cdots X A Y\cdots}_{\,\cdots a \,\nu\, b\cdots}(\cdots
p,q,r,\cdots)&=&U^{\cdots X Y\cdots}_{\,\cdots a\, b\cdots}(\cdots
p,q+r,\cdots)-U^{\cdots X Y\cdots}_{\,\cdots a  \,b\cdots}(\cdots
p+q,r,\cdots)\nonumber\\
\label{bosonwi}\\
q^{\nu}U^{\cdots X B Y\cdots}_{\,\cdots a \,\nu\, b\cdots}(\cdots
p,q,r,\cdots)&=&\pm U^{\cdots X \hat{Y}\cdots}_{\,\cdots a  \,b\cdots}(\cdots
p,q+r,\cdots)\mp U^{\cdots \hat{X} Y\cdots}_{\,\cdots a \, b\cdots}(\cdots
p+q,r,\cdots)\nonumber \\[3pt]
&&+2 U^{\cdots X D\sigma Y\cdots}_{\,\cdots a \ \ \ \,\,b\cdots}(\cdots
p,q,r,\cdots)\nonumber \\
\label{fermionwi}
\eea
The notation of Eqs.(\ref{bosonwi}) and (\ref{fermionwi}) needs a brief
explanation. $U$ represent any vertex either from the expansion of the
effective action or a wine vertex. $X$ and $Y$ are generic fields and the
indices $a$ and $b$, respectively referring to $X$ and $Y$, are either Lorentz
indices or nothing, according to the scalar or vectorial nature of $X$ and
$Y$. The ``$\wedge$'' on the fields in the RHS of eq.(\ref{fermionwi})
indicates a change in the spin-statistic nature of the fields, \ie
$\hat{A}_{\mu}=B_{\mu}$, $\hat{B}_{\mu}=A_{\mu}$ and so on. Finally,
the signs in the first line of eq.(\ref{bosonwi}) are $+$ and $-$ if there
is no $\sigma$ between $X$ and $B$ and $Y$ and $B$, and opposite sign in
the other case.

We write separately the identities for wine vertices where the point is at
the end of the line:
\bea\label{winewi}
p^{\mu_1}_1 W_{\,\mu_1\cdots\mu_n,\nu_1\cdots\nu_m}^{A X_2\cdots
X_n,Y_1\cdots Y_m,Z_1Z_2}(p_1,\dots,p_n;q_1,\dots,q_m;r,s)=&\\
 &&\!\!\!\!\!\!\!\!\!\!\!\!\!\!\!\!\!\!\!\!\!\!\!\!\!\!\!\!\!\!\!\!\!\!\!\!\!\!\!\!\!\!\!\!\!\!\!\!\!\!\!\!\!\!\!\!\!\!\!W_{\,\mu_2\cdots\mu_n,\nu_1\cdots\nu_m}^{X_2\cdots
X_n,Y_1\cdots Y_mZ_1Z_2}(p_1+p_2,p_3,\dots,p_n;q_1,\dots,q_m;r,s)\nonumber\\
 &&\!\!\!\!\!\!\!\!\!\!\!\!\!\!\!\!\!\!\!\!\!\!\!\!\!\!\!\!\!\!\!\!\!\!\!\!\!\!\!\!\!\!\!\!\!\!\!\!\!\!\!\!\!\!\!\!\!\!\!-W_{\,\mu_2\cdots\mu_n,\nu_1\cdots\nu_m}^{X_2\cdots
X_n,Y_1\cdots Y_mZ_1Z_2}(p_2,\dots,p_n;q_1,\dots,q_m;r+p_1,s)\nonumber
\eea
If the field on the wine hit by $p^{\mu_1}_1$ were a $B$ we would have had
a relation similar to \eq{fermionwi}. Similar identities would be obtained
hitting the vertex with momenta $p^{\mu_n}_n$, $q^{\nu_1}_n$ and
$q^{\nu_m}_m$ as is clear from \fig{fig:gi}.

We can now see them in some particular examples. For the two-point vertices
in the effective action expansion (the same applies to the $\hat{S}$ vertices), for example, we have:
\bea
p^{\mu} S^{AA}_{\,\mu\,\nu}(p)&=&0\label{wi1}\\
p^{\mu} S^{BB}_{\,\mu\,\nu}(p)&=&-2S^{BD\sigma}_{\,\nu}(p)\label{wi2}\\
p^{\mu} S^{BD\sigma}_{\,\mu}(p)&=&-S^{DD}(p)\label{wi3}
\eea
The first one, simply states that in the $A$ sector the gauge invariance
is not broken, and the $A$ propagator is still transverse. The last two,
relate vertices in the broken sector. Another observation
which can be done is that, since the seed action $\hat{S}$ must be gauge
invariant itself, (\ref{bosonwi}) and (\ref{fermionwi}) must apply to its
vertices. It is easy to check for the two-point ones, since we have written 
explicitly the expressions for their identities in (\ref{wi1}-\ref{wi3}). It is
straightforward to see that (\ref{wi1}) is true for (\ref{hataa}) since it
is proportional to the transverse tensor $\BBox_{\mu\nu}(p)$. Moreover, we
can see that, applying $p^{\mu}$ to (\ref{hatbb}) and comparing with (\ref{hatbds}), we get
\be
p^{\mu}\hat{S}^{BB}_{\,\mu\,\nu}(p)=p^{\mu}\left(\hat{S}^{AA}_{\,\mu\,\nu}(p)+4\frac{\Lambda^2}{\tilde{c}_p}\delta_{\mu\nu}\right)=
4\frac{\Lambda^2}{\tilde{c}_p}p_{\nu}=-2 \sh^{BD\sigma}_{\,\nu}(p)
\ee
and applying it to (\ref{hatbds}) and comparing with (\ref{hatdd}):
\be
p^{\mu} \sh^{BD\sigma}_{\,\mu}(p)=p^{\mu} \left( -2\frac{\Lambda^2
p_{\mu}}{\ct_p}\right)=-2\frac{\Lambda^2 p^2}{\ct_p}=-2\sh^{DD}(p)
\ee
Another check which can be done is on the two-point function
equations (\ref{treeq:aa}) and (\ref{bb})-(\ref{dd}). If everything is consistent the first
should give zero if contracted with $p^{\mu}$ and the others should be
connected by Ward identities. This is indeed the case, as it can be
verified applying (\ref{bosonwi}) and (\ref{fermionwi}) to them (the same
of course should apply for the higher point equations). Relations (\ref{bosonwi}) and (\ref{fermionwi}) will be useful for
calculations and other checks in the present section and in the next chapter.

\section{Tree level vertices}

\begin{figure}[h]
\psfrag{sum}{$\dsum_{f}$}
\psfrag{Xi}{$\tiny f$}
\psfrag{+}{$+$}
\psfrag{-}{$-$}
\psfrag{=}{$=$}
\psfrag{ldl}{$\Lam\de_{\Lam}$}
\psfrag{-1/2l}{$-\dfrac{1}{2}$}
\psfrag{A}{$\cal A$}
\psfrag{C}{$\cal C$}
\psfrag{S}{$S$}
\psfrag{Si}{$\Sigma_0$}
\begin{center}
\includegraphics[scale=.5]{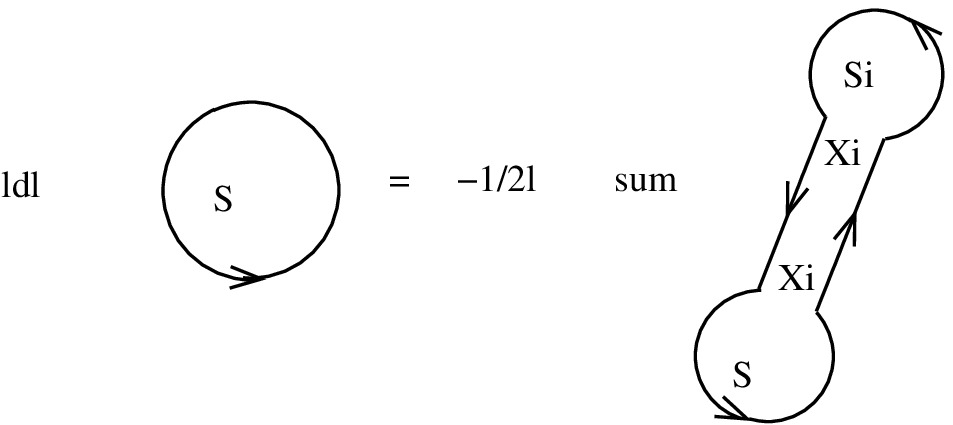}
\end{center}
\caption{Tree level equation; $f$ can be either an $A$, $B$, $C$ or $D$ field }\label{fig:treeq}
\end{figure}

After having discussed the importance of the symmetries of the flowing
effective action and their consequences on its vertices, we can now
concentrate on the equations which we need to extract the
$\beta$-function at one-loop from \eq{oneeq}. They are all listed in
Tab.\ref{tab:vertices}. Since they are all tree level vertices, let us
consider first the tree level equation (\ref{ergcl}) which we rewrite here in
its extended form:
\be\label{eq:tree}
\Lam\frac{\de S_0}{\de \Lam}=\sum_{f=\A,\C}{1\over2}\frac{\delta
S_0}{\delta f}\{\dDelta^{ff}\}\frac{\delta \Sigma_0}{\delta f}
\ee
where just in order to have a more compact equation the incorporated
$\C$-wines notation has been used. The previous equation is shown
diagrammatically in \fig{fig:treeq}.

\noindent We can now specify \eq{eq:tree} to the vertices of
Tab \ref{tab:vertices}.

\subsection{Two point tree level vertices and kernels}\label{twopoint&Kernels}

\begin{figure}[h]
\psfrag{sim}{$=$}
\psfrag{=}{$=$}
\psfrag{Wp}{$\dDelta^{AA}$}
\psfrag{Kp}{$\dDelta^{BB}$}
\psfrag{Hp}{$\dDelta^{CC}$}
\psfrag{Gp}{$\dDelta^{DD}$}
\psfrag{Am}{$A_{\mu}$}
\psfrag{Bm}{$B_{\mu}$}
\psfrag{C}{$C$}
\psfrag{D}{$D$}
\begin{center}
\includegraphics[scale=.5]{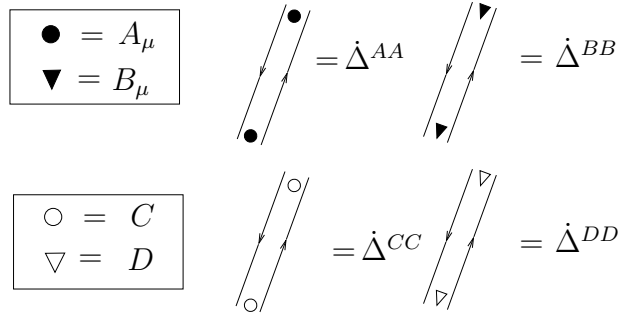}
\end{center}
\caption{Graphical representation of 0-point wines}\label{fig:0wines}
\end{figure}

We will start studying the equations for the two-point vertices. The kernels (zero-point wine
vertices) in the basis we are working (the ones listed in \eq{kernels}),
will be determined here through the request of the seed action two-point
vertices to equal the tree level ones. They are represented diagrammatically in
\fig{fig:0wines}. 

The first step is now, once the equations for the two-point vertices are
written, to assume the request expressed in \eq{s-shat}. Since the form of
the vertices of the two point is set by the argument described in section
\ref{sec:1loop}, this gives rise to a set of equations for the kernels (as
we have mentioned earlier). Because of the extra terms in the flow equation
that we have introduced through the ``decoration'' of the wines, we have
enough freedom to impose these constraint and the set of equations has
actually a solution that will be evaluated later in the present section.

\begin{figure}[h!]
\psfrag{S}{$\wedge$}
\psfrag{=}{$=$}
\psfrag{ldl}{$\Lam\de_{\Lam}$}
\psfrag{+}{$+$}
\psfrag{<->}{$\leftrightarrow$}
\psfrag{pa}{$\left(\right.$}
\psfrag{sis}{$\left.\right)$}
\psfrag{mu}{$\mu$}
\psfrag{nu}{$\nu$}
\begin{center}
\includegraphics[scale=.43]{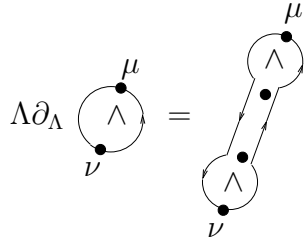}
\end{center}
\caption{$AA$ tree level equation}\label{fig:aa}
\end{figure}

The first two-point tree level equation to be studied here, is going to be
the two $A$'s. In the graphical representation, it is shown in \fig{fig:aa}. 

\noindent The equation in formulae reads:
\be\label{treeq:aa}
\Lam\de_{\Lam}\hat{S}^{AA}_{\,\mu\,\nu}(p)=\hat{S}^{AA}_{\,\mu\,\al}(p)\ \dDelta^{AA}_p \hat{S}^{AA}_{\,\al\,\nu}(p)
\ee
where $\sh^{AA}$ has the expression of \eq{hataa}. Before we solve the
equation for $\dDelta^{AA}$ we are going to list them all. 

\begin{figure}[h]
\psfrag{S}{$\wedge$}
\psfrag{Si}{$\Sigma$}
\psfrag{=}{$=$}
\psfrag{-1/2l2}{$-\dfrac{1}{2\Lam^2}$}
\psfrag{mu}{$\mu$}
\psfrag{nu}{$\nu$}
\psfrag{ldl}{$\Lam\de_{\Lam}$}
\psfrag{+}{$+$}
\psfrag{<->}{$\leftrightarrow$}
\psfrag{p}{$p$}
\psfrag{-}{$-$}
\psfrag{pa}{$\left(\right.$}
\psfrag{sis}{$\left.\right)$}
\begin{center}
\includegraphics[scale=.5]{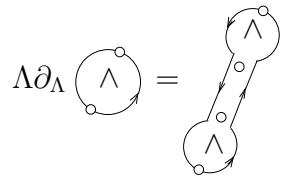}
\end{center}
\caption{$CC$ tree level equation}\label{fig:cc}
\end{figure}

The next one to be considered is then the two $C$'s
equation. Diagrammatically it is represented in \fig{fig:cc}.

\noindent While in formulae we have:
\be\label{treeq:cc}
\Lam\de_{\Lam} \hat{S}^{CC}(p)=\hat{S}^{CC}(p)\ \dDelta^{CC}_p\ \hat{S}^{CC}(p)
\ee
As one can notice, these first two equations are decoupled, and one can
solve for $\dDelta ^{AA}$ and $\dDelta^{CC}$. As we will shortly see, the
last three equations, for $BB$, $BD\sigma$ and $DD$ vertices, will be
coupled. The way they are connected, allows to write all of them in one,
introducing a compact notation for this sector, which will be
introduced in the next chapter for the gauge invariant calculation. We will
refer to this sector as the $BD\sigma$ sector. Here
they will be kept separate. 

\begin{figure}[h!]
\psfrag{S}{$\wedge$}
\psfrag{Si}{$\Sigma$}
\psfrag{=}{$=$}
\psfrag{-1/2l2}{$-\dfrac{1}{2\Lam^2}$}
\psfrag{+1/2l2}{$+\dfrac{1}{2\Lam^2}$}
\psfrag{mu}{$\mu$}
\psfrag{pa}{$\left(\right.$}
\psfrag{sis}{$\left.\right)$}
\psfrag{nu}{$\nu$}
\psfrag{ldl}{$\Lam\de_{\Lam}$}
\psfrag{-}{$+$}
\psfrag{<->}{$\leftrightarrow$}
\begin{center}
\includegraphics[scale=.5]{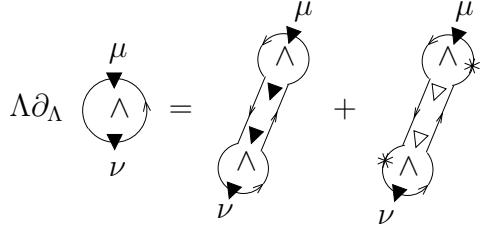}
\end{center}
\caption{$BB$ tree level equation; the ``star'' represents a $\sigma$.}\label{fig:bb}
\end{figure}

The first of the three is the equation for
the vertex $BB$. In \fig{fig:bb} it is displayed in its diagrammatic
notation.

\noindent Its analytical form is the following:
\be\label{bb}
\Lam\de_{\Lam}\hat{S}^{BB}_{\,\mu\,\nu}(p)=\hat{S}^{BB}_{\,\mu\,\al}(p)\
\dDelta^{BB}_p\hat{S}^{BB}_{\,\al\,\nu}(p)+\hat{S}^{BD\sigma}_{\,\mu}(p)\ \dDelta^{DD}_p\
\hat{S}^{BD\sigma}_{\,\nu}(p)
\ee
where for the only non-symmetric vertex in the change $p\leftrightarrow -p$, namely $\hat{S}^{BD\sigma}_{\,\mu}$,
the argument $+p$ refers to the momentum of the first field, and:
$S^{BD\sigma}_{\,\nu}(-p,p)=S^{D\sigma B}_{\,\nu}(p,-p)=S^{D\sigma B}_{\,\nu}(p)=-S^{BD\sigma}_{\,\nu}(p)$. 
The sign in the second term is then determined by an extra minus sign, from the fact that one of the
$\sigma$'s, in order to ``hit'' the other one and give the same
supertrace as the LHS ($\mbox{str}B_{\mu}B_{\nu}$), must pass through a
fermionic field. This result is more clear from its graphical representation
of \fig{fig:bb}. 

\begin{figure}[h]
\psfrag{S}{$\wedge$}
\psfrag{Si}{$\Sigma$}
\psfrag{=}{$=$}
\psfrag{-1/2l2}{$-\dfrac{1}{2\Lam^2}$}
\psfrag{mu}{$\mu$}
\psfrag{nu}{$\nu$}
\psfrag{ldl}{$\Lam\de_{\Lam}$}
\psfrag{+}{$+$}
\psfrag{<->}{$\leftrightarrow$}
\begin{center}
\includegraphics[scale=.5]{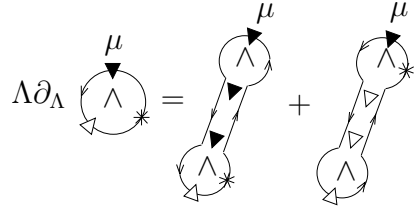}
\end{center}
\caption{$BD\sigma$ tree level equation}\label{fig:bds}
\end{figure}

The equation for the $BD\sigma$ vertex is represented graphically in
Fig.\ref{fig:bds}.

\noindent In formulae is:
\be\label{treeeq:bds}
\Lam\de_{\Lam}\hat{S}^{BD\sigma}_{\,\mu}(p)=\hat{S}^{BB}_{\,\mu\,\alpha}(p)\
\dDelta^{BB}_p\ \hat{S}^{BD\sigma}_{\,\al}(p)+\hat{S}^{BD\sigma}_{\,\mu}(p)\ \dDelta^{DD}_p\
\hat{S}^{DD}(p)
\ee

\begin{figure}[h]
\psfrag{S}{$\wedge$}
\psfrag{Si}{$\Sigma$}
\psfrag{=}{$=$}
\psfrag{-1/2l2}{$-\dfrac{1}{2\Lam^2}$}
\psfrag{mu}{$\mu$}
\psfrag{nu}{$\nu$}
\psfrag{ldl}{$\Lam\de_{\Lam}$}
\psfrag{+}{$+$}
\psfrag{<->}{$\leftrightarrow$}
\psfrag{p}{$p$}
\psfrag{-}{$-$}
\psfrag{pa}{$\left(\right.$}
\psfrag{sis}{$\left.\right)$}
\begin{center}
\includegraphics[scale=.5]{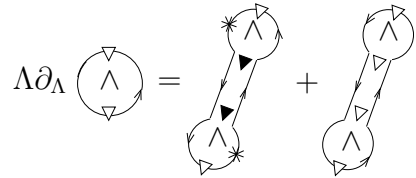}
\end{center}
\caption{$DD$ tree level equation}\label{fig:dd}
\end{figure}

Finally, the equation for the last two-point vertex, $DD$, is represented
graphically in Fig.\ref{fig:dd} 
and in formulae in:
\be\label{dd}
\Lam\de_{\Lam}\hat{S}^{DD}(p)=\hat{S}^{BD\sigma}_{\,\al}(p)\
\dDelta^{BB}_p\ \hat{S}^{BD\sigma}_{\,\al}(p)+\hat{S}^{DD}(p)\ \dDelta^{DD}_p\ \hat{S}^{DD}(p)
\ee

We are now ready to extract the four kernels from these five
equations. The first two equations give us two conditions to determine the
two $A$'s and two $C$'s kernels and the last three for the $B$ and $D$
ones. As one might think the set of three coupled equations is not
overconstrained because they are related to each other via Ward identities. It
is in fact possible to check that starting from the equation for the two
$B$'s vertex, one can obtain the other two via contracting respectively once or
twice with a momentum $p^{\mu}$ (respectively $-p^{\nu}$)\footnote{This and others were some of the
checks made possible by gauge invariance, that let us keep under control
the calculation}. They set then two conditions: one is given
by the equation for the transverse part of $\sh^{BB}_{\mu\nu}$, the other
one by the equation for its non transverse part, or by one of the last two
equations related via gauge invariance. What we find are function of
$x=p^2/\Lam^2$, written in terms of the two functions $c$ and $\tilde{c}$
introduced in Eqs.(\ref{hataa})-(\ref{hatbds}). They will be the momentum
space kernels, which will need to be covariantised as described in section
\ref{Covariantization}. Solving the two-point equations for the kernels, we
finally get:
\bea
\dDelta^{AA}_p&=&\frac{1}{\Lam^2}c'\label{ker:a}\\
\dDelta^{BB}_p&=&\frac{1}{\Lam^2}\left(\frac{xc\tilde{c}}{x\tilde{c}+2 c}\right)'\label{ker:b}\\
\dDelta^{CC}_p&=&\frac{1}{\Lam^4x}\left(\frac{x^2\tilde{c}}{x+2\lambda c}\right)'\label{ker:c}\\
\dDelta^{DD}_p&=&\frac{1}{\Lam^4x}\left(\frac{2x^2\tilde{c}^2}{x\tilde{c}+2 c}\right)'\label{ker:d}
\eea
where $c$ and $\tilde{c}$ are intended as functions of $x$ and the prime
stands for the derivative with respect to this argument.

We just spend a few words on how the equations may be solved. For the two decoupled ones, it was just a question of substituting
the two-point vertices and solve for the kernels. For the coupled ones, a
possible way was to contract the equation for the $BB$ vertex with the
combination $\BBox_{\mu\nu}$. In such a way, in \eq{bb} we are left with
the first member, containing only $\dDelta^{BB}$. It is then possible to
solve directly for that kernel, and substituting it in the equation for
the two $D$'s vertex, get the final kernel $\dDelta^{DD}$.

It is now time to move to the higher point vertex equations.

\subsection{Three and four point tree level vertices}\label{subsec:2.1.5}

The equations we finally have to find are the ones for the four-points and three points
vertices which are needed to extract the 1-loop $\beta$-function (plus the
three point needed to find the four point). Some of these vertices are directly needed to be substituted in the 1-loop equation.
Others will have to be used to calculate the latter. 

An important remark must be done at this point. Although we would expect
these tree level vertices to be finite, as will be discussed in the next
two sections, 
this is not true in general (for example if $\sh$
is the one chosen in \cite{su:pap}). However, this
problem can be fixed with an appropriate shift of the hatted vertices. This
will be shown for an explicit example in the last section of this
chapter. We will then have to add a further request on the seed action,
which is to ensure there are no classical divergences.

\begin{figure}[h]
\psfrag{pm}{\tiny $p_{\mu}$}
\psfrag{qa}{\tiny $q_{\al}$}
\psfrag{rb}{\tiny $r_{\beta}$}
\psfrag{Hm1}{$\dDelta^{A,CC}_{\mu}(p;q,r)$}
\psfrag{Hm2}{$\dDelta^{B,CD}_{\mu}(p;q,r)$}
\psfrag{Hm3}{$\dDelta^{B,DC}_{\mu}(p;q,r)$}
\psfrag{=}{$=$}
\psfrag{cpm1}{$\dDelta^{A,AA}_{\mu}(p;q,r)$}
\psfrag{cpm2}{$\dDelta^{B,AB}_{\mu}(p;q,r)$}
\psfrag{cpm3}{$\dDelta^{B,BA}_{\mu}(p;q,r)$}
\psfrag{Km}{$\dDelta^{A,BB}_{\mu}(p;q,r)$}
\psfrag{Gm}{$\dDelta^{A,DD}_{\mu}(p;q,r)$}
\psfrag{m1}{$\dDelta^{C\sigma,BB}(p;q,r)$}
\psfrag{m2}{$\dDelta^{\sigma D,BA}(p;q,r)$}
\psfrag{m3}{$\dDelta^{D\sigma,AB}(p;q,r)$}
\begin{center}
\includegraphics[scale=.5]{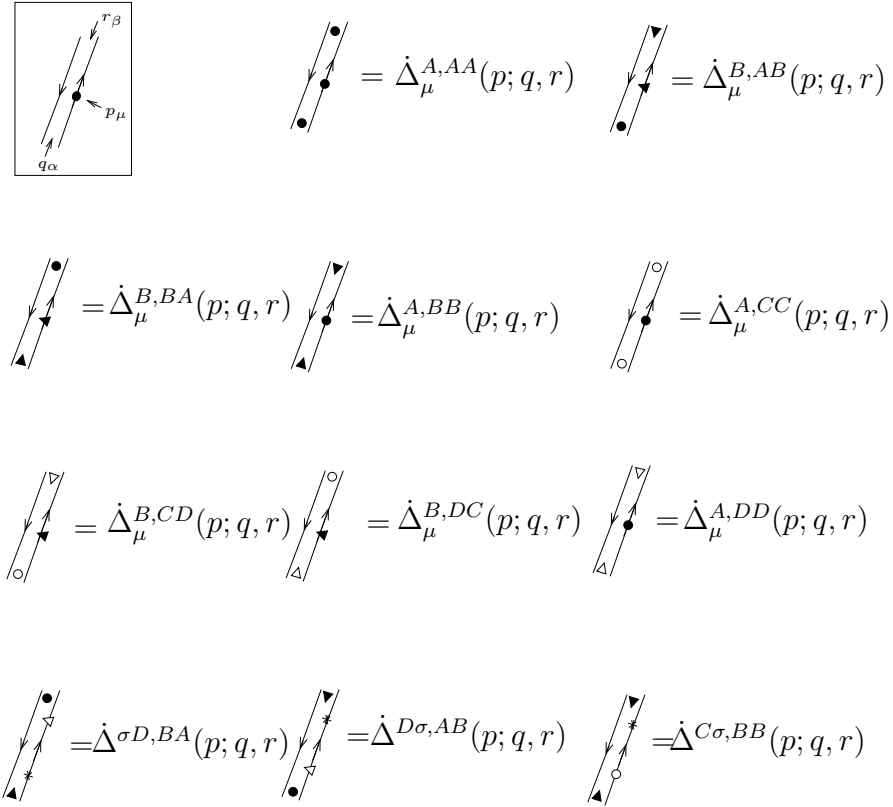}
\end{center}
\caption{Graphical representation of the 1-point Wines}\label{fig:1wines}
\end{figure}

While writing down the equations for these higher point vertices,
we will need the covariantised kernels described in section
\ref{Covariantization}. This did not happen for two-point vertices,
because had we considered a field on the wines, we would have been left
with one point vertices, which are ruled out from this theory. From
three point vertices onwards, we can instead consistently construct
supertraces of three fields, having one of them (or more) on the wines \ie
coming from the expansion in fields of the covariantisation of the
kernels. In \fig{fig:1wines} these one-point wines can be found in their
graphical representation.
Even though $\dDelta^{A,AA}_{\mu}(p;q,r)$, $\dDelta^{B,AB}_{\mu}(p;q,r)$ and
$\dDelta^{B,BA}_{\mu}(p;q,r)$ are all equal to $1/\Lam^2c'_{\mu}(p;q,r)$ (since they all come from the
covariantisation of $\dDelta^{AA}=c'/\Lam^2$) they will be indicated in the equations
as they appear in \fig{fig:1wines}, to make more clear the link between their
analytic and diagrammatic form. The same applies to the three
covariantisations of $\dDelta^{CC}$, namely $\dDelta^{A,CC}_{\mu}(p;q,r)$,
$\dDelta^{B,CD}_{\mu}(p;q,r)$ and $\dDelta^{B,DC}_{\mu}(p;q,r)$. One could also check that
$\dDelta^{\sigma D,BA}(p;q,r)=\dDelta^{\A\A}_{m\ q}/2$ and $\dDelta^{C\sigma,BB}(p;q,r)=(\dDelta^{\A\A}_{m\ q}+\dDelta^{\A\A}_{m\ r})/2$, from \eq{newa0}.

\begin{figure}[t!]
\psfrag{mu}{$\mu$}
\psfrag{nu}{$\nu$}
\psfrag{rho}{$\rho$}
\psfrag{si}{$\sigma$}
\psfrag{ldl}{$\Lam\de_{\Lam}$}
\psfrag{+}{$+$}
\psfrag{-}{$-$}
\psfrag{=}{$=$}
\psfrag{S}{$S$}
\psfrag{Si}{$\Sigma$}
\psfrag{hat}{$\wedge$}
\psfrag{cy}{Cycles}
\begin{center}
\includegraphics[scale=.5]{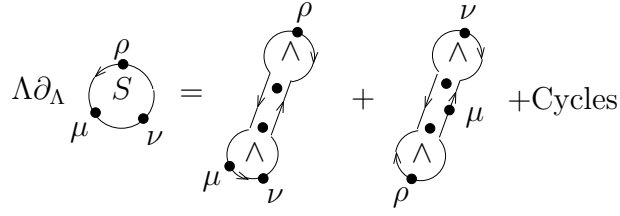}
\end{center}
\caption{Diagrammatic representation of the three $A$'s vertex's equation}\label{fig:aaa}
\end{figure}

We can start now with the first three point vertex we want
to consider which is the three $A$'s
vertex. Its equation is shown in the diagrammatic representation in
Fig.\ref{fig:aaa}.

\noindent In formulae is \eq{aaa}. The other 3-point vertex
equations follow it\footnote{If not stated otherwise, in all the formulae,
we will drop the label $(0)$ and $S$ will be intended as $S_0$, tree level action}.
\bea
S^{AAA}_{\,\mu\,\nu\,\rho}(p,q,r)=-\int_{\Lam}^{\infty}\dfrac{d\Lam_1}{\Lam_1}\left(\hat{S}^{AAA}_{\,\mu\,\nu\,\alpha}(p,q,r)\dDelta^{AA}_r+\hat{S}^{AA}_{\,\mu\,\alpha}(p)\dDelta^{A,AA}_{\,\nu}(q;p,r)\right)\hat{S}^{AA}_{\,\alpha\,\rho}(r)\nonumber\\[5pt]
+2(p_{\rho}\delta_{\mu\nu}-p_{\nu}\delta_{\rho\mu})+\mbox{cycles}\nonumber\\
\label{aaa}\\[10pt]
S^{BBA}_{\,\mu\,\nu\,\rho}(p,q,r)=-\int_{\Lam}^{\infty}\dfrac{d\Lam_1}{\Lam_1}\left\{\hat{S}^{BBA}_{\,\mu\,\nu\,\alpha}(p,q,r)\dDelta^{AA}_r\hat{S}^{AA}_{\,\alpha\,\rho}(r)+\hat{S}^{BBA}_{\,\alpha\,\nu\,\rho}(p,q,r)\dDelta^{BB}_p\hat{S}^{BB}_{\,\alpha\,\mu}(p)\right.\nonumber\\[5pt]
+\hat{S}^{BBA}_{\,\mu\,\alpha\,\rho}(p,q,r)\dDelta^{BB}_q\hat{S}^{BB}_{\,\alpha\,\nu}(q)+\hat{S}^{BB}_{\,\mu\,\alpha}\dDelta^{A,BB}_{\,\rho}(r;q,p)\hat{S}^{BB}_{\,\alpha\,\nu}(q)\nonumber\\[5pt]
+\hat{S}^{BB}_{\,\mu\,\alpha}(p)\dDelta^{B,BA}_{\,\nu}(q;p,r)\hat{S}^{AA}_{\,\alpha\,\rho}(r)+\hat{S}^{BB}_{\,\rho\,\alpha}(r)\dDelta^{B,AB}_{\,\mu}(p;q,r)\hat{S}^{AA}_{\,\alpha\,\nu}(q)\nonumber\\[5pt]
-\hat{S}^{BAD\sigma}_{\,\nu\,\rho}(q,r)\dDelta^{DD}_p\hat{S}^{BD\sigma}_{\,\mu}(p)-\hat{S}^{ABD\sigma}_{\,\rho\,\mu}(r,p)\dDelta^{DD}_q\hat{S}^{BD\sigma}_{\,\nu}(q)\nonumber\\[5pt]
\left.-\hat{S}^{BD\sigma}_{\,\mu}(p)\dDelta^{A,DD}_{\,\rho}(r;q,p)\hat{S}^{BD\sigma}_{\,\nu}(q)\right\}+\mbox{Int.Const.}\nonumber\\\label{bba}
\eea
\bea
S^{ACC}_{\mu}(p,q,r)=-\int_{\Lam}^{\infty}\dfrac{d\Lam_1}{\Lam_1}\left\{\sh^{ACC}_{\,\alpha}(p,q,r)\dDelta^{AA}_p\sh^{AA}_{\,\alpha\,\mu}(p)+\sh^{ACC}_{\,\mu}(p,q,r)\dDelta^{CC}_r\sh^{CC}(r)\right.\nonumber\\[5pt]
\left.+\sh^{ACC}_{\,\mu}(p,q,r)\dDelta^{CC}_q\sh^{CC}(q)+\sh^{CC}(r)\dDelta^{A,CC}_{\mu}(p;r,q)\sh^{CC}(q)\right\}+I.C.\nonumber\\
\label{acc}\\[10pt]
S^{ADD}_{\,\mu}(p,q,r)=-\int_{\Lam}^{\infty}\dfrac{d\Lam_1}{\Lam_1}\left\{\sh^{AA}_{\,\mu\,\alpha}(p)\dDelta^{AA}_p\sh^{ADD}_{\,\alpha}(p,q,r)+\sh^{BAD\sigma}_{\,\alpha\,\mu}(r,p,q)\dDelta^{BB}_r\sh^{BD\sigma}_{\,\alpha}(r)\right.\nonumber\\[5pt]
+\sh^{ABD\sigma}_{\,\mu\,\alpha}(p,q,r)\dDelta^{BB}_q\sh^{BD\sigma}_{\,\alpha}(q)-\sh^{BD\sigma}_{\,\alpha}(r)\dDelta^{A,BB}_{\,\mu}(p;r,q)\sh^{BD\sigma}_{\,\alpha}(q)\nonumber\\[5pt]
-\sh^{AA}_{\,\mu\,\alpha}(p)\dDelta^{D\sigma,AB}(q;p,r)\sh^{BD\sigma}_{\,\alpha}(r)+\sh^{BD\sigma}_{\,\alpha}(q)\dDelta^{D\sigma,BA}(r;q,p)\sh^{AA}_{\,\alpha\,\mu}(p)\nonumber\\[5pt]
+\sh^{ADD}_{\,\mu}(p,q,r)\dDelta^{DD}_r\sh^{DD}(r)+\sh^{ADD}_{\,\mu}(p,q,r)\dDelta^{DD}_q\sh^{DD}(q)\nonumber\\[5pt]
\left.+\sh^{DD}(r)\dDelta^{A,DD}_{\,\mu}(p;r,q)\sh^{DD}(q)\right\}+I.C.\nonumber\\
\label{add}\\[10pt]
S^{ABD\sigma}_{\,\mu\,\nu}(p,q,r)=-\int_{\Lam}^{\infty}\dfrac{d\Lam_1}{\Lam_1}\left\{-\sh^{BBA}_{\,\nu\,\alpha\,\mu}(q,r,p)\dDelta^{BB}_r\sh^{BD\sigma}_{\,\alpha}(r)+\sh^{ABD\sigma}_{\,\alpha\,\nu}(p,q,r)\dDelta^{AA}_p\sh^{AA}_{\,\alpha\,\mu}(p)\right.\nonumber\\[5pt]
+\sh^{ABD\sigma}_{\,\mu\,\alpha}(p,q,r)\dDelta^{BB}_q\sh^{BB}_{\,\alpha\,\nu}(q)-\sh^{AA}_{\,\mu\,\alpha}(p)\dDelta^{B,AB}_{\,\nu}(q;p,r)\sh^{BD\sigma}_{\,\alpha}(r)\nonumber\\[5pt]
-\sh^{BB}_{\,\nu\,\alpha}(q)\dDelta^{A,BB}_{\,\mu}(p;r,q)\sh^{BD\sigma}_{\,\alpha}(r)+\hf\sh^{AA}_{\,\mu\,\alpha}(p)\dDelta^{D\sigma,BA}(r;q,p)\sh^{BB}_{\,\,\alpha\,\nu}(q)\nonumber\\[5pt]
+\sh^{ABD\sigma}_{\,\mu\,\nu}(p,q,r)\dDelta^{DD}_r\sh^{DD}(r)+\sh^{ADD}_{\,\mu}(p,q,r)\dDelta^{DD}_q\sh^{BD\sigma}_{\,\nu}(q)\nonumber\\[5pt]
\left.+\sh^{BD\sigma}_{\,\nu}(q)\dDelta^{A,DD}_{\,\mu}(p;r,q)\sh^{DD}(r)\right\}+I.C.\nonumber\\\label{abds}
\eea
\begin{figure}[h]
\psfrag{mu}{$\mu$}
\psfrag{nu}{$\nu$}
\psfrag{rho}{$\rho$}
\psfrag{si}{$\sigma$}
\psfrag{ldl}{$\Lam\de_{\Lam}$}
\psfrag{+}{$+$}
\psfrag{-}{$-$}
\psfrag{=}{$=$}
\psfrag{S}{$S$}
\psfrag{cy}{Cycles}
\psfrag{Si}{$\Sigma$}
\psfrag{1/l2}{$\dfrac{1}{\Lam^2}$}
\psfrag{1/2l2}{$\dfrac{1}{2}$}
\psfrag{hat}{$\wedge$}
\begin{center}
\includegraphics[scale=.5]{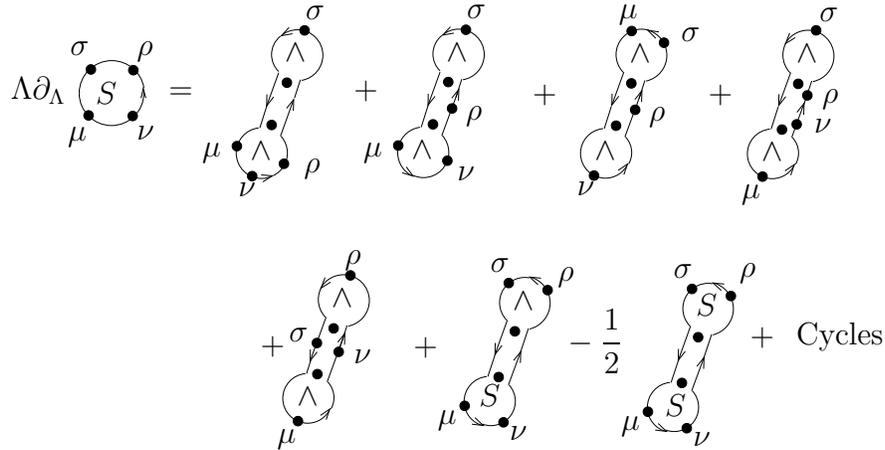}
\end{center}
\caption{Diagrammatic representation of the four $A$'s vertex's equation}\label{fig:aaaa}
\end{figure}
The four point needed are four: AAAA, AABB, AADD, AACC. The equation for
the 4-$A$'s vertex is shown in its diagrammatic representation in
Fig.\ref{fig:aaaa} and in formulae:
\bea
S^{AAAA}_{\,\mu\,\nu\,\rho\,\sigma}(p,q,r,s)=-\int_{\Lam}^{\infty}\dfrac{d\Lam_1}{\Lam_1^3}\left\{\sh^{AAAA}_{\,\mu\,\nu\,\rho\,\alpha}(p,q,r,s)\dDelta^{AA}_s\sh^{AA}_{\,\alpha\,\sigma}(s)\right.\nonumber\\[5pt]
-\hf S^{AAA}_{\,\mu\,\nu\,\alpha}(p,q,r+s)\dDelta^{AA}_{p+q}S^{AAA}_{\,\alpha\,\rho\,\sigma}(p+q,r,s)\nonumber\\[5pt]
+\hf\sh^{AAA}_{\,\mu\,\nu\,\alpha}(p,q,r+s)\dDelta^{AA}_{p+q}S^{AAA}_{\,\alpha\,\rho\,\sigma}(p+q,r,s)\nonumber\\[5pt]
+\hf S^{AAA}_{\,\mu\,\nu\,\alpha}(p,q,r+s)\dDelta^{AA}_{p+q}\sh^{AAA}_{\,\alpha\,\rho\,\sigma}(p+q,r,s)\nonumber\\[5pt]
+\sh^{AAA}_{\,\mu\,\nu\,\alpha}(p,q,r+s)\dDelta^{A,AA}_{\,\rho}(r;p+q,s)\sh^{AA}_{\,\alpha\,\sigma}(s)\nonumber\\[5pt]
+\sh^{AAA}_{\,\sigma\,\mu\,\alpha}(s,p,q+r)\dDelta^{A,AA}_{\,\rho}(r;q,p+s)\sh^{AA}_{\,\alpha\,\nu}(q)\nonumber\\[5pt]
+\sh^{AA}_{\,\mu\,\alpha}(p)\dDelta^{AA,AA}_{\,\nu\,\rho}(q,r;p,s)\sh^{AA}_{\,\alpha\,\sigma}(s)\nonumber\\[5pt]
\left.+\hf\sh^{AA}_{\,\mu\,\alpha}(p)\dDelta^{A,A,AA}_{\,\nu,\,\sigma}(q;s;p,r)\sh^{AA}_{\,\alpha\,\rho}(r)+\mbox{cycles}\right\}+I.C.\nonumber\\
\label{aaaa}
\eea
In the previous equations, I.C. stands for ``Integration Constant'' and was
not written explicitly here but in the three $A$'s equation (\ref{aaa}). For the
other vertices, the divergent parts of the integrating constants will be discussed in the next two sections. Their finite
parts are given in Appendix \ref{App:bare}.

The equations for the four-point vertices left out here, are listed in Appendix \ref{app:4eq}

\subsection{Enforcing universality of $\beta_1$ (and $\beta_2$)}
 \label{Enforcing}

 Before we discuss the finiteness of the tree level vertices from the
 previous section, let us review in this context, the standard argument for why we should
 expect to get the same value for $\beta_1$, and indeed $\beta_2$, in
 the $\beta$ function \eq{betafn} as in other methods, despite the
 fact that our renormalisation scheme for $g(\Lambda)$ differs from that of 
 the corresponding coupling $\gt(\mu\mapsto\Lambda)$ defined by these other 
 methods. 

 In principle we can extract from \eq{Sloope}, 
 by computing quantum corrections, the value of the other coupling as a 
 function of ours, and thus match the two couplings perturbatively:
 \be
 \label{gt}
 {1/\gt^2} = {1/g^2} +\gamma +O(g^2),
 \ee
 where the classical agreement is guaranteed by the standard normalisations
 of the fields and kinetic term in \eq{defg}, after scaling $g$ back to its
 usual position, and $\gamma$ is a one-loop matching coefficient. 
 Differentiating with respect to $\Lambda$ and using \eq{betafn},
 the corresponding $\beta$ function for $\gt$,  and \eq{gt}, we have
 \be
 \label{betas}
 \bt_1 +\bt_2\, g^2 = \beta_1 +\beta_2 \,g^2 +\ldl\gamma +O(g^4).
 \ee
 Since $\gamma$ is dimensionless, it cannot depend upon $\Lambda$, there being
 no other scale to form the necessary dimensionless combination. Thus
 $\ldl\gamma=0$ in \eq{betas},  and we immediately recover the standard
 facts that $\bt_1=\beta_1$ and $\bt_2=\beta_2$.

 Clearly this argument fails if some other scale has been introduced,
 for example the standard arbitrary finite physical scale $\mu$, or if
 other running couplings get introduced. (After solving for their flows, \ie
 solving their corresponding $\beta$ functions, this becomes equivalent
 to the first failure since by dimensional transmutation a new finite
 physical scale has been introduced.) Importantly, $\ldl\gamma$ can then have
 an $O(g^2)$ one-loop contribution or in extreme cases even a tree-level
 $O(g^0)$ contribution. From \eq{betas} one sees that a one-loop contribution
 to the running of $\gamma$ destroys $\beta_2$ agreement, whilst a 
 tree-level running would even modify $\beta_1$.

 As we will see shortly, a generic $\sh$, including the simple form used for
 the bare action in ref. \cite{su:pap}, can lead to such tree-level
 corrections. Fortunately, there is also an infinite class of seed actions
 that cannot. As with the earlier constraints discussed,
 since we never specify $\sh$,
 it is not the solution that matters, only knowing that one exists.

 To get agreement with the standard $\beta$ function at the two-loop level,
 one needs to confirm that there are no further couplings hidden, that
 run at one loop, and to take into account contributions from 
 $g_2(\Lambda)$. This can be done \cite{us}.

 Even with a non-vanishing $\ldl\gamma$, one could still recover 
 the usual $\beta$ function coefficients, by defining a standard low 
 energy --or infrared-- coupling $\gt(\mu)$ at some scale $\mu<\Lambda$,  
 this coupling being distinguished from the `ultraviolet' coupling 
 $g(\Lambda)$ in the effective action 
 $S_\Lambda$ \cite{bonini,litim}. We want to avoid this because the
 introduction of $\mu$  would destroy, or at least obscure, the
 power and elegance of self-similarity (\cf \sec{Necessary}).

\subsection{Finiteness at tree level}

As we have anticipated earlier in this chapter, it is now time to prove
that it is possible to have all the integrals for the tree level vertices, UV
regulated for a specific choice of $\sh$. One should not usually expect
divergences at the classical level, but the incorporation of Pauli-Villars
fields directly into an exact RG, can cause them. In this section we do not intend to
give a proof, but we just want to show that all the vertices we have to
deal with in order to perform our calculation (two, three and four point
tree level vertices), are convergent for an
appropriate choice of the seed action. We start recalling the two main
requirements on $\sh$, namely to be supergauge invariant, and to have the
two-point vertices of the form of Eqs.(\ref{hataa})-(\ref{hatbds}). A
possible choice in the unbroken phase is:
\bea\label{s-hat}
\hat{S}&=&\dfrac{1}{2}{\cal F}^{\mu\nu}\{c^{-1}\}{\cal F}_{\mu\nu}+\hf\nabla^{\mu}\cdot{\cal C}\{\tilde{c}^{-1}\}\nabla_{\mu}\cdot{\cal C}+\nonumber\\[.4cm]
&&+\dfrac{\lambda}{4}\mbox{str}\left({\cal C}^2-\Lambda^2\right)^2
\eea
 Now, once the ${\cal C}$ field has been shifted ($SU(N|N)$ broken phase), and it
has been redefined in order to be dimensionless (${\cal C}\to\Lam{\cal C}$), the
seed action chosen here has the following form (the bare action in ref. \cite{su:pap}):
\bea
\hat{S}&=&\hf{\cal F}_{\mu\nu}\{c^{-1}\}{\cal
F}_{\mu\nu}+\frac{\Lambda^2}{2}\nabla_{\mu}\cdot{\cal
C}\{\tilde{c}^{-1}\}\nabla_{\mu}\cdot{\cal C}-\frac{\Lambda^2}{2}[{\cal A}_{\mu},\sigma]\{\tilde{c}^{-1}\}[{\cal A}_{\mu},\sigma]\nonumber\\
&&-i\Lambda^2[{\cal A}_{\mu},\sigma]\{\tilde{c}^{-1}\}\nabla_{\mu}\cdot{\cal
C}+\frac{\lambda}{4}\Lambda^4 \mbox{str}\int_x(\{\sigma,{\cal C}\}+{\cal C})^2
\eea
We can, at this point split the fields in the diagonal and off-diagonal
(bosonic + fermionic) components:
\bea
{\cal A}_{\mu}=A_{\mu}+B_{\mu}\label{a}\\
{\cal C}=C+D\label{c}
\eea
Since $[A_{\mu},\sigma]=0$ and $[B_{\mu},\sigma]=2B_{\mu}\sigma$, our
vertices simplify because now they have either one $\sigma$ (which for
convention we decide to put at the end), or no $\sigma$'s in them.
The seed action in terms of split fields is:
\bea
\hat{S}&=&\hf{\cal F}_{\mu\nu}\{c^{-1}\}{\cal
F}_{\mu\nu}+\frac{\Lambda^2}{2}\nabla_{\mu}\cdot{\cal
C}\{\tilde{c}^{-1}\}\nabla_{\mu}\cdot{\cal C}-2\Lambda^2
B_{\mu}\sigma\{\tilde{c}^{-1}\}B_{\mu}\sigma\nonumber\\
&&-2i\Lambda^2B_{\mu}\sigma\{\tilde{c}^{-1}\}\nabla_{\mu}\cdot{\cal
C}+\frac{\lambda}{4}\Lambda^4 \mbox{str}\int_x (C^2+D^2+CD+DC+2C\sigma)^2\nonumber\\
\eea
One can easily check that the two-point vertices are exactly the one
listed in section \ref{sec:1loop}. The higher vertices (three and four-point), which are a
possible covariantisation of the two-point ones, are:
\begin{itemize}
\item Three point:
\bea
\hat{S}_{\,\mu\,\nu\,\rho}^{AAA}(p,q,r)&=&\frac{2}{c_p}(p_{\rho}\delta_{\mu\nu}-p_{\nu}\delta_{\mu\rho})\nonumber\\
&&+2c^{-1}_{\nu}(q;p,r)(p_{\rho}r_{\mu}-p\cdot
r\delta_{\rho\mu})+\ \mbox{cycles}\nonumber\\
&\equiv&\hat{S}_{\mu\nu\rho}\\[5pt]
\hat{S}_{\,\mu\,\nu\,\rho}^{BBA}(p,q,r)&=&\hat{S}_{\mu\nu\rho}+4\Lambda^2\ct^{-1}_{\rho}(r;q,p)\delta_{\mu\nu}\\[5pt]
\hat{S}^{ACC}_{\,\mu}(p,q,r)&=&\Lambda^2(\frac{q_{\mu}}{\ct_q}-\frac{r_{\mu}}{\ct_r}-q\cdot
r\ct^{-1}_{\mu}(p;r,q))\nonumber\\
&=&\hat{S}^{ADD}_{\,\mu}(p,q,r)\\[5pt]
\hat{S}^{BAD\sigma}_{\,\mu\,\nu}(p,q,r)&=&2\Lambda^2(\ct^{-1}_p\delta_{\mu\nu}-r_{\mu}\ct^{-1}_{\nu}(q;r,p))\nonumber\\
&=&-\hat{S}^{ABD\sigma}_{\nu\mu}(q,p,r)
\eea
\item Four point:
\bea
\hat{S}^{AAAA}_{\,\mu\,\nu\,\rho\,\lambda}(p,q,r,s)&=&\frac{1}{c_{p+q}}(\delta_{\lambda\mu}\delta_{\rho\nu}-\delta_{\rho\mu}\delta_{\nu\lambda})+2\ct^{-1}_{\nu}(q;p,r+s)(p_{\lambda}\delta_{\rho\mu}-p_{\rho}\delta_{\lambda\mu})\nonumber\\
&&+2c^{-1}_{\lambda}(s;p,r+q)(p_{\nu}\delta_{\mu\rho}-p_{\rho}\delta_{\mu\nu})\nonumber\\
&&+2c^{-1}_{\nu\rho}(q,r;p,s)(p_{\lambda}s_{\mu}-p\cdot
s\delta_{\lambda\mu}\nonumber)\\
&&c^{-1}_{\nu,\lambda}(q;s;p,r)(p_{\rho}r_{\mu}-p\cdot r\delta_{\rho\mu})+\
\mbox{cycles}\equiv \hat{S}_{\mu\nu\rho\lambda}\\[5pt]
\hat{S}^{AABB}_{\,\mu\,\nu\,\rho\,\lambda}(p,q,r,s)&=&\hat{S}_{\mu\nu\rho\lambda}+4\Lambda^2\delta_{\rho\lambda}\ct^{-1}_{\mu\nu}(p,q;r,s)\\[5pt]
\hat{S}^{AADD}_{\,\mu\,\nu}(p,q,r,s)&=&\Lambda^2(\ct^{-1}_{p+s}\delta_{\mu\nu}+s_{\nu}\ct^{-1}_{\mu}(p;q+r,s)+r_{\mu}\ct^{-1}_{\nu}(q;p+s,r)\nonumber\\
&&-r\cdot s\ \ct^{-1}_{\mu\nu}(p,q;s,r))\\[5pt]
\hat{S}^{AACC}_{\,\mu\,\nu}(p,q,r,s)&=&\hat{S}^{AADD}_{\mu\nu}(p,q,r,s)\\[5pt]
\hat{S}^{AABD\sigma}_{\,\mu\,\nu\,\rho}(p,q,r,s)&=&2\Lambda^2(s_{\rho}\ct^{-1}_{\mu\nu}(p,q;s,r)+\delta_{\mu\nu}\ct^{-1}_{\nu}(q;r,p+s))\\[5pt]
\hat{S}^{BAAD\sigma}_{\,\mu\,\nu\,\rho}(p,q,r,s)&=&\hat{S}^{AABD\sigma}_{\rho\nu\mu}(r,q,p,s)
\eea
\end{itemize}
One could check that indeed all the Ward identities are satisfied, by
contracting the vertices with the proper momenta. 

In order to control the
divergences, we have first to regulate the integrals with a cutoff
$\Lam_0$, to make the divergences explicit.  We can then substitute the
seed action vertices in the three-point tree-level integrals, and, then in the four-point ones, and see if there are any logarithmic or power-like
divergences.

What we expect is to find that many divergences are connected via  Ward
identities, some of them related back to two point vertices. Since they
present only power divergences, we are not worried about those. These are
canceled by term that must be present for gauge invariance, in the action
at tree level when $\Lam=\Lam_0$, leaving the constant which are listed in
Appendix \ref{App:bare}. More worrying are possible logarithmic
divergencies of the form $\alpha \ln{\frac{\Lam_0}{\Lam}}$. Terms of this
kind would have to be transverse since in the two-point vertices such
divergences are not present. To cure those, we would have to add
marginal operators to the action at $\Lam_0$, proportional to
$\ln{\frac{\Lam_0}{\mu}}$, $\mu$ being another finite scale (this way was
the one followed in \cite{mor:erg1,mor:erg2}). Although this is a possible
solution, introducing a new scale would result in a loss of
self-similarity, as it was discussed in the previous section, and causing problems through the
calculation carried on in the next chapter, which relies on the fact that we are
dealing with only one finite scale $\Lam$. We will then see that a possible
way out is to redefine the vertices of the seed action in order to tune to
zero these logarithmic divergences, without fixing a new scale. In other
words, it seems reasonable to infer that it is possible to choose a seed action
which keeps all the integrals in which it appears UV finite . 

The first vertex to be considered is the three-point pure $A$ vertex. It is
easy to see that it does not present any divergence as we would expect. The
corresponding two-point vertex is in fact finite and, by dimensions and
Lorentz invariance, a transverse term in
three different momenta cannot be constructed, which would carry a
divergent factor of $\Lam_0$.

Let us analyse the other divergences, dividing them in two sets. The first
set includes the two vertices $AAC$ and $AACC$. Their divergences are:
\be\label{ACCd}
\left.S^{ACC}_{\,\mu}(p,q,r)\right|_{DIV}=\al_{\mu}(p,q,r)\ln{\frac{\Lam_0}{\Lam}}+\beta_{\mu}(p,q,r)\Lam_0^2
\ee
where
\bea
&&\alpha_{\mu}(p,q,r)=-4(\cp_0+2\ctp_0)r_{\al}\BBox_{\al\mu}(p)\\[3pt]
&&\beta_{\mu}(p,q,r)=2(q-r)_{\mu}
\eea
and
\bea
\left.S^{AACC}_{\,\mu\,\nu}(p,q,r,s)\right|_{DIV}&=&4\ln{\frac{\Lam_0}{\Lam}}(c^{\prime}_0+2\tilde{c}^{\prime}_0)(r_{\mu}s_{\nu}-r_{\nu}s_{\mu}+p_{\nu}s_{\mu}\nonumber\\
&&+q_{\mu}r_{\nu}-(p\cdot
s +q\cdot r)\delta_{\mu\nu})\label{AACC1d}\\
&&+4\ln{\frac{\Lam_0}{\Lam}}(c^{\prime}_0-2\tilde{c}^{\prime}_0)(p_{\nu}q_{\mu}-p\cdot
q\delta_{\mu\nu})\label{AACC2d}\\
&&+2\Lam_0^2\delta_{\mu\nu}
\eea
respectively. For the first one, the three-point $ACC$, one can notice that the quadratic divergence is exactly cancelled by a term already
contained in $S^{ACC}_{\,\mu bare}$ and the logarithmic one is transverse in the
A-field momentum, $p_{\mu}$, as expected. For the four-point one, the same
can be said about the power divergences, but there are two independent
logarithmic divergent terms. We will see that one of the two, namely
(\ref{AACC1d}), can be cancelled via the same counterterm that cancels the one in
(\ref{ACCd}), in which it can be transformed via a Ward identity. For the
other one, transverse in  both the $A$ momenta, $p$ and $q$, we will have
to add an independent term to the seed action.

The second set of vertices includes $ABB$, $ADD$,
$ABD\sigma$, $AABB$ and $AADD$. Their divergences are:
\begin{itemize}
\item Three point:
\bea
\left.S^{BBA}_{\,\mu\,\nu\,\rho}(p,q,r)\right|_{DIV}&=&16\ln{\frac{\Lam_0}{\Lam}}(r_{\mu}\delta_{\nu\rho}-r_{\nu}\delta_{\rho\mu})\label{BBAd}\\[5pt]
\left.S^{ADD}_{\,\mu}(p,q,r)\right|_{DIV}&=&12\Lam_0^2(q-r)_{\mu}-4\ln{\frac{\Lam_0}{\Lam}}
r_{\al}\BBox_{\al\mu}(p)\label{ADDd}\\[5pt]
\left.S^{ABD\sigma}_{\,\mu\,\nu}(p,q,r)\right|_{DIV}&=&-4\Lam_0^2\delta_{\mu\nu}+8\ln{\frac{\Lam_0}{\Lam}}(p\cdot
r\delta_{\mu\nu}-p_{nu}r_{\mu})\label{ABDSd}
\eea 
\item Four point:
\bea
\left.S^{AABB}_{\,\mu\,\nu\,\rho\,\sigma}(p,q,r,s)\right|_{DIV}&=&-16\ln{\frac{\Lam_0}{\Lam}}(\delta_{\mu\rho}\delta_{\nu\sigma}-\delta_{\mu\sigma}\delta_{\nu\rho})\label{AABBd}\\[5pt]
\left.S^{AADD}_{\,\mu\,\nu}(p,q,r,s)\right|_{DIV}&=&4\ln{\frac{\Lam_0}{\Lam}}(r_{\mu}
s_{\nu}-r_{\nu}s_{\mu}+p_{\nu}s_{\mu}+q_{\mu}r_{\nu}\nonumber\\
&&-(p\cdot s +q\cdot
r)\delta_{\mu\nu})+2\Lam_0^2\delta_{\mu\nu}\label{AADDd}
\eea
\end{itemize}
Here the situation is simpler: all the three-point divergences have the
logarithmic part which is transverse in the corresponding $A$ momentum, as
one can expect, and all the four point have their logarithmic divergences,
falling in those of their corresponding coefficients related by gauge transformation. In this case it
will then be possible to cancel all of them with the addition of only one
counterterm.

Let us start with the former set of vertices. Following
\cite{mor:erg1,mor:erg2}, we can add to the tree-level action at $\Lam_0$
the terms:
\bea
S_0^1&=&-\frac{i}{16}\gamma^{ACC}\mbox{str}\{{\cal C}, \nabla_{\mu}\cdot {\cal C}\}{\cal F}_{\mu\nu}\{{\cal
C},\nabla_{\nu}\cdot {\cal C}\}\label{ACC}\\
&&+\frac{\gamma}{4}^{AACC}\mbox{str}[{\cal C}^2,{\cal
F}_{\mu\nu}]^2\label{AACC}
\eea
It can be shown that both terms (\ref{ACC}) and (\ref{AACC}) are $SU(N|N)$ invariant
and no-$\A^0$ symmetric. Moreover, choosing the two constants as follows:
\bea
&&\gamma^{ACC}=16(c^{\prime}_0+2\tilde{c}^{\prime}_0)\ln{\frac{\Lam_0}{\mu}\label{gammaAAC}}\\
&&\gamma^{AACC}=2(c^{\prime}_0-2\tilde{c}^{\prime}_0)\ln{\frac{\Lam_0}{\mu}}\label{gammaAACC},
\eea
one can check that all the logarithmic divergences in eq.(\ref{ACCd}) and
\eqs{AACC1d}{AACC2d} are cancelled.

For the second set of terms, it is possible to show that adding to the tree
level action at $\Lam_0$ the following counterterm\footnote{$SU(N|N)$
invariant and no-$\A^0$ symmetric as well.}
\be\label{BBA}
S_0^2=
-\frac{i}{16}\gamma^{BBA}\mbox{str}[{\cal C},(\nabla_{\mu}\cdot\cc)]{\cal
F}_{\mu\nu}[{\cal C},(\nabla_{\nu}\cdot\cc)],
\ee
can cancel all the logarithmic divergences of
Eqs.(\ref{BBAd})-(\ref{AADDd}), for the following choice of the constant:
\be\label{gammaBBA}
\gamma^{BBA}=16\ln{\frac{\Lam_0}{\mu}}
\ee

As we have mentioned in the previous section, though, even if this
procedure cures all the divergences at tree level, it introduces a new
scale $\mu$. In the next chapter, we will see that throughout the calculation, we will often rely on the fact that there is
only one finite scale $\Lam$. This argument is no longer true if we
regularise the tree level vertices by modifying the integration constant in the way we just mentioned. Fortunately,
it is possible to overcome this problem. Instead of adding terms to the
tree level action at $\Lam_0$, we can in fact add
terms to the seed action which appears inside the integrals, without
introducing any new scale and, thus, making sure logarithmic divergences do not
appear. In other words, we can choose an $\sh$ which does not produce
any logarithmic divergences, leaving the power divergences unchanged which
must appear by gauge invariance. In the specific case we are considering,
for example, it is enough to add to the seed action, the same terms we
wanted to add to the bare action (\ref{ACC}), (\ref{AACC}) and (\ref{BBA}), with different coefficients. If one
chooses here coefficients (similarly defined) to be:
\bea\label{coeff}
\hat{\gamma}^{BBA}&=&-4\\
\hat{\gamma}^{ACC}&=&(c'_0+2\tilde{c}'_0)/2\nonumber\\
\hat{\gamma}^{AACC}&=&(c'_0-2\tilde{c}'_0)/2\nonumber
\eea
it is easy to check\footnote{This check was done with a script in FORM.}
that all the logarithmic divergences disappear.

A way to see this issue from a more general perspective, as can be found in
\cite{us}, is to realise as we have already mentioned earlier, that the
problem of classical divergences is associated to the Pauli-Villars
sector. These terms have a classical divergent action as $\Lam\to\infty$, having a divergent mass, causing logarithmic divergences along the
marginal directions. The solution which avoids the introduction of a new scale, is
precisely to shift the $\sh$ vertices along those directions, in order to
tune the logarithmic divergences to zero. If the shift for each vertex is, in our
particular case, a coefficient as the ones listed in (\ref{coeff}) times
the structure of the divergence of the corresponding vertex, then the
potential logarithmic divergent term is removed from the integrand. Since the structure of the classical flow equations \eq{ergcl} 
 is such that the flow of each vertex 
 $S^{\ph0X_1\cdots X_n}_{0\,a_1\,\cdots\,a_n}$ has the corresponding
 $\sh^{X_1\cdots X_n}_{\,a_1\,\cdots\,a_n}$ as its highest-point $\sh$
 contribution,
 contracted with kernels and with the 
 appropriate two point vertices (\viz $\dDelta^{XX}\sh^{XX}$ where $X=$
 $A,C,B$ or $D$) \cite{mor:erg1,mor:erg2}, 
 and since these $\dDelta^{XX}\sh^{XX}$ terms are 
 non-vanishing at zero momentum precisely when $X$ is a massive Pauli-Villars field, 
 it follows that we can always remove the divergence associated with 
 these marginal directions by tuning 
 $\sh^{X_1\cdots X_n}_{\,a_1\,\cdots\,a_n}$ in the same direction.

Once this last check has been performed it is now possible to move onto the
gauge invariant calculation of the $SU(N)$ Yang-Mills $\beta$-function at
one loop. This calculation was mostly inspired by the scalar field case,
which gave us all the necessary hints and uncovered the whole machinery
that made it possible: above all, the use of the flow equations to
eliminate the hatted vertices in favour of the effective  ones, and the
integrated wine technique. This will all be explained in detail
in the next chapter.


\psfrag{d+R}{$\one+B$}
\psfrag{R1}{$R_1$}
\psfrag{R2}{$R_2$}
\psfrag{R3}{$R_3$}
\psfrag{sim}{$\sim$}
\psfrag{Hp}{$\Delta^{CC}(p)$}
\psfrag{1/2}{$\displaystyle \frac{1}{2}$}
\psfrag{O(p3)}{$O(p^3)$}
\psfrag{2}{$2$}
\psfrag{Fig.id}{fig.\ref{fig:id}}
\psfrag{=}{$=$}
\psfrag{mu}{$\mu$}
\psfrag{nu}{$\nu$}
\psfrag{Si}{$\Sigma_0$}
\psfrag{F}{\tiny$f$}
\psfrag{+}{$+$}
\psfrag{R34}{$R^f_{3,4}$}
\psfrag{R1314}{$R^f_{13,14}$}
\psfrag{R15}{$R^f_{15}$}
\psfrag{-}{$-$}
\psfrag{S0}{$S_0$}
\psfrag{ldl}{$\Lambda\partial_{\Lambda}$}
\psfrag{Sum}{\tiny $\displaystyle \sum_{\begin{array}{r}
                          f=A,B\\
                          C,D
                     \end{array}}$}
\psfrag{1}{$1$}
\psfrag{-4b1}{$-4\beta_1$}
\psfrag{(p)}{$(p)$}
\psfrag{2/l2}{$\displaystyle \frac{2}{\Lam^2}$}
\chapter{Gauge Invariant calculation}

\begin{figure}[h!]
\begin{center}
\includegraphics[scale=.5]{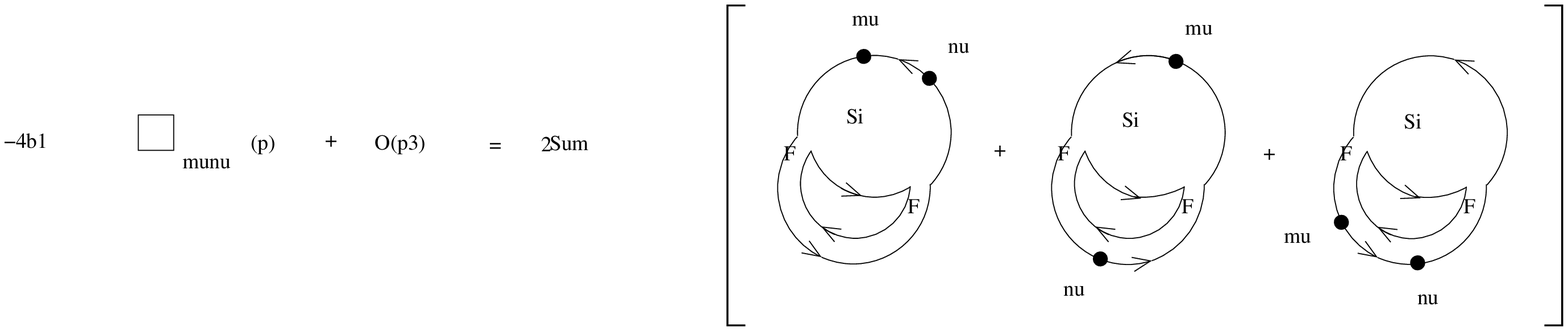}
\caption{Graphical representation of the equation for $\beta_1$}\label{fig:beta1}
\end{center}
\end{figure}

Let us first consider the equation for $\beta_1$ (\ref{5.19}). This can
be expressed in diagram, as it is represented in fig.\ref{fig:beta1}. 
$f$ in the figure, can represent any of the fields present in the
theory: $A$, $B$, $C$ or $D$. The content of fields and the symmetries,
allow to separate the problem in three different sectors: the $A$ and $C$
ones, which are possible to study independently and the $B-D$ which can be 
considered together. The reason we must study the $B$ and $D$ sectors together,
comes from the fact that unlike the $A$ and the $C$ ones, their equations
appear coupled trough the vertex $BD\si$. In the $A-C$ case, due to the
fact that $AC\si$ is not allowed (see section \ref{simsim}), this
mix does not occur and the study of the two sectors can be carried out
separately. Moreover, it is possible to avoid the study of a complicated
set of coupled equations, recognising that all of them can be recast into
only one equation via the introduction of a compact notation which will be
introduced in the next section. The crucial step in order to obtain a set
of only one equation for each generalised vertex is the introduction of a
doublet field $F=(B_{\mu},D\sigma)$. This will also allow to have equations of
the same form of the corresponding ones in the $A$ and $C$ sectors. 
This further observation allows us  to write all the equations for each
vertex in term of an even more generalised field multiplet, $f$, containing
also the $A$ and $C$ fields. Of course this is not a fundamental choice as
it is for the $BD\sigma$ sector, because the former two being decoupled fields, it will just amount to having to deal with block diagonal
matrices, but it will help indeed to keep the calculation neater. 

In the next section, the compact notation for the $BD\sigma$ sector will be
introduced as well as the corresponding equations and Ward identities. In the
second section, the actual calculation will be carried out in the 5-fields
notation as far as it possible and it will be finished by splitting down to components in the final part. Finally in the last section, will be considered the potentially
non-universal terms proportional to $\dDelta^{AA}_0$ and it will be shown that, as
one can expect, they do not contribute to the final result.

\section{Compact notation for the $BD\sigma$ sector}\label{sub:compact}

As can be recognised from Eqs.(\ref{bb})-(\ref{dd}) it is possible to
combine them together introducing a compact notation as follows. Let us
first define the matrices:
\bea
S^{FF}_{MN}(p)&=&\left(\ba{cc}
                     S^{BB}_{\,\mu\,\nu}(p)&S^{BD\sigma}_{\,\mu}(p)\\
                     S^{D\sigma B}_{\,\nu}(p)&S^{D\sigma D\sigma}(p)
                     \ea\right)\label{sFF}\\[5pt]
\dDelta^{FF}_{MN}(p)&=&\left(\ba{cc}
              \dDelta^{BB}_p\delta_{\mu\nu}&0\\
              0&-\dDelta^{DD}_p
              \ea\right) \label{cprimo}
\eea
where  $F$ refers to the doublet $(B,D\si)$ and in the eq.(\ref{sFF}), the momenta are referred to the first field. Making
use of Eqs.(\ref{sFF}) and (\ref{cprimo}) we can now rewrite
Eqs.(\ref{bb})-(\ref{dd}) in the following compact form:
\be\label{FF}
\Lam\de_{\Lam}S^{FF}_{MN}(p)=S^{FF}_{ML}(p)\dDelta^{FF}_{LS}(p)S^{FF}_{SN}(p)
\ee
Equation (\ref{FF}) has now exactly the same form as the
two-point $A$ and $C$ vertices of Eqs. \eqs{treeq:aa}{treeq:cc}.
Extending the idea it is possible to rewrite all the equations for the
three and four point vertices in the $BD\si$ sector. First of all we
have to group together the three point vertices in the following tensor representation
\be
S^{AFF}_{\,\mu RS}(p,q,r)=\left(\ba{cc}
                     S^{ABB}_{\,\mu\,\rho\,\sigma}&S^{ABD\sigma}_{\,\mu\,\rho}\\
                     S^{AD\sigma B}_{\,\mu\ \ \ \,\sigma}&S^{AD\sigma D\sigma}_{\,\mu}
                     \ea\right)\label{aff}
\ee
and similarly for the four point ones:
\be
S^{AAFF}_{\,\mu\,\nu\, RS}(p,q,r,s)=\left(\ba{cc}
                     S^{AABB}_{\,\mu\,\nu\,\rho\,\sigma}&S^{AABD\sigma}_{\,\mu\,\nu\,\rho}\\
                     S^{AAD\sigma B}_{\,\mu\,\nu\ \ \ \,\sigma}&S^{AAD\sigma D\sigma}_{\,\mu\,\nu}
                     \ea\right)\label{aaff}
\ee
(where, in both, the momentum dependence on the right hand side is omitted since it is
the same); then a one point wine must be defined as:
\be
\dDelta^{A,FF}_{\,\mu\, RS}(p;q,r)=\left(\ba{cc}
              \dDelta^{A,BB}_{\,\mu}(p;q,r)\delta_{\rho\sigma}&0\\
              0&-\dDelta^{A,DD}_{\,\mu}(p;q,r)
              \ea\right) \label{cprimomu}
\ee
and its natural two point extension:
\be
\dDelta^{AA,FF}_{\,\mu\,\nu\, RS}(p,q;r,s)=\left(\ba{cc}
              \dDelta^{AA,BB}_{\,\mu\,\nu}(p,q;r,s)\delta_{\rho\sigma}&0\\
              0&-\dDelta^{AA,DD}_{\,\mu\,\nu}(p,q;r,s)
              \ea\right) \label{cprimomunu}
\ee
Finally in order to be able to write all the three point equations of the
$BD\si$ sector in a compact form, it is necessary to define the
following two objects:
\bea
\dDelta^{F,AB}_{R}(p;q,r)=\left(\ba{c}
                           \dDelta^{B,AB}_{\rho}(p;q,r)\\
                           \dDelta^{D\sigma,AB}(p;q,r)
                           \ea\right)\\ 
\dDelta^{F,BA}_{R}(p;q,r)=\left(\ba{c}
                           \dDelta^{B,BA}_{\rho}(p;q,r)\\
                           \dDelta^{D\sigma,BA}(p;q,r)
                                 \ea\right)
\eea 
These are one point wine vertices which
have either a $B$ or a $D$ field. All the usual wine vertices rules and
properties apply to them. Their two point extensions (covariantisations)
with an $A$ field is then:
\bea
\dDelta^{AF,AB}_{\,\mu R}(p,q;r,s)=\left(\ba{c}
                                  \dDelta^{AB,AB}_{\,\mu\,\rho}(p,q;r,s)\\
                                  \dDelta^{AD\sigma,AB}_{\,\mu}(p,q;r,s)
                                 \ea\right)\\ 
\dDelta^{AF,BA}_{\,\mu R}(p,q;r,s)=\left(\ba{c}
                                  \dDelta^{AB,BA}_{\,\mu\,\rho}(p,q;r,s)\\
                                  \dDelta^{AD\sigma,BA}_{\mu}(p,q;r,s)
                                 \ea\right)
\eea 
where $\dDelta^{AD\sigma,BA}(p,q;r,s)=\dDelta^{\A\A}_{m\ \mu}(p;r,q+s)/2$.
These are two point wine vertices, one of whose field is an $A$ and the
other one can be either a $B$ or a $D$. Also in this case, all the usual
rules and properties for wine
vertices are valid. It is now possible with this
notation to write the equations for the three point functions (\ref{bba}),
(\ref{add}) and (\ref{abds}) collected together in the following form:
\bea
\Lam\de_{\Lam}S^{AFF}_{\,\mu
RS}(p,q,r)&=&\sh^{AA}_{\,\mu\,\alpha}(p)\dDelta^{AA}_p\sh^{AFF}_{\,\alpha
RS}(p,q,r)+\sh^{AFF}_{\,\mu TS}(p,q,r) \dDelta^{FF}_{TU}(q)\sh^{FF}_{UR}(-q)\nonumber\\
&&+\sh^{FF}_{ST}(r)\dDelta^{FF}_{TU}(r)\sh^{AFF}_{\,\mu RU}(p,q,r)\nonumber\\
&&+\sh^{FF}_{ST}(r)\dDelta^{A,FF}_{\,\mu TU}(p;r,q)\sh^{FF}_{UR}(-q)\nonumber\\
&&+\sh^{AA}_{\,\mu\,\alpha}(p)\dDelta^{F,AB}_{R}(q;p,r)\sh^{BF}_{\,\alpha
S}(-r)\nonumber\\
&&+\sh^{AA}_{\,\mu\,\alpha}(p)\dDelta^{F,BA}_{S}(r;q,p)\sh^{BF}_{\,\alpha R}(q)\label{AFF}
\eea
With this formalism all the four point equation of the $BD\sigma$ sector can be as well recast in only one compact equation that reads:
\bea
&&\!\!\!\!\!\!\!\!\!\Lam\de_{\Lam}S^{AAFF}_{\,\mu\,\nu\, RS}(p,q,r,s)=\sh^{AAFF}_{\,\mu\,\nu\, RT}(p,q,r,s)\dDelta^{FF}_{TU}(s)\sh^{FF}_{US}(-s)+\sh^{AAFF}_{\,\mu\,\nu\, TS}(p,q,r,s)\dDelta^{FF}_{TU}(r)\sh^{FF}_{UR}(-r)\nonumber\\
&&\!\!\!\!\!\!\!\!\!+\sh^{AAFF}_{\,\mu\,\alpha\, RS}(p,q,r,s)\dDelta^{AA}_q\sh^{AA}_{\,\alpha\,\nu}(q)+\sh^{AAFF}_{\,\alpha\,\nu\, RS}(p,q,r,s)\dDelta^{AA}_p\sh^{AA}_{\,\alpha\,\mu}(p)\nonumber\\
&&\!\!\!\!\!\!\!\!\!-S^{AAA}_{\,\mu\,\nu\,\alpha}(p,q,r+s)\dDelta^{AA}_{p+q}S^{AFF}_{\,\alpha RS}(p+q,r,s)+\sh^{AAA}_{\,\mu\,\nu\,\alpha}(p,q,r+s)\dDelta^{AA}_{p+q}S^{AFF}_{\,\alpha RS}(p+q,r,s)\nonumber\\
&&\!\!\!\!\!\!\!\!\!+S^{AAA}_{\,\mu\,\nu\,\alpha}(p,q,r+s)\dDelta^{AA}_{p+q}\sh^{AFF}_{\,\alpha RS}(p+q,r,s)-S^{AFF}_{\,\mu TS}(p,q+r,s)\dDelta^{FF}_{TU}(q+r)S^{AFF}_{\,\nu RU}(q,r,p+s)\nonumber\\
&&\!\!\!\!\!\!\!\!\!+\sh^{AFF}_{\,\mu
TS}(p,q+r,s)\dDelta^{FF}_{TU}(q+r)S^{AFF}_{\,\nu RU}(q,r,p+s)\nonumber\\
&&\!\!\!\!\!\!\!\!\!+S^{AFF}_{\,\mu TS}(p,q+r,s)\dDelta^{FF}_{TU}(q+r)\sh^{AFF}_{\,\nu RU}(q,r,p+s)\nonumber\\
&&\!\!\!\!\!\!\!\!\!+\sh^{AFF}_{\,\alpha RS}(p+q,r,s) \dDelta^{A,AA}_{\,\mu}(p;r+s,q)\sh^{AA}_{\,\alpha\,\nu}(q)+\sh^{AFB}_{\,\nu R\,\alpha}(q,r,s+p)\dDelta^{F,BA}_{S}(s;q+r,p)\sh^{AA}_{\,\alpha\,\mu}(p)\nonumber\\
&&\!\!\!\!\!\!\!\!\!+\sh^{AAA}_{\,\mu\,\nu\,\alpha}(p,q,r+s)\dDelta^{F,AB}_{R}(r;p+q,s)\sh^{BF}_{\,\alpha S}(-s)+\sh^{AFF}_{\,\mu TS}(p,q+r,s) \dDelta^{A,FF}_{\,\nu TU}q;p+s,r)\sh^{FF}_{UR}(-r)\nonumber\\
&&\!\!\!\!\!\!\!\!\!+\sh^{AA}_{\,\mu\,\alpha}(p)\dDelta^{A,AA}_{\,\nu}(q;p,r+s)\sh^{AFF}_{\,\alpha RS}(p+q,r,s)+\sh^{FF}_{ST}(s)\dDelta^{A,FF}_{\,\mu TU}(p;s,q+r)\sh^{AFF}_{\,\nu RU}(q,r,p+s)\nonumber\\
&&\!\!\!\!\!\!\!\!\!+\sh^{fB}_{R\alpha}(r)\dDelta^{F,BA}_{S}(s;r,p+q)\sh^{AAA}_{\,\alpha\,\mu\,\nu}(r+s,p,q)+\sh^{AA}_{\,\nu\,\alpha}(q)\dDelta^{F,AB}_{R}(r;q,s+p)\sh^{ABF}_{\,\mu\,\alpha S}(p,q+r,s)\nonumber\\
&&\!\!\!\!\!\!\!\!\!+\sh^{AA}_{\,\mu\,\alpha}(p)\dDelta^{AF,AB}_{\,\nu R}(q,r;p,s)\sh^{BF}_{\,\al S}(-s)+\sh^{FF}_{ST}(s) \dDelta^{AA,FF}_{\,\mu\,\nu\ TU}(p,q;s,r)\sh^{FF}_{UR}(-r)\nonumber\\
&&\!\!\!\!\!\!\!\!\!+\sh^{FB}_{R\al}(r)\dDelta^{FA,BA}_{S\,\mu}(s,p;r,q)\sh^{AA}_{\,\alpha\,\nu}(q)+\sh^{AA}_{\,\nu\,\alpha}(q)\dDelta^{FF,AA}_{RS}(r,s;q,p)\sh^{AA}_{\,\alpha\,\mu}(p)\nonumber\\
&&\!\!\!\!\!\!\!\!\!+\sh^{FB}_{S\al}(s)\dDelta^{A,F,BA}_{\,\mu,R}(p;r;s,q)\sh^{AA}_{\,\alpha\,\nu}(q)+\sh^{AA}_{\,\mu\,\alpha}(p)\dDelta^{A,F,AB}_{\,\nu,S}(q;s;p,r)\sh^{FB}_{R\,\al}(r)\label{AAFF}
\eea 
In this form it is easy to notice a similarity with the $A$ (and $C$)
sector. In fact the Ward identities obtained acting on the $A$ momenta of
the vertices defined in Eqs.(\ref{aaff}) and (\ref{aff}) are the same as
the ones for the pure $A$ case. We can also recognise generalised Ward
identities for the remaining momenta, carrying the indices in capital
letters, which show that gauge invariance is fully preserved in this
sector. Considering the $B$ and $D$ sectors separately, as we can recall
from the discussion carried out in section \ref{sub:bwi}, we had instead to
work with broken Ward identities. Moreover \eq{AFF} has the same form of the
equation for the three $A$'s vertex (\ref{aaa}) and of the three-point vertex
$ACC$ of \eq{acc}, with $A$ ($C$) replaced by $F$, and \eq{AAFF} has the same of the equations for the four $A$'s vertex (\ref{aaaa}) and the four-point
vertex $AACC$ (\ref{app:aacc}). This observation will allow us in the next
section to use an even more compact notation, with the introduction of the
label $f$, representing all the fields of the theory.

Before we start this analysis in detail, let us first define 
some more elements which will be useful later, the two {\it generalised momenta}:
\bea
k_R=(k_{\rho},-2)\label{genmom1}\\
k'_R=(f_k k_{\rho}/\Lam^2,-g_k),\label{genmom2}
\eea
where $f_k \equiv f(x={k^2\over \Lam^2}) = \frac{\tilde{c}(x)}{x
\tilde{c}(x)+2 c(x)}$ and $g_k \equiv g(x={k^2\over \Lam^2}) =
\frac{c(x)}{x \tilde{c}(x)+2 c(x)}$.  We can now recognise that the former
of the two generalised momenta acts on the vertices as though it was a standard gauge transformation. On the four point vertices, for example:
\bea\label{genward1}
k_R S^{AAFF}_{\,\mu\,\nu\, RS}(p,-p,k,-k)&=&S^{AA\vec{F}}_{\,\mu\,\nu S}(p,-p,0) -S^{ABF}_{\,\mu\,\nu S}(p,k-p,-k)\\
(-k)_S S^{AAFF}_{\,\mu\,\nu\, RS}(p,-p,k,-k)&=&S^{BAF}_{\,\mu\,\nu R}(p-k,-p,k)-S^{AA\tiny\la{F}}_{\,\mu\,\nu R}(p,-p,0)
\eea
and on the three point ones:
\bea\label{genward2}
k_R S^{AFF}_{\,\mu RS}(p,k,-p-k)&=&S^{A\vec{F}}_{\,\mu S}(p)-S^{BF}_{\,\mu S}(p+k)\\
(-k)_S S^{AFF}_{\,\mu RS}(p,-p+k,-k)&=&S^{BF}_{\,\mu R}(p-k)-S^{A\tiny\la{F}}_{\,\mu R}(p)
\eea
where $\ra{F}=(A, C\sigma)$ and $\la{F}=(A,-C\sigma)$. The action of the generalised momentum (\ref{genmom1}) on the two point vertices is finally a further gauge invariance statement since, as for the pure $A$ case a generalised transversality is underlined:
\bea
k_R S^{FF}_{RS}(k,-k)=0\\
(-k)_S S^{FF}_{RS}(k,-k)=0\label{genward3}
\eea
It is now time to move to the actual calculation.

\section{Calculation}

Due to the notation introduced in the previous section, comparing the set
of equations for the 2,3, and 4-point vertices in the $BD\sigma$ sector
with the corresponding equations in the $A$ and $C$ sectors, it is possible
to notice some similarities. The label $F$, which in the previous section
was introduced to represent the field doublet $(B,D\sigma)$ and which
simplified drastically the set of equations, could be
replaced by $A$ or $C$ and introducing some new Feynman rules for the new
wine vertices, from (\ref{FF}), (\ref{AFF}) and (\ref{AAFF}) we can get
the corresponding equations for the other two sectors. It is possible then
to introduce a field multiplet $f=(A,C,B,D\sigma$), to represent all
the equations, which could then be specified for each sector with the right
wine rules. As it was mentioned at the beginning of this section, this
further grouping of fields is not necessary since the $A$ and $C$ sectors
are decoupled, but since the calculation to be done is similar in all the
sectors, it is worth doing it in this notation to avoid repetitions and
specifying the components only at a later stage. In this new notation, (\ref{5.19}) gets the form:
\bea
-4\beta_1\BBox_{\mu\nu}(p^2)+ O(p^3)=2\int_k &&\sum_{f=A,C,F}(-)^{s_f}\left\{\dDelta^{ff}_{SR}(k)\Sigma^{ffAA}_{RS\mu\nu}(-k,k,p,-p)\right.\nonumber\\[3pt]
&&\phantom{\{}+\dDelta^{A,ff}_{\mu SR}(p;k-p,-k)\Sigma^{ffA}_{RS\nu}(p-k,k,-p)\nonumber\\[3pt]
&&\phantom{\{}+\left.\dDelta^{AA,ff}_{\mu\nu SR}(p,-p;k,-k)\Sigma^{ff}_{RS}(k)\right\}\nonumber\\[3pt]\label{complete}
\eea
where $s_A=s_C=0$ and $s_F=1$. The $\Sigma$'s here are $S_0-2\hat{S}$, the equations for the
$S_0$'s are now (\ref{FF}), (\ref{AFF}) and (\ref{AAFF}), with $F$ replaced
by the new $f$. The wine vertices are defined in the previous section, when
$f=F$ and for $A$ and $C$ are the ones listed in sections
\ref{twopoint&Kernels} and \ref{subsec:2.1.5}.

First of all, once we have defined the zero point wine (kernel) $\dDelta^{ff}_{RS}(p)$, we
can define its integrated form, as it was anticipated in section \ref{manifestly} (see also App.\ref{App:intwine} for details):
\be\label{wine-intwine}
\dDelta^{ff}_{SR}(p)=-\Lam\de_{\Lam}\Delta^{ff}_{SR}(p)
\ee
being now $\Delta^{ff}_{SR}(p)$ the {\it integrated wine}. Let us now consider the two point equation for the generalised
field multiplet $f$:
\be\label{ffeq}
\Lam\de_{\Lam}S^{ff}_{MN}(p)=S^{ff}_{ML}(p)\dDelta^{ff}_{LS}(p)S^{ff}_{SN}(p)
\ee
We can recognise the following relation (App.\ref{App:intwine}):
\be
S^{ff}_{RS}(k)\Delta^{ff}_{ST}(k)=\delta_{RT}-B^{ff}_{RT}(k)\label{inverse}
\ee

\begin{figure}[h!]
\begin{center}
\includegraphics[scale=.4]{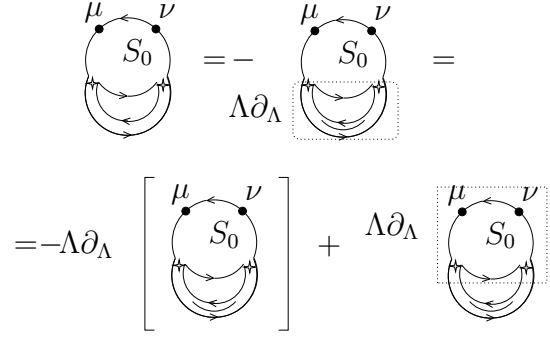}
\caption{Graphical representation of the method used to eliminate $\hat{S}$
                          in the equation for $\beta_1$. The stars
                          represent the field multiplet $f$}\label{fig:trick1f}
\end{center}
\end{figure}

\begin{figure}[h!]
\begin{center}
\includegraphics[scale=.5]{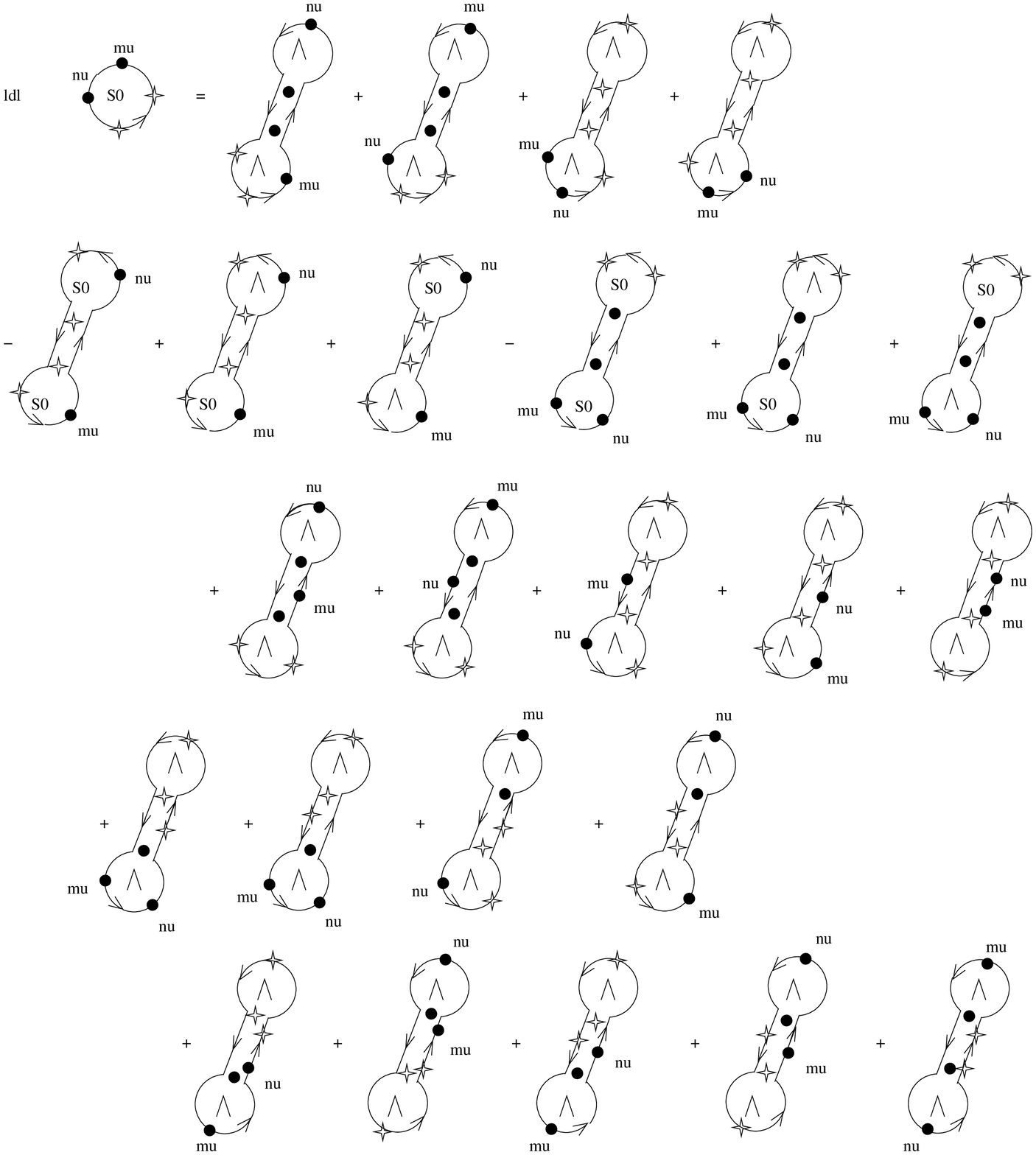}
\caption{Graphical representation of the flow of the tree level vertex AAff}\label{fig:aaff}
\end{center}
\end{figure}

Let us now consider the equation for $\beta_1$ (\ref{complete}). This is
expressed in diagram in fig.\ref{fig:beta1}. Following the steps of the the
scalar field case, shown in section \ref{sec:1.4} and more extensively in \cite{scaus}, we will try to
use the flow equations of the effective vertices in order to eliminate the
$\hat{S}$ ones. Consider then the first line on the RHS in (\ref{complete}):
\be\label{termA}
\dDelta^{ff}_{SR}(k)\Sigma^{ffAA}_{RS\mu\nu}(-k,k,p,-p)=\dDelta^{ff}_{SR}(k)\left[S^{ffAA}_{RS\mu\nu}-2\hat{S}^{ffAA}_{RS\mu\nu}\right]
\ee
Let us take only the effective vertex term (the first one in the previous
equation) and recalling eq.(\ref{wine-intwine}) we can write it as:
\bea
\dDelta^{ff}_{SR}(k) S^{ffAA}_{RS\mu\nu}(-k,k,p,-p)&=&-(\Lam\de_{\Lam}\Delta^{ff}_{SR}(k))
S^{ffAA}_{RS\mu\nu}(-k,k,p,-p)\nonumber\\[3pt]
&=&-\Lam\de_{\Lam}\left[\Delta^{ff}_{SR}(k) S^{ffAA}_{RS\mu\nu}(-k,k,p,-p)\right]\nonumber\\[3pt]
&\phantom{=}&+\Delta^{ff}_{SR}(k)
\Lam\de_{\Lam} S^{ffAA}_{RS\mu\nu}(-k,k,p,-p)\label{subeqmotion}
\eea
This is represented diagrammatically in fig.\ref{fig:trick1f}. 
We can now use the equation for the effective vertex $ffAA$ at tree-level, substituting it into the previous one. The graphical
representation of the equation for the effective vertex $ffAA$ is illustrated in fig.\ref{fig:aaff}

\noindent Substituting this equation into (\ref{subeqmotion}), which correspond to
substitute all the diagrams of fig.\ref{fig:aaff} in fig.\ref{fig:trick1f},
leads to the following equation:
\bea                    
\dDelta^{ff}_{SR}(k) S^{ffAA}_{RS\mu\nu}(-k,k,p,-p)=-\Lam\de_{\Lam}\left[\Delta^{ff}_{SR}(k) S^{ffAA}_{RS\mu\nu}(-k,k,p,-p)\right.]&&(0)\nonumber\label{bigeq}\\
+\Delta^{ff}_{SR}(k)\left\{\sh^{ffAA}_{RS\mu\alpha}(k,-k,p,-p)\dDelta^{AA}_p\sh^{AA}_{\,\alpha\,\nu}(p)\right.&&(1)\nonumber\\
+\sh^{ffAA}_{RS\alpha\nu}(k,-k,p,-p)\dDelta^{AA}_p\sh^{AA}_{\,\alpha\,\mu}(p)&&(2)\nonumber\\
+\sh^{ffAA}_{RT\mu\nu}(k,-k,p,-p)\dDelta^{ff}_{TU}(k)\sh^{ff}_{US}(-k)&&(3)\nonumber\\
+\sh^{ffAA}_{TS\mu\nu}(k,-k,p,-p)\dDelta^{ff}_{TU}(k)\sh^{ff}_{UR}(k)&&(4)\nonumber\\
-S^{ffA}_{TS\mu}(k-p,-k,p)\dDelta^{ff}_{TU}(k-p)S^{Aff}_{\,\nu RU}(-p,k,p-k)&&(5)\nonumber\\
+S^{ffA}_{TS\mu}(k-p,-k,p)\dDelta^{ff}_{TU}(k-p)\sh^{Aff}_{\,\nu RU}(-p,k,p-k)&&(6)\nonumber\\
+\sh^{ffA}_{TS\mu}(k-p,-k,p)\dDelta^{ff}_{TU}(k-p)S^{Aff}_{\,\nu RU}(-p,k,p-k)&&(7)\nonumber\\
-S^{AAA}_{\,\mu\,\nu\,\alpha}(p,-p,0)\dDelta^{AA}_{0}S^{Aff}_{\,\alpha RS}(0,k,-k)&&(8)\nonumber\\
+S^{AAA}_{\,\mu\,\nu\,\alpha}(p,-p,0)\dDelta^{AA}_{0}\sh^{Aff}_{\,\alpha RS}(0,k,-k)&&(9)\nonumber\\
+\sh^{AAA}_{\,\mu\,\nu\,\alpha}(p,-p,0)\dDelta^{AA}_{0}S^{Aff}_{\,\alpha RS}(0,k,-k)&&(10)\nonumber\\
+\sh^{ffA}_{RS\alpha}(k,-k,0)\dDelta^{A,AA}_{\,\mu}(p;0,-p)\sh^{AA}_{\,\alpha\,\nu}(p)&&(11)\nonumber\\
+\sh^{AA}_{\,\mu\,\alpha}(p)\dDelta^{A,AA}_{\,\nu}(-p;p,0)\sh^{Aff}_{\,\alpha RS}(0,k,-k)&&(12)\nonumber\\
+\sh^{ff}_{ST}(-k)\dDelta^{A,ff}_{\,\mu TU}(p;-k,k-p)\sh^{Aff}_{\,\nu RU}(-p,k,p-k)&&(13)\nonumber\\
+\sh^{Aff}_{\,\mu TS}(p,k-p,-k) \dDelta^{A,ff}_{\,\nu TU}(-p;p-k,k)\sh^{ff}_{UR}(-k)&&(14)\nonumber\\
+\sh^{ff}_{ST}(-k) \dDelta^{AA,ff}_{\,\mu\,\nu\ TU}(p,-p;-k,k)\sh^{ff}_{UR}(-k)&&(15)\nonumber\\
+\sh^{AAA}_{\,\mu\,\nu\,\alpha}(p,-p,0)\dDelta^{f,AB}_{R}(k;0,-k)\sh^{Bf}_{\,\alpha S}(k)&&(16)\nonumber\\
+\sh^{fB}_{R\alpha}(k)\dDelta^{f,BA}_{S}(-k;k,0)\sh^{AAA}_{\,\alpha\,\mu\,\nu}(0,p,-p)&&(17)\nonumber\\
+\sh^{AfB}_{\,\nu R\,\alpha}(-p,k,p-k)\dDelta^{f,BA}_{S}(-k;k-p,p)\sh^{AA}_{\,\alpha\,\mu}(p)&&(18)\nonumber\\
+\sh^{AA}_{\,\nu\,\alpha}(p)\dDelta^{f,AB}_{R}(k;-p,p-k)\sh^{BfA}_{\,\al S\,\mu}(k-p,-k,p)&&(19)\nonumber\\
+\sh^{AA}_{\,\mu\,\alpha}(p)\dDelta^{Af,AB}_{\,\nu R}(-p,k;p,-k)\sh^{Bf}_{\,\al S}(k)&&(20)\nonumber\\
+\sh^{fB}_{R\al}(k)\dDelta^{fA,BA}_{S\mu}(-k,p;k,-p)\sh^{AA}_{\,\alpha\,\nu}(p)&&(21)\nonumber\\
+\sh^{AA}_{\,\mu\,\alpha}(p)\dDelta^{A,f,AB}_{\,\nu,S}(-p;-k;p,k)\sh^{fB}_{R\al}(-k)&&(22)\nonumber\\
+\sh^{fB}_{S\al}(-k)\dDelta^{A,f,BA}_{\,\mu,R}(p;k;-k,-p)\sh^{AA}_{\,\alpha\,\nu}(p)&&(23)\nonumber\\
\left.+\sh^{AA}_{\,\nu\,\alpha}(p)\dDelta^{ff,AA}_{RS}(k,-k;-p,p)\sh^{AA}_{\,\alpha\,\mu}(p)\right\}&&(24)\nonumber\\
\eea
The terms inserted follow the order of fig.\ref{fig:aaff}, and the momenta
are specialised to the ones needed in the present case.
\begin{figure}[h!]
\begin{center}
\includegraphics[scale=.35]{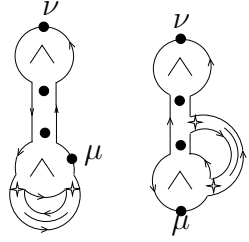}
\caption{Examples of potentially universal diagram}\label{fig:onef}
\end{center}
\end{figure}
Since we want only the order $p^2$ on the RHS of eq.(\ref{complete}), the
previous equation can be greatly simplified. Noticing that the two point
$A$ vertex is already order momentum squared, in all the terms in which it
appears carrying momentum $p$, we can set all the other $p$ dependences to
zero. This will cause many terms to be either not contributing to the
wanted order of $p$, or, due to gauge invariance, to have simplified
expressions. Terms of this kind, depending only on seed action two-point
vertices and their associated zero-point kernels (integrated or otherwise),
will be addressed as {\it potentially
universal}: since the seed action two-point vertices and the kernels derived from them,
 are the only things that we have 
 explicitly prescribed, for the result to be universal,
 it must be that we can reduce everything to such potentially universal
 terms or to total $\Lambda$ derivatives as in \eq{subeqmotion}. In turn,
 potentially universal terms must, and do, collect into total $k$
 derivatives, whose boundary terms on integration,
 are universal as a result of 
 restrictions on the large momentum behaviour, \eg \eq{inequalities},
 and the renormalization condition \eq{defg}. (Actually, 
 since $\dDelta^{AA}_0\propto c'_0$, by \eq{ker:a}, and $1/c'_0$ is
 never produced, terms such as \eq{here} are universal only because they
 combine to give boundary terms that vanish, as it is proved in section \ref{sub:c'0}). Examples of these terms are presented in the diagrammatic form
in fig.\ref{fig:onef}. As it can be noted, in fact, the upper blobs are two
$A$'s point vertices with momentum $p$.

Before starting to consider any of the terms of eq.(\ref{bigeq}), it is
important to point out some relations, due to gauge invariance, for
vertices evaluated at special momenta:
\bea
&&S^{Aff}_{\,\mu RS}(0,k,-k)=-S^{Aff}_{\,\mu RS}(0,-k,k)=\de^k_{\mu}S^{ff}_{RS}(k)\label{tre}\\
&&S^{AAff}_{\,\mu\,\nu\, RS}(0,0,k,-k)+S^{AAff}_{\,\nu\,\mu\, RS}(0,0,k,-k)=\de^k_{\mu}\de^k_{\nu}\dDelta^{ff}_{RS}\label{quattro}
\eea
(The second one, is symmetric in $k\to -k$). For the derivation of these equations see Appendix \ref{App:specmom}.
It is now time to perform the analysis of eq.(\ref{bigeq}) term by term. 
First of all we notice that the first two terms of eq.(\ref{bigeq}), have a
factor containing the two point $A$ vertex, evaluated at momentum $p$. At
order $p^2$ it becomes then proportional to $\dDelta^{AA}_0$:
\be\label{here}
2\dDelta^{AA}_0\BBox_{\na}(p)\Delta^{ff}_{SR}(k)\de_{\mu}^k\de_{\al}^kS^{ff}_{RS}(k)
\ee
(the extra factor 2 is for using $\mu\leftrightarrow \nu$ invariance and eq.(\ref{quattro}) has been used). A detailed study of this terms is left to section\ref{sub:c'0}.
\begin{figure}[h!]
\begin{center}
\includegraphics[scale=.4]{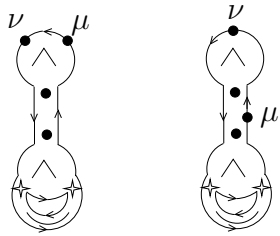}
\caption{These diagrams do not contribute to the order $p^2$}\label{fig:threef}
\end{center}
\end{figure}
There is now another group of terms to be considered together, which are those diagrams
not contributing to the order $p^2$. Two examples of this type are
represented diagrammatically in fig.\ref{fig:threef}. As one can notice, in
fact, one of them has the factor:
$S^{AAA}_{\,\mu\,\nu\,\al}(p,-p,0)$ (the left one in fig.\ref{fig:threef}). Making use of eq.(\ref{tre}), it is possible to
see that this term becomes:
\be
S^{AAA}_{\,\mu\,\nu\,\al}(p,-p,0)=\de^p_{\al}S^{AA}_{\,\mu\,\nu}(p)=\de^p_{\al}\left(\frac{2\BBox_{\mu\nu}(p)}{c_p}\right)
\ee
which is of order odd in $p$ and can not contribute to the wanted order.The
terms ruled out by this observation are (8)-(10) and (16)-(17). We can now
make a further reduction of the terms. Among the group of the potentially
universal the terms (11) and (12) in eq.(\ref{bigeq}) are left with
$\dDelta^{A,AA}_{\,\mu}(0;0,0)$ (the right one in fig.\ref{fig:threef}) which is
zero (clearly by Lorentz invariance). Finally, term (24) has two factors
$S^{AA}_{\,\mu\,\al}(p)$ therefore it is of order $p^4$. We are now ready to
proceed evaluating the terms left which will eventually give the
$\beta$-function at 1-loop.

We still have to evaluate (3)-(7), (13)-(15) and (18)-(23) of
eq.(\ref{bigeq}). Let us then start with (3) and (4):
\bea
\Delta^{ff}_{SR}(k)\left\{\sh^{ffAA}_{RT\mu\nu}(k,-k,p,-p)\dDelta^{ff}_{TU}(k)\sh^{ff}_{US}(-k)\right.&&\nonumber\\
\left.+\sh^{ffAA}_{TS\mu\nu}(k,-k,p,-p)\dDelta^{ff}_{TU}(k)\sh^{ff}_{UR}(k)\right\}&&\label{3-4}
\eea
\begin{figure}[h!]
\begin{center}
\includegraphics[scale=.45]{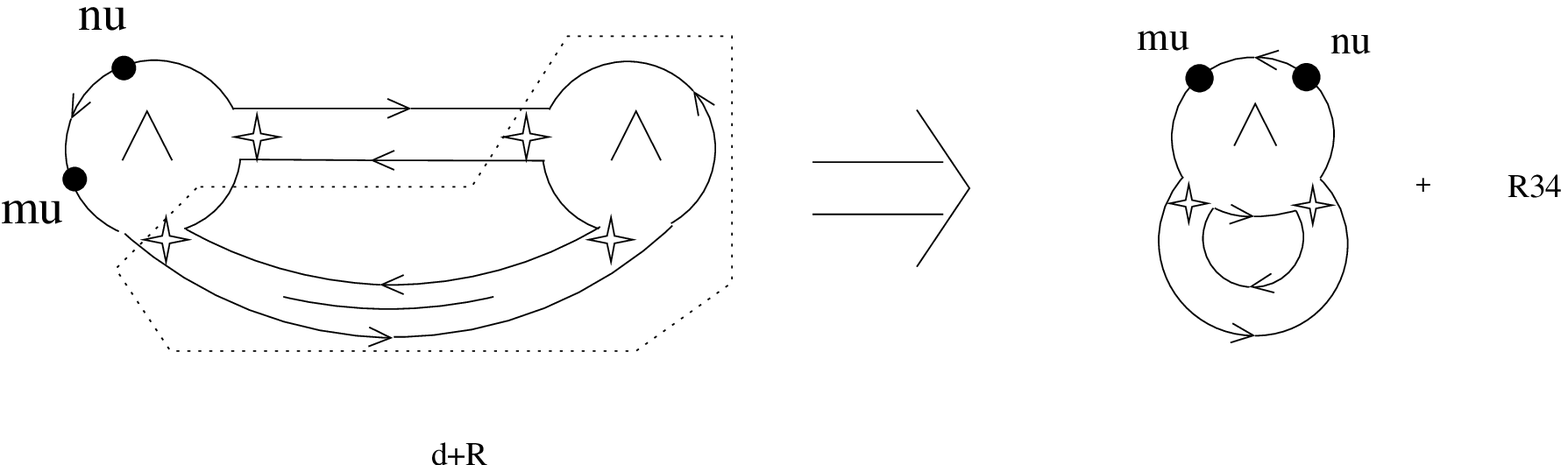}
\caption{Diagrammatical representation of the mechanism responsible of the
cancellation of the first $\hat{S}$-term of fig.\ref{fig:beta1} (first line of eq.(\ref{complete}))}\label{fig:twof}
\end{center}
\end{figure}
Making use of eq.(\ref{inverse}) it is possible to write the previous
equation as:
\be\label{twof}
2\sh^{ffAA}_{TR\mu\nu}(k,-k,p,-p)\dDelta^{ff}_{RT}(k)+R^f_{3,4}
\ee
since the wine vertices are diagonal. Where $R_{3,4}$ is the remainder, since
the contraction between the two point vertex and the integrated
wine, eq.(\ref{inverse}), does not give just $\one$, and it will be considered later. For now, let us compare the expressions in
eq.(\ref{twof}) and the second term in eq.(\ref{termA}). As it is shown
diagrammatically in fig.\ref{fig:twof}, the two terms (3) and (4) get the
same form and opposite sign of the hatted term of fig.\ref{fig:beta1} up to
the rest $R^f_{3,4}$, so they cancel out.

\begin{figure}[h!]
\begin{center}
\includegraphics[scale=.5]{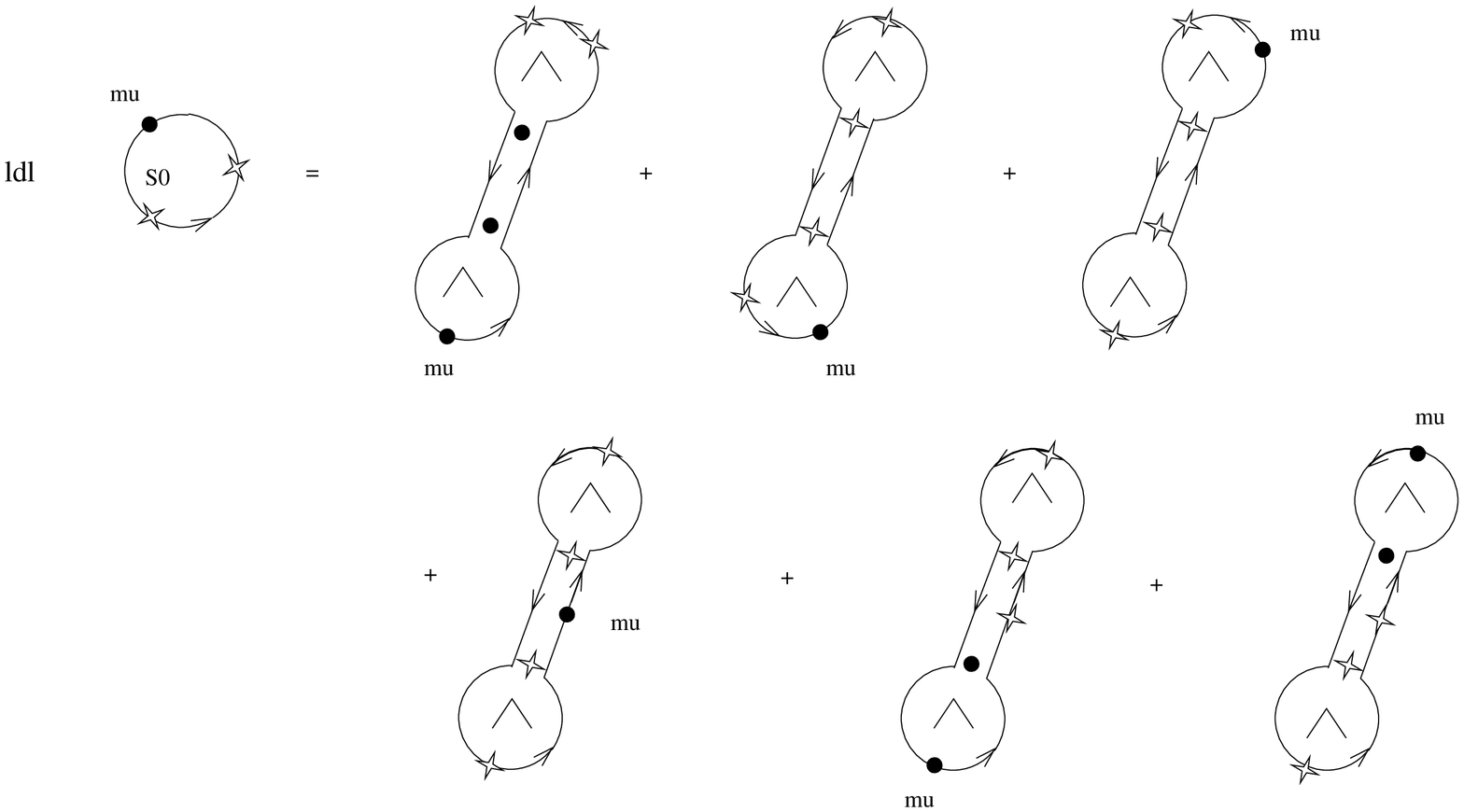}
\caption{Flow equation for the three point vertex Aff}\label{fig:aff}
\end{center}
\end{figure}

\begin{figure}[h!]
\begin{center}
\includegraphics[scale=.5]{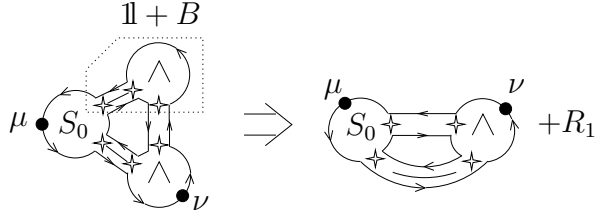}
\caption{Term arising from (5) in eq.(\ref{bigeq}), once the equation for $Aff$ is used, canceling (6)}\label{fig:term1}
\end{center}
\end{figure}

\begin{figure}[h!]
\begin{center}
\includegraphics[scale=.5]{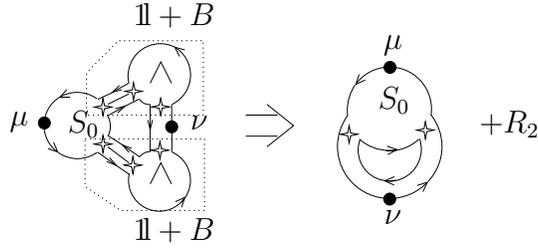}
\caption{Term arising from (5) in eq.(\ref{bigeq}), once the equation for $Aff$ is used, canceling the three point effective action term in eq.(\ref{complete})}\label{fig:term2}
\end{center}
\end{figure}

\begin{figure}[h!]
\begin{center}
\includegraphics[scale=.45]{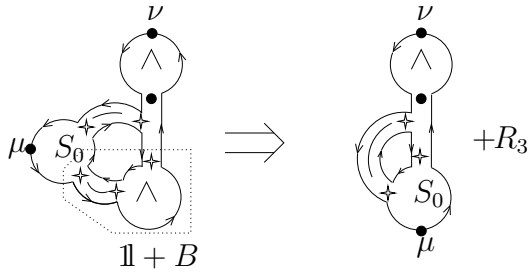}
\caption{Term arising from (5) in eq.(\ref{bigeq}), once the equation for $Aff$ is used,
canceling (18) and (19) in the same equation.}\label{fig:term3}
\end{center}
\end{figure}

\begin{figure}[h!]
\begin{center}
\includegraphics[scale=.45]{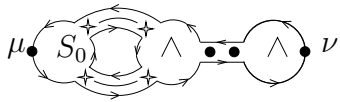}
\caption{Potentially universal terms left from (5) of eq.(\ref{bigeq})}\label{fig:terms}
\end{center}
\end{figure}

The next terms which will be considered are (13) and (14) of
eq.(\ref{bigeq}):
\bea
\Delta^{ff}_{SR}(k)\left\{\sh^{ff}_{ST}(-k)\dDelta^{A,ff}_{\,\mu TU}(p;-k,k-p)\sh^{Aff}_{\,\nu RU}(-p,k,p-k)\right.&&\nonumber\\
\left.+\sh^{Aff}_{\,\mu TS}(p,k-p,-k) \dDelta^{A,ff}_{\,\nu TU}(-p;p-k,k)\sh^{ff}_{UR}(-k)\right\}&&
\eea
\begin{figure}[h!]
\begin{center}
\includegraphics[scale=.45]{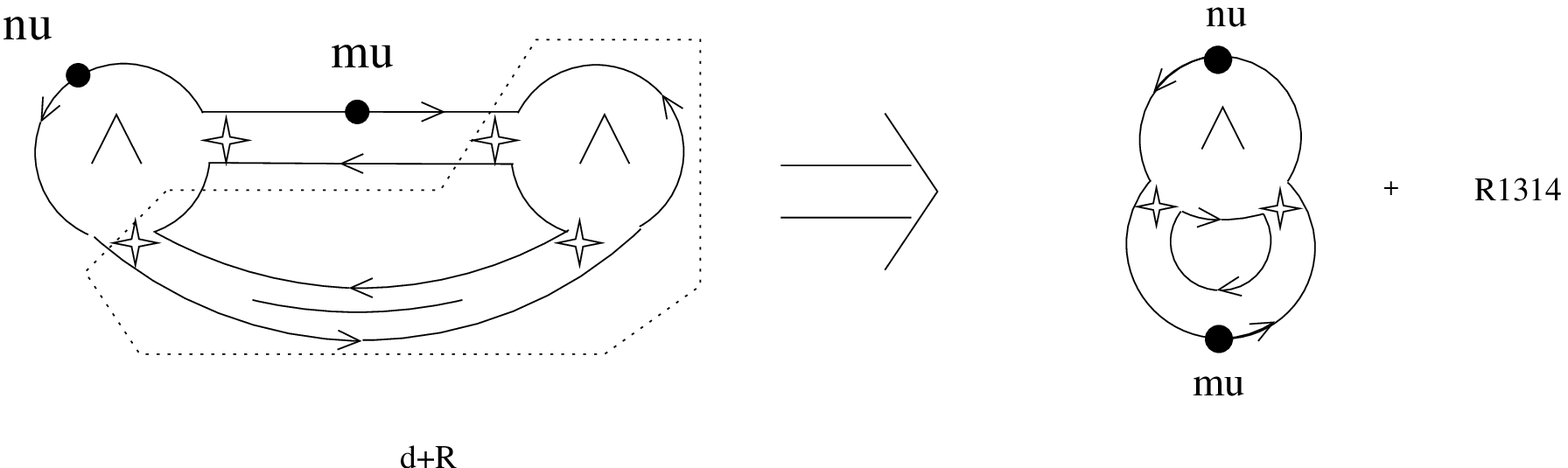}
\caption{Diagrammatical representation of the mechanism responsible of the cancellation of the second
$\hat{S}$-term of fig.\ref{fig:beta1} (second line of eq.(\ref{complete}))}\label{fig:fourf}
\end{center}
\end{figure}
Applying also in this case eq.(\ref{inverse}) one can recast the equation
above in the following form:
\be\label{fourf}
2\dDelta^{A,ff}_{\,\mu TU}(p;-k,k-p)\sh^{Aff}_{\,\nu TU}(-p,k,p-k)+R^f_{13,14}
\ee
Here we made use of the fact that $\Delta^{ff}_{RS}(k)$ is diagonal in $RS$
and is an even function of $k$ and we used the $\mu\leftrightarrow\nu,\
p\leftrightarrow -p$ symmetry. We also have considered the fact that the
above term had to be integrated over $k$ and used the translation
invariance for the integrated variable. It is possible also here to notice
that the first term in Eq(\ref{fourf}) exactly cancels the term containing
$\sh$ in the second line of eq.(\ref{complete}). This is diagrammatically
expressed in fig.\ref{fig:fourf}. The two terms
(13) and (14) of eq.(\ref{bigeq}) cancel the three-point hatted vertex from
eq.\ref{complete}) and what is left is just $R^f_{13,14}$ which will be
considered later. We will now move to term (15) of eq.(\ref{bigeq}). Its expression is the following:
\be
\Delta^{ff}_{SR}(k)\sh^{ff}_{ST}(-k) \dDelta^{AA,ff}_{\mu\nu TU}(p,-p;-k,k)\sh^{ff}_{UR}(-k)
\ee
Using the usual (\ref{inverse}) relation, we get:
\be\label{fivef}
\dDelta^{AA,ff}_{\,\mu\,\nu\ RU}(p,-p;-k,k)\sh^{ff}_{UR}(-k)+R^f_{15}
\ee
\begin{figure}[h!]
\begin{center}
\includegraphics[scale=.4]{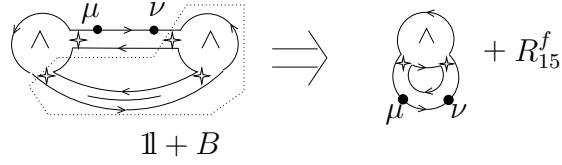}
\caption{Diagrammatical representation of the mechanism which causes the cancellation of the third
$\hat{S}$-term of fig.\ref{fig:beta1} (third line of eq.(\ref{complete}))}\label{fig:fivef}
\end{center}
\end{figure}
In this case it is already clear that the first term in eq.(\ref{fivef})
cancels the term containing $\sh$ in the third line of
eq.(\ref{complete}). The mechanism is again described by the diagrams of
fig.\ref{fig:fivef}. What is left of term (15) from eq.(\ref{bigeq}) after
the cancellation is the usual reminder, which will be indicated with $R^f_{15}$
and considered later. We now move to study the terms from (5) to (7) of
eq.(\ref{bigeq}). Also here the method adopted involves the use of the flow
equations for the effective vertices. Let us consider (5):
\be\label{(5)}
-\Delta^{ff}_{SR}(k)S^{ffA}_{TS\mu}(k-p,-k,p)\dDelta^{ff}_{TU}(k-p)S^{Aff}_{\nu RU}(-p,k,p-k)
\ee
\begin{figure}[hh]
\begin{center}
\psfrag{S0}{$S_0$}
\includegraphics[scale=.55]{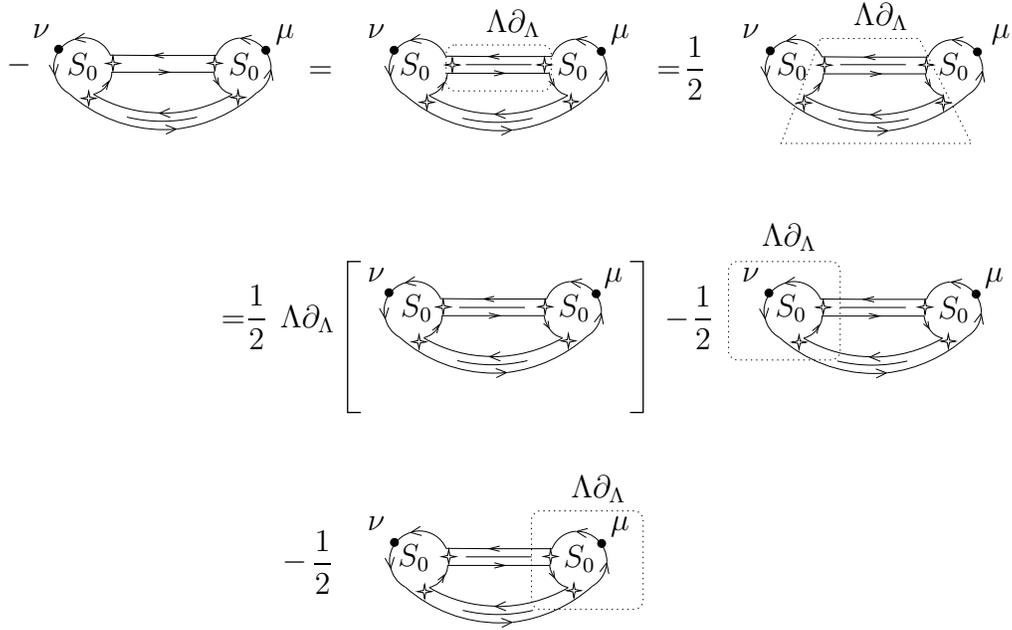}
\caption{Method used for the diagrams (5)-(7) of equation (\ref{bigeq}). The flow equation for the 3-point Aff is now needed}\label{fig:trick2f}
\end{center}
\end{figure}
As we can see from fig.\ref{fig:trick2f}, eq.(\ref{(5)}) can be rewritten
as:
\bea
\frac{1}{2}\Lam\de_{\Lam}\left[\Delta^{ff}_{SR}(k)\Delta^{ff}_{TU}(k-p)S^{ffA}_{TS\mu}(k-p,-k,p)S^{Aff}_{\,\nu
RU}(-p,k,p-k)\right]\nonumber\\
-\frac{1}{2}\Delta^{ff}_{SR}(k)\Delta^{ff}_{TU}(k-p)S^{Aff}_{\,\nu
RU}(-p,k,p-k)\Lam\de_{\Lam}S^{ffA}_{TS\mu}(k-p,-k,p)+p_{\mu}\leftrightarrow p_{\nu}\label{term5}
\eea
The first term will be considered later, together
with (0) in eq.(\ref{bigeq}), while the second term needs now the
introduction of the equation for the three point function $ffA$. This equation is expressed in diagrams in fig.\ref{fig:aff}.
If now one substitutes the terms of fig.\ref{fig:aff} in
fig.\ref{fig:trick2f} there would be terms such as those represented in
figs.\ref{fig:term1}, \ref{fig:term2} and \ref{fig:terms}. Making use of the
inverse relation eq.(\ref{inverse}) many cancellations occur. To start with, as one can notice from fig.\ref{fig:term1}, term (6) is
cancelled when the last two terms of the first line in fig.\ref{fig:aff} are
substituted in fig.\ref{fig:trick2f}. In the same way (7) is canceled when
the $\mu\leftrightarrow\nu$ term of fig.\ref{fig:trick2f} is considered.
Moreover, the effective three point vertex term (the last left in
eq.(\ref{complete})) is canceled when the first term on the second line of
fig.\ref{fig:aff} is substituted in fig.\ref{fig:trick2f} (as it is
explained diagrammatically in fig.\ref{fig:term2}) and when the same is
done for its analogous in the swop $\mu\leftrightarrow\nu$. This can be
checked writing them in formulae using the rules for wines, effective and
hatted vertices. A further cancellation occurs due to (5), and it is the one
described in fig.\ref{fig:term3}. Due to the inverse relation
(\ref{inverse}), the last two terms of the three point vertex $Aff$ when
substituted in eq.(\ref{term5}), cancel exactly terms (18) and (19) of
eq.(\ref{bigeq}). What we are left with, after the analysis of term (5),
are then three reminders from the three cancellations which occurred, which will be
considered later on with all the others, and the term
described in diagrams in fig.\ref{fig:terms}.
This term comes from the first term of the equation for $Aff$. Its
analytical form comes directly from the diagram in
fig.\ref{fig:terms}. Since it contains the two point $A$ vertex evaluated at
momentum $p$, which as we noted previously has already order $p^2$, once
the other momenta $p$ are set to zero to get only the wanted order in $p$
we get:
\be
-2\dDelta^{AA}_0\BBox_{\nu\al}(p)\Delta^{ff}_{RU}(k)\Delta^{ff}_{ST}(k)\de^k_{\mu}S^{ff}_{RS}(k)\de^k_{\al}S^{ff}_{TU}(k)
\ee
(the extra factor of $2$ comes from Lorentz invariance, considering it
together with the analogous $\mu\leftrightarrow\nu$ term and eq.(\ref{tre}) has
been used). This is another term proportional to $\dDelta^{AA}_0$ and will be considered
separately in section \ref{sub:c'0}.
What we are left with from eq.(\ref{bigeq}) are then the terms (20)-(23), the
six remainders and the two total derivative terms. 
Let us consider (20)-(23) first. Since they all are proportional to the
two-point $A$ vertex at momentum $p$, using the inverse relation (\ref{inverse}), they can be grouped together as follows:
\be\label{20-23}
4\BBox\na(p)\left[B^{ff}_{\al R}(-k)\dDelta^{Af,BA}_{\,\mu
R}(0,k;-k,0)-\dDelta^{Af,fA}_{\,\mu R,R\alpha}(0,k;-k,0)\right]
\ee
To get \eq{20-23}, we have also used a consequence of the coincident lines
identity, which relates wine vertices, with fields on different lines; in
particular, for our purposes, we have considered:
\be
\dDelta^{A,f,BA}_{\,\mu, R}(p;q;r,s)=-\dDelta^{Af,BA}_{\,\mu
R}(p,q;r,s)-\dDelta^{fA,BA}_{R\mu}(q,p;r,s)
\ee
It is also important to notice, that the second term in \eq{20-23}, will
not appear when $f$ is specified to the $C$ sector, since there is no
$\dDelta^{\A\C}$ kernel. For the $A$ and $F=(B,D\sigma)$ sectors,
it will be respectively $\dDelta^{AA,AA}_{\,\mu\,\alpha}(0,k;-k,0)$ and
$\dDelta^{AB,BA}_{\,\mu\,\alpha}(0,k;-k,0)=\dDelta^{AA,AA}_{\,\mu\,\alpha}(0,k;-k,0)$
(again because there is no $\dDelta^{\A\C}$ kernel). These contributions,
as it can be proved, cancel between the $A$ and $F$ sectors, being of opposite
sign. However, since they represent Wines biting their tail diagrams, which
are excluded by eqs. (\ref{noATailBiting}) and (\ref{noTailBiting}), they
will not be included in the rest of the calculation. 
We are now ready to summarise and rewrite
eq.(\ref{complete}) as:
\bea
-4\beta_1\BBox_{\mu\nu}(p)=N\int_k
\left\{\Lam\de_{\Lam}\left[\Delta^{ff}_{SR}(k)\Delta^{ff}_{TU}(k-p)S^{ffA}_{TS\mu}(k-p,-k,p)S^{Aff}_{\,\nu
RU}(-p,k,p-k)\right.\right.\nonumber\\
\left.-2\Delta^{ff}_{SR}(k) S^{ffAA}_{RS\mu\nu}(-k,k,p,-p)\right]_{p^2}&(1)\nonumber\\
-2\dDelta^{ff}_{TU}(k)\left[\sh^{ffAA}_{RT\mu\nu}(k,-k,p,-p)B^{ff}_{UR}(k)+\sh^{ffAA}_{TS\mu\nu}(k,-k,p,-p)B^{ff}_{US}(-k)\right]&(2)\nonumber\\
-2\dDelta^{A,ff}_{\,\mu TU}(p;-k,k-p)\sh^{Aff}_{\,\nu
RU}(-p,k,p-k)B^{ff}_{TR}(k)\nonumber\\
-2\dDelta^{A,ff}_{\,\nu
TU}(-p;p-k,k)\sh^{Aff}_{\,\mu TR}(p,k-p,-k) B^{ff}_{UR}(-k)&(3)\nonumber\\
-2\dDelta^{AA,ff}_{\,\mu\,\nu\ TU}(p,-p;-k,k)\sh^{ff}_{UR}(-k)B^{ff}_{TR}(k)&(4)\nonumber\\
+2S^{Aff}_{\,\nu RU}(-p,k,p-k)\left[\sh^{Aff}_{\,\mu MS}(p,k-p,-k)\Delta^{ff}_{SR}(k)\dDelta^{ff}_{MN}(k-p)B^{ff}_{NU}(p-k)+\right.&\nonumber\\
\left.+\sh^{Aff}_{\,\mu TM}(p,k-p,-k)\Delta^{ff}_{TU}(p-k)\dDelta^{ff}_{MN}(k)B^{ff}_{NR}(k)\right]&(5)\nonumber\\
+2S^{Aff}_{\,\mu RT}(p,k-p,-k)\left[\dDelta^{A,ff}_{\,\nu RU}(-p;p-k,k)B^{ff}_{UT}(-k)+\dDelta^{A,ff}_{\,\nu UT}(-p;p-k,k)B^{ff}_{UR}(k-p)\right.\nonumber\\
\left.-\dDelta^{A,ff}_{\,\nu VU}(-p;p-k,k)B^{ff}_{UT}(-k)B^{ff}_{VR}(k-p)\right]&(6)\nonumber\\
+8\BBox\na(p)\dDelta^{f,Af}_{U,\alpha V}(k;0,-k)B^{ff}_{VR}(k)\de^k_{\mu}\sh^{ff}_{\,RT}(k)\Delta^{ff}_{TU}(k)&(7)\nonumber\\
+8\BBox\na(p)B^{ff}_{\,VR}(k)\dDelta^{Af,fA}_{\,\mu
R,V\alpha}(0,k;-k,0)&(8)\nonumber\\
\left.+4\dDelta^{AA}_0\BBox\na(p)\left[\Delta^{ff}_{SR}(k)\de_{\mu}^k\de_{\al}^kS^{ff}_{RS}(k)-\Delta^{ff}_{RU}(k)\Delta^{ff}_{ST}(k)\de^k_{\mu}S^{ff}_{RS}(k)\de^k_{\al}S^{ff}_{TU}(k)\right]\right\}&(9)\nonumber\\\label{final}
\eea

\subsection{The $A$ sector}

Let us start, with the first contribution to be evaluated in this sector
which is $R^f_{3,4}$. The
general expression is represented by term (2) in \eq{final}. Recalling the form
of $B^{ff}_{RS}(k)$ in the pure $A$ case from eq.(\ref{inversea}), we can
rewrite this term as follows:
\be\label{r34a}
R^A_{3,4}=-4\dDelta^{AA}(k)\frac{k_{\sigma}k_{\rho}}{k^2}\sh^{AAAA}_{\,\rho\,\sigma\,\mu\,\nu}(k,-k,p,-p).
\ee
Making use of the usual Ward identities for the pure gauge field, and
recalling eq.(\ref{tre}), we can write the previous equation as:
\be
R^A_{3,4}=-\frac{4\dDelta^{AA}(k)}{k^2}\left(k_{\sigma}\de^p_{\sigma}
S^{AA}_{\,\mu\,\nu}(p)+S^{AA}_{\,\mu\,\nu}(p-k)-S^{AA}_{\,\mu\,\nu}(p)\right)
\ee
Since the first term in the above equation has only odd powers of $p$, it does not
contribute to the order $p^2$ and it will not be taken into account for the
rest of the calculation. We are then left with:
\be\label{r34ap2}
\left.R^A_{3,4}\right|_{p^2}=-\frac{4\dDelta^{AA}(k)}{k^2}\left[\left.S^{AA}_{\,\mu\,\nu}(p-k)\right|_{p^2}-2\BBox_{\mn}(p)\right]
\ee
The next term to be considered is now $R^f_{13,14}$ (term (3) in
\eq{final}) which in this sector gets the following form:
\bea
R^A_{13,14}&=&-2\dDelta^{A,AA}_{\mu}(p;-k,k-p)\frac{k_{\rho}k_{\sigma}}{k^2}\sh^{AAA}_{\,\nu\,\rho\,\sigma}(-p,k,p-k)\nonumber\\
&&-2\dDelta^{A,AA}_{\,\nu}(-p;p-k,k)\frac{(-k)_{\rho}(-k)_{\sigma}}{k^2}\sh^{AAA}_{\,\mu\,\rho\,\sigma}(p,k-p,-k)\label{r1314a}
\eea
Applying the Ward identities and making use of the symmetry $\mu\leftrightarrow\nu$, it is possible to get the final expression (at order $p^2$):
\be\label{r1314afinal} 
\left.R^A_{13,14}\right|_{p^2}=\frac{4k_{\rho}}{k^2}\left[\left.\dDelta^{A,AA}_{\,\nu}(-p;p-k,k)\sh^{AA}_{\,\mu\,\rho}(p-k)\right|_{p^2}-2\de^k_{\mu}\dDelta^{AA}(k)\BBox_{\nu\rho}(p)\right]
\ee
We now have to analyse term (3) of \eq{final}, which is $R^f_{15}$. Its expression in the sector $A$ is:
\be\label{r15a}
R^A_{15}=\dDelta^{AA,AA}_{\,\mu\,\nu}(p,-p;,-k,k)\frac{k_{\al}k_{\beta}}{k^2}S^{AA}_{\al\beta}(k)=0,
\ee
by gauge invariance.

We have now to evaluate the last three reminders (5)-(7) of eq.(\ref{final}). They have in the $A$
sector the following form:
\bea
R_1^A&=&-\frac{16\dDelta^{AA}(k)}{k^2}\BBox\na(p)\left(\delta\am-\dfrac{k_{\al}k_{\mu}}{k^2}\right)+4\frac{\dDelta^{AA}(k)}{k^2}\left.\sh^{AA}_{\,\mu\,\nu}(p-k)\right|_{p^2}\label{r1a}\\
R_2^A&=&\frac{4 k_{\rho}}{k^2}\left[-\dDelta^{A,AA}_{\,\nu}(-p;k,p-k)\left.\sh^{AA}_{\,\mu\,\rho}(p-k)\right|_{p^2}+\de^k_{\nu}\dDelta^{AA}(k)\BBox_{\mu\rho}(p)\right]\label{r3a}\\
R_3^A&=&\frac{-8k_{\al}}{k^2}\BBox\na(p)\left[\dDelta^{A,AA}_{\,\mu}(k;0,-k)+\frac{k_{\mu}}{k^2}(\dDelta^{AA}(k)-\dDelta^{AA}_0)\right]=0\label{r2a}
\eea
where Ward identities and the inverse relation (\ref{inverse}) have been used,
and the order $p^2$ taken. Before starting with the total derivative term (1)
and the one proportional to $\dDelta^{AA}_0$ (9), let us consider term (8) in
eq.(\ref{final}). In this sector, it takes the form:
\be\label{9}
8\BBox\na(p)\left[\frac{k_{\al}}{k^2}\de^k_{\mu}\dDelta^{AA}(k)-\frac{\k_{\mu}k_{\alpha}}{k^4}\dDelta^{AA}(k)\right]
\ee
It is possible to notice now that adding up all these contributions, the
non-universal terms cancel out and that many other cancellations
occur. 
Collecting the terms which are left, recalling the expression of the $A$-kernel from \eq{ker:a} and considering the integral over
$k$ in (\ref{final}), after the
average over the $k$ components is taken, we finally get:
\be\label{allbutc'andtd-a}
-\frac{8N}{\Lam^4}\BBox\mn(p)\dints \frac{d^D k}{(2\pi)^D}\ \left[\left(1-\dfrac{1}{D}\right)\dfrac{\Lam^2c'_k}{k^2}-\dfrac{1}{D}c''_k\right]
\ee
Setting now $x=k^2/\Lam^2$, and with some algebra, we find the previous
equation becomes:
\bea
2N\dirac{\Omega}_D\BBox\mn(p)\dints_0^{\infty} dx\ \left\{\left[\left(\frac{4}{D}-3\right)c(x)\
x^{D/2-2}+\frac{2}{D}c'(x)\ x^{D/2-1}\right]'\right.\\
\left.-\frac{(D-4)(4-3D)}{D}\ \frac{x^{D/2-3}}{2}c(x)\right\}\nonumber
\eea
where $\dirac{\Omega}_D$ is the integration over the angles divided by
$(2\pi)^D$. (For the other calculations, we will not specify this last
step, but we will just write the result in $D=4$.) Since this integral is
regular in $D$ around $D=4$, we can specify to that value and get a total derivative:
\be\label{remaindercontr-a}
\ba{r}
N\dirac{\Omega}_4\BBox\mn(p)\dints_0^{\infty}dx\left[-4c(x)+x\ c'(x)\right]'\\
=4N\dirac{\Omega}_4\BBox\mn(p)
\ea
\ee

In \eq{allbutc'andtd-a} we did not include the total derivative term and the
ones proportional to $\dDelta^{AA}_0$. The total contribution from the $A$ sector to
this latter set of terms is:
\be
 4\dDelta^{AA}_0
    \BBox_{\mu\al}(p)\left[\Delta^{AA}(k)\de^k_{\al}\de^k_{\nu}S^{AA}_{\,\beta\,\beta}(k)-\left(\Delta^{AA}(k)\right)^2\de^k_{\al}S^{AA}_{\,\lambda\,\rho}(k)\de^k_{\nu}S^{AA}_{\,\lambda\,\rho}(k)\right] 
\ee
which is term (9) of \eq{final} in this sector, and:
\be
8\dDelta^{AA}_0\BBox_{\mu\al}(p)\frac{k_{\nu}k_{\al}}{k^4}
\ee
a further term of this kind to be added to the previous one, which comes
from eqs.(\ref{r3a}) and (\ref{9}). These terms will be considered later
on in section \ref{sub:c'0}. 

It is now time to consider (1) of \eq{final}, \ie the total derivative term.
After substituting in it the expression for the integrated $A$-kernel from
\eq{app:aa}, a first observation we can make is that if the term was UV and IR regulated
we could pull the $\Lam$-derivative out of the integral, as in the scalar field case, and the
result would be identically zero. The integrand is in fact dimensionless
and once the integration is carried out the result would be a
$\Lam$-derivative of a scale independent quantity. 
Since the UV finiteness is ensured by the regularisation, the non zero
contribution of this term, if there is one, must come from its IR poles. We
want then to study the $k\to 0$ behaviour of the following term:
\be\label{td}
\frac{c_k}{2k^2}\left[\frac{c_{p-k}}{2(p-k)^2}S^{AAA}_{\,\mu\,\rho\,\sigma}(p,k-p,-k)S^{AAA}_{\,\nu
\,\sigma\,\rho}(-p,k,p-k)-2S^{AAAA}_{\,\mu\,\nu\,\al\,\al}(p,-p,-k,k)\right]
\ee
and take its order $p^2$. Eq.(\ref{td}) has to be integrated in $d^Dk$, and
the second term of the two does not contribute at all since its integral is
IR convergent. The first one instead has to be studied. The first step
is to expand the factor that might carry extra poles (the second integrated
wine) up to the order $p^2$:
\be\label{exp}
\left.\frac{c_{p-k}}{(p-k)^2}\right|_{p^2}=\frac{c_k}{k^2}-2\frac{c_k}{k^4}k\cdot
p-\frac{c_k}{k^4}\left(p^2-4\frac{(k\cdot
p)^2}{k^2}\right)+\frac{c'_k}{k^2\Lam^2}\left(p^2-4\frac{(k\cdot
p)^2}{k^2}\right)
\ee
The last term cannot contribute because the product of the two three
point vertices gives at least as $k^2$, and thus results in an integrable term. We are then left with three
terms, a $p^0$ term, a term linear in $p$ and a term of order $p^2$. The
crucial observation to handle this total derivative term is now that even
though it is not potentially universal, its IR contribution happens to be,
as we will describe. Let us take for example the $p^0$ term in the
expansion (\ref{exp}). In $D=4$, substituting it in \eq{td}, we can see
that the integral is IR convergent unless in both of the three point vertices appearing in the product,
we take the order zero in $k$. This means that using \eq{tre} we get a
potentially universal term. If we now consider the term linear in $p$, we
must get another power of $p$ and this can come from either of the two
three-point vertices in the product. This means that the other one must be
taken at order $p^0$ and this makes it potentially universal. Moreover, in
$D=4$, the first three point function must be taken at order zero in $k$
since otherwise is at least linear in $k$ and the term would be integrable. This makes the term potentially universal. Finally, there is
the term of \eq{exp} which is already of order $p^2$. It is obvious in this
latter case that we do not need any more powers of $p$ and the three point
functions must be taken at $p=0$. The whole total derivative term is then
potentially universal in its IR-divergent regime. The only contribution
from expression (\ref{td}) takes the following form:
\bea
\dfrac{c_k^2}{2k^4}\left[\de^p_{\sigma}S^{AA}_{\,\mu\,\rho}(p)\de^p_{\sigma}S^{AA}_{\,\nu\,\rho}(p)-\dfrac{4}{k^2}(k\cdot
p)\de^k_{\mu}S^{AA}_{\,\rho\,\sigma}(k)\de^p_{\sigma}S^{AA}_{\,\nu\,\rho}(p)\right.&&\nonumber\\
\left.-\dfrac{1}{k^2}\left(p^2-4\dfrac{(k\cdot p)^2}{k^2}\right)\de^k_{\mu}S^{AA}_{\,\rho\,\sigma}(k)\de^k_{\nu}S^{AA}_{\,\rho\,\sigma}(k)\right]&&
\eea
Considering now the expression for the two point function $2\BBox\mn/c_p$,
taking the order $p^2$ of the previous equation and the average on the $k$
momenta, we get (in $D=4$ and recalling the factor N in (\ref{final})):
\be\label{tdcontr-a}
N\dirac{\Omega}_4\left(\frac{19}{6}p^2\delta\mn-\frac{11}{3}p_{\mu}p_{\nu}\right)
\ee
where the following identity has effectively been used:
\be
\lim_{D\to 4^+}\Lam\de_{\Lam}\int\frac{d^D k}{(2\pi)^D\ k^4}=\lim_{D\to 4^+}\frac{\dirac{\Omega}_D}{D-4}\Lam\de_{\Lam}\Lam^{D-4}=\dirac{\Omega}_4
\ee
(Here, only the IR divergent part is shown).
This was the contribution from the total derivative term of the $A$ sector.
In order to get the total contribution to the $\beta$ function at one loop
from this sector we have to sum to this, the result obtained from
the rest in \eq{remaindercontr-a}. As one can notice this partial result
is not transverse as it would be expected (the LHS of \eq{final} is transverse and so the RHS ought to be). However, the
contribution from the $BD\sigma$ sector will contribute with another non
transverse term and give a transverse final result.

\subsection{The $C$ sector}

In the present sector there is no contribution expected from terms (2)-(7)
in \eq{final}, since as can be seen from eq.(\ref{inversec}) $B^{CC}(k)=0$. What we have is then:
\be\label{rc}
R^C_{3,4}=R^C_{13,14}=R^C_{15}=R_1^C=R_2^C=R_3^C=0
\ee
Moreover term (8) from \eq{final} is zero too, since it comes from (20)-(23)
in \eq{complete} and these two point wines are zero when $f=C$ (as was
pointed out above \eq{final}). As far as
this sector is concerned, the only contributions are from (1) and (9) of
\eq{final}, the total derivative term and the one proportional to
$\dDelta^{AA}_0$. Let us analyse the former first.The analysis to be carried out here is
similar to that one for the $A$ sector. We will be looking for IR
poles. The total derivative term in this sector has the following form:
\bea
\Delta^{CC}(k)\left[\Delta^{CC}(k-p)S^{ACC}_{\,\mu}(p,k-p,-k)S^{ACC}_{\,\nu}(-p,k,p-k)\right.\nonumber\\
\left.-2S^{AACC}_{\,\mu\,\nu}(p,-p,-k,k)\right]\label{tdc}
\eea
From the explicit expression of $\Delta^{CC}$, \eg \eq{deltacc}, one can
notice that it is regular as $k\to 0$, as its derivatives are. The
expression in \eq{tdc} is then not only UV regular (after adding in the
other sectors), but also IR regular, and the $\Lam$-derivative can be pulled out of the integral
giving a vanishing contribution.

To finish the analysis of this sector we are then left with the set of
terms proportional to $\dDelta^{AA}_0$. Here we will just write term (9) of
\eq{final} specialised in the sector $C$:
\be
4\dDelta^{AA}_0\BBox\na(p)\left[\Delta^{CC}(k)\de_{\mu}^k\de_{\al}^kS^{CC}(k)-(\Delta^{CC}(k))^2\de^k_{\mu}S^{CC}(k)\de^k_{\al}S^{CC}(k)\right]
\ee
This term will be considered in section \ref{sub:c'0}.

\subsection{The $BD\sigma$ sector}

Let us finish this analysis considering the last sector. The first term that
has to be evaluated is again $R^f_{3,4}$. In this particular case, from the
expression of $B^{FF}_{RS}(k)$ of eq.(\ref{inversebds2}) and recalling that
there is an extra minus sign for all the terms when $f=F$ (as one can see
comparing eqs.(\ref{5.19}) and (\ref{complete})), we can write it as:
\be
R^F_{3,4}=2\dDelta^{FF}_{TU}(k)\left[ k'_U k_R\sh^{FFAA}_{RT\,\mu\,\nu}(k,-k,p,-p)+(-k')_U (-k)_S\sh^{FFAA}_{TS\,\mu\,\nu}(k,-k,p,-p)\right]
\ee
where $k$ and $k'$ are the generalised momenta of eqs.(\ref{genmom1}) and (\ref{genmom2}). It is
now possible to see that, in this case, one can apply the
generalised Ward identities described in section \ref{sub:compact}, namely
Eqs.(\ref{genward1})-(\ref{genward3}). Due to a similar argument to the one used in the pure
$A$ field, the wanted expression at order $p^2$ is:
\be\label{r34bds}
\left.R^F_{3,4}\right|_{p^2}=-4\dDelta^{FF}_{TU}(k)k'_U\left.\sh^{ABF}_{\,\mu\,\nu T}(p,k-p,-k)\right|_{p^2}
\ee
the components will be specified once the other terms are considered. The
next term which has to be studied is then $R^f_{13,14}$. Its general
expression is term (3) of \eq{final} and in the present sector, gets the form:
\bea
R^F_{13,14}&=&2\dDelta^{A,FF}_{\,\mu TU}(p;-k,k-p)k'_T k_R\sh^{AFF}_{\,\nu RU}(-p,k,p-k)\nonumber\\
&=&2\dDelta^{A,FF}_{\,\nu\, TU}(-p;p-k,k)(-k)'_U (-k)_R\sh^{AFF}_{\,\mu\, TR}(p,k-p,-k) \label{r1314bds}
\eea
Making use of the generalised Ward identities, using the properties of the wine vertices and with some algebra, it is possible to recognise the final result to be:
\be\label{r1314bdsfinal}
\left.R^F_{13,14}\right|_{p^2}=4k'_U\left.\dDelta^{A,FF}_{\,\mu\,
TU}(p;k-p,-k)\sh^{BF}_{\,\nu
T}(k-p)\right|_{p^2}+8\BBox\ma(p)k'_{\al}\de^k_{\nu}\dDelta^{BB}(k)
\ee
Also in this case we keep for now the compact notation before we consider all the other terms, in order to see all the simplifications which occur already at this level. The last term of this kind to consider is $R^f_{15}$, which in the $BD\sigma$ sector gets the form:
\be\label{r15bds}
R^F_{15}=2\dDelta^{AA,FF}_{\,\mu\,\nu\, TU}(p,-p;-k,k)k'_T k_R\sh^{FF}_{UR}(-k,k)
\ee
Due to the generalised Ward identity (\ref{genward3}) this is identically
zero as in the case of the pure $A$.

The last three remainders in the $BD\sigma$ sector have the following form:
\bea
\ba{r}
\left.R^F_1\right|_{p^2}=-8\BBox\ma(p)\left[\dDelta^{BB}(k)k'_{\al}k'_{\nu}+\dDelta^{FF}_{MN}(k)\Delta^{BB}(k)k'_N\de^k_{\nu}\sh^{BF}_{\al
M}(k)\right]\\[5pt]
\ \ \left.+4\dDelta^{FF}_{SN}(k)k'_N (k-p)'_{\nu}\sh^{BF}_{\,\mu
S}(k)\right|_{p^2}\\[5pt]
\ \ +\left.4\dDelta^{FF}_{MN}(k)k'_N\sh^{ABF}_{\,\mu
\,\nu M}(p,k-p,-k)\right|_{p^2}
\ea\nonumber&&\\
&&\\
\left.R^F_2\right|_{p^2}=-4\BBox\ma(p)\left[2k'_{\al}\de^k_{\nu}\dDelta^{BB}(k)+k_{\al}k'_U(-k')_V\de^k_{\nu}\dDelta^{FF}_{UV}(k)\right]\nonumber&&\\
\left.-4\dDelta^{A,FF}_{\,\mu RU}(p;k-p,-k)k'_U\sh^{BF}_{\,\nu R}(k-p)\right|_{p^2}&&\nonumber\\
&&\\
\left.R^F_3\right|_{p^2}=-8k'_{\al}\BBox\ma(p)\left[\dDelta_{\,\nu}^{B,BA}(k;-k,0)+k'_{\nu}\left(\dDelta^{AA}_0-\dDelta^{BB}(k)\right)\right]&&\label{rf3}
\eea
From the two terms (8) of \eq{final}, and recalling the comment just above
eq.(\ref{final}), we have this final expression:
\be\label{lastterm}
-8\BBox\ma(p)k'_{\al}\left(\de^k_{\nu}\dDelta^{BB}(k)-\dDelta^{B,BA}_{\,\nu}(k;-k,0)\right)
\ee
It is possible already to notice many cancellations, in particular of the
non-potentially universal terms of eqs. (\ref{r34bds}) and
(\ref{r1314bdsfinal}) respectively due to $R^F_1$ and $R^F_2$, plus many
others as in the $A$ sector. Unlike from the pure gauge field case there is
a residual term left, from $R^F_1$, which is not clearly transverse. As we will
analyse it, it will be clear that it is not transverse and in fact will
restore the transversality of the pure gauge result. The other
terms left are the total derivative, the ones proportional to $\dDelta^{AA}_0$ and all
the ones which are not cancelled between the equations above but are already
potentially universal. Let us consider the latter group of terms first. 
After specifying the components and grouping them together it is possible
integrating by parts and with some algebra to recognise that in $D=4$, we are left
with a total derivative, which becomes a surface term to be evaluated between the boundaries:
\bea
-\left.\dirac{\Omega}_4\BBox_{\mu\nu}(p)\ x^2f^2(x)[xc'(x)-4c(x)]\right|^{\infty}_0
\eea
where $x=k^2/\Lam^2$, and the two functions $f(x)$ and $g(x)$ were defined below  \eq{genmom2}.
Considering again the behaviour of $c$ and $\tilde{c}$ at the boundaries,
it is possible to recognise that this surface term does not contribute.

It is now time to consider the non-transverse term mentioned earlier:
\be
\left.4\dDelta^{FF}_{SN}(k)k'_N(k-p)'_{\nu}\sh^{BF}_{\,\mu
S}(k)\right|_{p^2}
\ee 
Specifying the components, taking the order $p^2$ and with some algebra it
is possible to show it gets the following form:
\be
\frac{16 N}{D(D+2)}\Lam^{D-4}\dirac{\Omega}_D(p^2\delta\mn+2p_{\mu}p_{\nu})\int^{\infty}_0dx\
 g'(x)\left(x^{D/2+1}f'(x)\right)'
\ee
where $f$ and $g$ are the two functions defined above. Integrating by parts
and making use of the properties of $c$ and $\tilde{c}$ at the boundaries, we get to the following final result for this term (in $D=4$):
\be\label{nt-term}
\frac{N}{6}\dirac{\Omega}_4(p^2\delta\mn+2p_{\mu}p_{\nu})
\ee
As one can notice, this term is not transverse, as it was announced, but we
will see that it is the right combination which added to the contribution
(\ref{tdcontr-a}), makes the total result transverse. 

We have now to consider the total derivative term in this section, term (1) in
\eq{final}. Expanding in components to have a clearer view, its expression
in this sector is: 
\bea
2\Delta^{BB}(k)S^{AABB}_{\,\mu\,\nu\,\al\,\al}(p,-p,k,-k)+2\Delta^{DD}(k)S^{AADD}_{\,\mu\,\nu}(p,-p,k,-k)&&\nonumber\\
+2\Delta^{BB}(k)\Delta^{DD}(p-k)S^{ABD\sigma}_{\,\nu\,\rho}(-p,k,p-k)S^{BAD\sigma}_{\,\rho\,\mu}(-k,p,k-p)&&\nonumber\\
-\Delta^{BB}(k)\Delta^{BB}(p-k)S^{ABB}_{\,\nu\,\rho\,\sigma}(-p,k,p-k)S^{ABB}_{\,\mu\,\sigma\,\rho}(p,k-p,-k)&&\nonumber\\
-\Delta^{DD}(k)\Delta^{DD}(p-k)S^{ADD}_{\,\nu}(-p,k,p-k)S^{ADD}_{\,\mu}(p,k-p,-k)&&
\eea
In this case, as in the case of the $C$ sector, we can notice that due to
the regular behaviour of $\Delta^{BB}(k)$ and $\Delta^{DD}(k)$
and their derivatives as $k\to 0$ (see eqs.(\ref{deltabb}) and
(\ref{deltadd})), there are no IR poles and therefore no contribution comes
from this term either.

Before we add up all the contribution collected so far, we have finally to
recognise the terms proportional to $\dDelta^{AA}_0$ from this sector. The first one
is the same in common with all the three sectors, and is term (9) of
\eq{final}. In this sector it gets the following form: 
\be
-4\dDelta^{AA}_0\BBox\na(p)\left[\Delta^{FF}_{SR}(k)\de_{\mu}^k\de_{\al}^kS^{FF}_{RS}(k)-\Delta^{FF}_{RU}(k)\Delta^{FF}_{ST}(k)\de^k_{\mu}S^{FF}_{RS}(k)\de^k_{\al}S^{FF}_{TU}(k)\right]
\ee
There are in this case two more contributions to this set of terms. The
first one comes from $R^F_3$ (\eq{rf3}) and the second from the third term
of \eq{lastterm}. Their expression is:
\be
-8\dDelta^{AA}_0\BBox_{\mu\al}(p)k'_{\al}k'_{\nu}
\ee
These terms will be treated with all the ones proportional to $c'_0$, from
the other sectors in the next section. In particular it will be proved
that there is no contribution from them. As we would expect in fact, a contribution from those terms would make the result dependent upon the choice of the cutoff.

We have now all the contribution to the $\beta$-function at one loop, from
\eq{final} and we are ready to sum them up. First of all, adding
(\ref{tdcontr-a}), from the total derivative term of the pure gauge sector,
and (\ref{nt-term}) from the remainder $R^F_1$ of the $BD\sigma$ sector
(which is the only contribution from this sector), we get:
\be
\frac{N}{6}\dirac{\Omega}_4(p^2\delta\mn+2p_{\mu}p_{\nu})+\dirac{\Omega}_4N\left(\frac{19}{6}p^2\delta\mn-\frac{11}{3}p_{\mu}p_{\nu}\right)=\frac{10}{3}N\dirac{\Omega}_4\BBox\mn(p)
\ee
This makes the RHS of \eq{final} transverse just as it is on the LHS. Adding finally
(\ref{remaindercontr-a}), from the remainders of the $A$ sector and
substituting everything in \eq{final}:
\be\label{betaatoneloop}
-4\beta_1\BBox\mn(p)=\frac{22}{3}N\dirac{\Omega}_4\BBox\mn(p),\ \ie\ \beta_1=-\frac{11}{3}\frac{N}{16\pi^2}
\ee
Which is the $\beta$-function at one loop for $SU(N)$ Yang-Mills, evaluated
with the use of a gauge invariant flow equation.

\section{$\dDelta^{AA}_0$-terms}\label{sub:c'0}

In all the three sectors it is possible to recognise terms which are
proportional to $\dDelta^{AA}_0$. Since a final answer dependent on the cutoff
function at zero momentum would make the result dependent on its choice, it
must be possible to collect all of them and recognise a surface term
in any dimension $D$, vanishing at the boundary. That is indeed the case, as we will prove in
this section.

Let us first consider the class of these terms, coming from the pure $A$
sector. They are:
\be\label{c-10a}
\ba{lr}
4\dDelta^{AA}_0
    \BBox_{\mu\al}(p)\left[\Delta^{AA}(k)\de^k_{\al}\de^k_{\nu}S^{AA}_{\,\beta\,\beta}(k)-\left(\Delta^{AA}(k)\right)^2\de^k_{\al}S^{AA}_{\,\lambda\,\rho}(k)\de^k_{\nu}S^{AA}_{\,\lambda\,\rho}(k)\right] &(1)\\[5pt]
8\dDelta^{AA}_0\BBox_{\mu\al}(p)\dfrac{k_{\al}k_{\nu}}{k^4}&(2)
\ea
\ee
Let us now write the corresponding ones from the $C$ sector:
\be\label{c-10c}
4\dDelta^{AA}_0\BBox_{\mu\al}(p)\left[\Delta^{CC}(k)\de^k_{\al}\de^k_{\nu}S^{CC}(k)-\left(\Delta^{CC}(k)\right)^2\de^k_{\al}S^{CC}(k)\de^k_{\nu}S^{CC}(k)\right]
\ee
Finally the ones from the $F$ sector are:
\be\label{c-10bds}
\ba{lr}
-4\dDelta^{AA}_0\BBox_{\mu\al}(p)\left[\Delta^{FF}_{RS}(k)\de^k_{\al}\de^k_{\nu}S^{FF}_{RS}(k)-\Delta^{FF}_{TU}(k)\Delta^{FF}_{RS}(k)\de^k_{\al}S^{FF}_{RU}(k)\de^k_{\nu}S^{FF}_{TS}(k)\right]&(1)\\[5pt]
-8\dDelta^{AA}_0\BBox_{\mu\al}(p)k'_{\al}k'_{\nu}&(2)
\ea
\ee
Let us consider the first term of each group. Modulo a factor
$4c'_0\BBox_{\mu\al}(p)$, we can indicate them with (term (9) of \eq{final}
after some relabelling):
\be\label{c-10all}
\Delta^{ff}_{RS}(k)\de^k_{\al}\de^k_{\nu}S^{ff}_{RS}(k)-\Delta^{ff}_{TU}(k)\Delta^{ff}_{RS}(k)\de^k_{\al}S^{ff}_{RU}(k)\de^k_{\nu}S^{ff}_{TS}(k)\label{prima}
\ee
where $f$ can represent either $A$, $C$ or $F$. If we now integrate the first term by parts, the term in brackets becomes;
\bea
&&\de^k_{\nu}\left[\Delta^{ff}_{RS}(k)\de^k_{\al}S^{ff}_{RS}(k)\right]-(\de^k_{\nu}\Delta^{ff}_{RS}(k)\de^k_{\al}S^{ff}_{RS}(k))\nonumber\\
&&-\Delta^{ff}_{TU}(k)\Delta^{ff}_{RS}(k)\de^k_{\al}S^{ff}_{RU}(k)\de^k_{\nu}S^{ff}_{TS}(k)\label{bypart}
\eea
Remembering the relation between the two point functions and their integrated zero point wines, eq.(\ref{inverse}), the last two terms of the three of eq.(\ref{bypart}) can be rearranged as follows:
\be
\Delta^{ff}_{TU}(k)\left[\de_{\nu}^kB^{ff}_{US}(-k)\right]\de^k_{\al}S^{ff}_{TS}(k)-\de^k_{\nu}\Delta^{ff}_{RS}(k)B^{ff}_{RT}(k)\de^k_{\al}S^{ff}_{TS}(k)
\ee
We now remember that:
\be
B^{ff}_{RT}(k)S^{ff}_{TS}(k)=0
\ee
We use this to transfer $\partial_\alpha$ in the second term
 to $B_{RT}$, and integrate by parts the $\partial_\nu$ onto $S_{TS}$,
 similarly to above. Redefining the indices, changing $k\to -k$ and using
 the $\al\leftrightarrow\nu$ symmetry (which automatically has, since it is
 just a function of $k$), we can rewrite eq.(\ref{prima}) as:
\be\label{finalc'0}
\de^k_{\nu}\left[\Delta^{ff}_{RS}(k)\de^k_{\al}S^{ff}_{RS}(k)\right]-(\de^k_{\nu}B^{ff}_{RT}(k))(\de^k_{\al}B^{ff}_{TR}(k))
\ee
Before continuing, we can simplify further the first term of the equation
above. We can in fact notice via a dimensional analysis that the expression
in square brackets can be written as:
\be
\frac{2k_{\al}}{\Lam^2}F^f(k^2/\Lam^2)
\ee
The full term with all the coefficients as it appears in the equation for
$\beta_1$ has to be integrated in $d^D k$ and it has then the following
form:
\be\label{ffff}
\frac{4c'_0}{\Lam^4}N\BBox_{\mu\al}(p)\int \frac{d^D
k}{(2\pi)^D}\de^k_{\nu}[2k_{\al}F^f(k^2/\Lam^2)],
\ee
(recalling that $\dDelta^{AA}_0=c'_0/\Lam^2$).
Performing the derivative, taking the average on the $k$ components and
defining $x=k^2/\Lam^2$ we finally find the expression:
\be\label{fff}
\frac{16c'_0}{D}N\BBox_{\mn}(p)\dirac{\Omega}_D\Lam^{D-4}\int^{\infty}_{0} dx\left[x^{D/2}F^f(x)\right]'
\ee 
which allows us to extract from the first term of eq.(\ref{finalc'0}) the surface term.
We can now specify for each different sector these expressions in order to
evaluate the contribution of this set of terms to the one loop
$\beta$-function. 

\subsection{$A$ sector}

In this sector, we recall from eq.(\ref{inversea}) that:
\be
B^{AA}_{\mu\nu}(p)=\frac{p_{\mu}p_{\nu}}{p^2}
\ee
and eq.(\ref{finalc'0}) becomes:
\be
\de^k_{\nu}\left[\Delta^{AA}(k)\de^k_{\al}S^{AA}_{\,\beta\,\beta}(k)\right]-\left(\de^k_{\nu}\frac{k_{\rho}k_{\lambda}}{k^2}\right)\left(\de^k_{\al}\frac{k_{\lambda}k_{\rho}}{k^2}\right)
\ee
The second term evaluates to:
\be
-\frac{2}{k^2}\left(\delta_{\na}-\frac{k_{\nu}k_{\al}}{k^2}\right)
\ee
Remembering now the factor $\frac{4c'_0}{\Lam^2}\BBox_{\mu\al}(p)$ in front
of it, we can collect it together with (2) in eq.(\ref{c-10a}), to get:
\be\label{finefine}
-8\frac{c'_0}{\Lam^2}\BBox_{\mu\alpha}(p)\de_{\alpha}\frac{k_{\nu}}{k^2}
\ee
This is a total derivative term and it will be considered later.
The last term proportional to $c'_0$ in the $A$ sector, yet to be
considered, is another total derivative which we are going to
evaluate now. First of all we recall that:
\bea
\Delta^{AA}(k)=\frac{1}{2\Lam^2}\frac{c(x)}{x}\\
S^{AA}_{\beta\beta}(k)=2(D-1)\Lam^2\frac{x}{c(x)}
\eea
where $x=k^2/\Lam^2$. In this case then we have:
\be
F^A(x)=(D-1)\left(\ln{\frac{x}{c(x)}}\right)'
\ee
In order to find the contribution of this term we must evaluate at the
boundary the quantity:
\be\label{fa}
\left.x^{D/2}\left(\ln{\frac{x}{c(x)}}\right)'\right|_0^{\infty}
\ee
with $c(x)\to 1\ \mbox{as}\ x\to 0$ and $c(x)\to 0\ \mbox{as}\
x\to\infty$. This term is regular in the infrared and it does not
contribute at $x=0$, but it has a divergent
ultraviolet behaviour. As we shall see this infinity will be fixed when we take into account the corresponding term coming from the $BD\sigma$
sector.

\subsection{C sector}

In this sector as we can see from eq.(\ref{inversec}), $B^{CC}(k)=0$, and
the only possible contribution from this sector to the $c'_0$ terms comes
from the total derivative. In this case this term takes the form:
\be
\de^k_{\nu}\left[\Delta^{CC}(k)\de^k_{\al}(\Delta^{CC}(k))^{-1}\right]
\ee
where (here $x=k^2/\Lam^2$): 
\be\label{deltacc}
\Delta^{CC}(k)=\frac{1}{2\Lam^4}\ \frac{\tilde{c}(x)}{x+2\lam \tilde{c}(x)}
\ee
referring to the notation of eq.(\ref{fff}), we can recognise:
\be
F^C(x)=\left[\ln{\left(\frac{x}{\tilde{c}(x)}+2\lam\right) }\right]'
\ee
The term we have to evaluate at the boundaries is now:
\be\label{fc}
\left.x^{D/2}\left[\ln{\left(\frac{x}{\tilde{c}(x)}+2\lam\right) }\right]'\right|_0^{\infty}
\ee
Also in this case, due to the properties of $\tilde{c}(x)$ at $0$ and
$\infty$ (similar to those of $c(x)$), there is no contribution at $x=0$ and a
divergence at $x\to\infty$. Nevertheless, the corresponding term of the
$BD\sigma$ sector will cure also this problem.

\subsection{$BD\sigma$ sector}

In the present sector, from eqs.(\ref{app:ush}) and
(\ref{app:Art}), we can see there are two types of remainder, depending
on the order the two point vertex and the integrated zero point wine are placed:
\bea
A^{FF}_{RS}(k)&=&(-k)'_R (-k)_S\nonumber\\
B^{FF}_{RS}(k)&=&k'_R k_S
\eea
Nevertheless, we can always  express everything in term of one of the two,
since relation (\ref{a-brel}) holds between them. The second term of eq.(\ref{finalc'0})
can then be written in terms of $B^{FF}_{RS}(k)$ and considering the
explicit expression of the generalised momenta of eqs.(\ref{genmom1}) and
(\ref{genmom2}), we get
\be\label{pino}
-(\de^k_{\nu}B^{FF}_{RT}(k))(\de^k_{\al}B^{FF}_{TR}(k))=-2k'_{\al}k'_{\nu}-2\de^k_{\al}k'_{\nu}
\ee
Adding the right factors in front, the first term on the RHS, cancels
exactly (2) in eq.(\ref{c-10bds}). The second one on the RHS, once it is
considered with all its factors and in the $d^Dk$ integral, is of the form
of the one in eq.(\ref{ffff}). It is therefore a total derivative and it
has to be considered together with the first term of
eq.(\ref{finalc'0}). Let us first consider the latter. Once we split in
components we can apply the same arguments of the two previous sections and
it gets the following form:
\be
\de^k_{\nu}\left[\Delta^{BB}(k)\de^k_{\al}S^{BB}_{\beta\beta}(k)+\Delta^{DD}(k)\de^k_{\al}S^{DD}(k)\right]
\ee
Let us now consider (see Appendix \ref{App:intwine}):
\bea
\Delta^{BB}(k)&=&\frac{1}{2\Lam^2}\ \frac{\tilde{c}(x)c(x)}{x\tilde{c}(x)+2
c}\label{deltabb}\\
\Delta^{DD}(k)&=&\frac{1}{\Lam^4}\ \frac{\tilde{c}^2(x)}{x\tilde{c}(x)+2
c}\label{deltadd},
\eea
($x=k^2/\Lam^2$ here), and
\bea
S^{BB}_{\,\beta\,\beta}(k)&=&2\Lam^2\left((D-1)\frac{x}{c(x)}+\frac{2D}{\tilde{c}(x)}\right)\\
S^{DD}(k)&=&\Lam^4\frac{x}{\tilde{c}(x)}
\eea
In this last case it is possible to see that we can write:
\bea
F^B(x)=(D-1)\left[\ln\left(\frac{x}{c(x)}+\frac{2}{\tilde{c}(x)}\right)\right]'-\frac{2c(x)\tilde{c}(x)}{x\tilde{c}(x)+2c}\left(\frac{1}{\tilde{c}(x)}\right)'\label{fb}\\[5pt]
F^D(x)=\left[\ln\left(\frac{x}{\tilde{c}(x)}+\frac{2c(x)}{\tilde{c}^2(x)}\right)\right]'-\frac{2\tilde{c}^2(x)}{x\tilde{c}(x)+2
c}\left(\frac{c(x)}{\tilde{c}^2(x)}\right)'\label{fd}
\eea
What we have to calculate is now:
\be
\left.x^{D/2}F^F(x)\right|_0^{\infty}
\ee
It is possible to notice that the second terms of both eqs.(\ref{fb})
and (\ref{fd}) do not give contribution either at $x=0$ or when
$x\to\infty$. On the contrary, even if the first terms of the two equations
do not give a contribution for $x=0$ they do give one for $x\to\infty$, but
it cancels the divergent contribution of respectively the two terms of
eqs.(\ref{fa}) and (\ref{fb}).
In order to finish the check of the contributions to the 1-loop $\beta$
function of the terms proportional to $c'_0$, we just have now to evaluate
the second term in the RHS of eq.(\ref{pino}), with the right factor in
front, and the term in eq. (\ref{finefine}). They can be collected together into the form:
\bea
-\frac{8c'_0}{\Lam^2}N\BBox_{\mu\al}(p)\int \frac{d^Dk}{(2\pi)^D}\
\de^k_{\al}\left(\frac{k_{\nu}}{k^2}-k'_{\nu}\right)&\!\!\!\!\!=&\!\!\!\!\!-\dfrac{8c'_0}{\Lam^2}N\BBox_{\mu\al}(p)\dints \dfrac{d^Dk}{(2\pi)^D}\ \de^k_{\al}\left(\dfrac{2c}{x (x\tilde{c}+2
c)}\ k_{\nu}\right)\nonumber\\[3pt]
&\!\!\!\!\!=&\!\!\!\!\!-\dfrac{16 c'_0}{D}N\dirac{\Omega}_D\Lam^{D-4}\BBox_{\mu\nu}\left[x^{D/2}\dfrac{2c}{x(x\tilde{c}+2
c)}\right]_0^{\infty}\nonumber\\[3pt]
&\!\!\!\!\!=&\!\!\!\!\!0 
\eea
Since neither of these last two terms give any contribution, we can conclude
that the final result will be independent from $c'_0$. This result does not
surprise us and we were expecting it from the beginning, since the $\beta$
function at one loop is a universal quantity, independent from the choice
of the cutoff function.

\chapter{Conclusions}\label{conclusions}

To summarise the work that has been developed in this thesis, we can say
that we started by revisiting an exact RG flow equation for a massless scalar field
theory as it was described by Polchinski in \cite{pol}. Making use of the
ideas about scheme independence introduced in \cite{latmor}, we considered a
generalisation of it, and computed the
effective mass and wavefunction renormalisation at one
loop. Combining these results with the flow equation for the one-loop four-point
vertex at zero momenta, we calculated the first coefficient of the
$\beta$-function, obtaining the expected universal answer expressed in
\eq{betaf}.

This result, achieved with a Polchinski-type flow equation with a general
kernel is a proof of universality for $\beta$ at one loop for the massless
scalar field, beyond the independence of the choice of the cutoff
function. The totally generic form of the kernel, in fact, includes not only
a general form of the cutoff, but also the introduction in the game of an
auxiliary action, the {\it seed action}, which can contain all
sorts of extra vertices compatible with the symmetry. (In the case of
Polchinski it is recognised to coincide with the kinetic term only). The
presence of the seed action can complicate the calculation as long as the
scalar field is concerned, but the freedom on its choice (with just some
mild constraint on its vertices) turns out to be useful for the gauge case.

The independence of universal quantities, such as $\beta$ at one loop, upon
the unphysical vertices of the seed action, introduced by hand, come out as
an expected result from the calculation performed for the scalar field
case, but the way these extra parameters were eliminated in favour of the
physical ones, was crucial to set up a powerful method to deal with these
generalised flow equations. It was through the calculation in the scalar
case that it becomes clear the need to use the flow equations, in order
to eliminate the seed action vertices in favour of the effective ones. 

Since the main aim of this work was to set up a flow equation for a
$SU(N)$ Yang-Mills gauge theory which preserved the symmetry, in the second
part of the thesis, we started by revisiting a scheme in which this theory was regularised in a gauge invariant
way considering it as a sector of a bigger graded group, known as
$SU(N|N)$, which is broken spontaneously (see
\cite{su:pap}). This was the first step towards our goal, since with a gauge
invariant regularised action, we could then build a flow equation with
the right features in order to preserve this symmetry through the
flow. With the wide choice given by scheme independence, it was possible to
write a flow equation for $SU(N|N)$, see \eq{sunnfl}, which was supergauge invariant, via
the introduction of the covariantisation of the kernels, described in section
\ref{sub:1.1.1}. The extra choice on the seed action vertices comes out to
be another crucial point, since it was through the properties we could set
on them, that the calculation was simplified. Moreover, the finiteness at
tree level could be also ensured, without the introduction of an extra scale,
and without loosing the selfsimilar flow property, not only elegant but crucial for our calculation.

In the last part of the thesis, we could eventually perform a check on the
flow equation just built, evaluating the first coefficient of the
$\beta$-function for the physical $SU(N)$ Yang-Mills sector of
$SU(N|N)$.  

The method used for the scalar field case was implemented and
adapted to the present case, with the extra aid of the graphical
interpretation. The power of the diagrams, representing the vertices can be
appreciated through the whole calculation, as they could often be used
instead of specifying their analytical expressions. The symmetries present
in the effective action allowed also to use, for most part of the
calculation, a compact notation, in which all the fields present in the
theory (the basis used in the broken phase), were collected in a
supermultiplet $f$ and all the equations could be written in terms of it. This greatly simplified the task and
avoided repetitions. 

The expected universal answer for the first
coefficient of the $\beta$-function was
eventually calculated for the $SU(N)$ Yang-Mills sector of $SU(N|N)$ as it
is expressed in \eq{betaatoneloop}. This resulted not
only in a check of universality for this quantity beyond the change of the
cutoff, as for the scalar case, but also represented the first time the
finite $N$ value was extracted in a gauge invariant way. 

We are now trying to adapt this method,
to perform other gauge invariant calculations making use of this flow
equation \eg second coefficient of the $\beta$-function. Even though we
made many progresses and all this machinery seems to be not just
exploitable in the relatively simple calculation of $\beta_1$, we will not
develop this further analysis in this thesis and leave this discussion to
future references (see \cite{sequel}).

\appendix
\chapter{Basis for $SU(N|N)$ algebra, and (anti)commutation
relations}\label{app:bla}

In order to write more explicitly the commutation and anticommutation
relations for the $SU(N|N)$ algebra, we have to choose a basis. We want to
choose the one that allows us to write all the relations of the algebra in
terms of the structure constants of $SU(N)$. A possible choice is then the
following:
\bea\label{app:basis}
&\tau^{(1)}_{\alpha}=\left(\begin{array}{cc} 
                           \tau_a&0\\   
                            0&\tau_a
                           \ea\right)&\tau^{(2)}_{\alpha}=\left(\ba{cc} 
                                                                \tau_a&0\\   
                                                                0&-\tau_a
                                                                \ea\right)\nonumber\\
&s^{(1)}_{\alpha}=\left(\ba{cc} 
                           0&\tau_a\\   
                            \tau_a&0
                           \ea\right)&s^{(2)}_{\alpha}=\left(\ba{cc} 
                                                             0&-i\tau_a\\   
                                                             i\tau_a&0
                                                             \ea\right)\nonumber\\
&s^{(1)}_0=\left(\ba{cc} 
                           0&\one_N\\   
                            \one_N&0
                           \ea\right)&s^{(2)}_0=\left(\ba{cc} 
                                                      0&-i\one_N\\   
                                                      i\one_N&0
                                                      \ea\right)
\eea
The index $\alpha$ runs here from $1$ to $N^2-1$, and they are $4N^2-2$ ($2N\times 2N$) traceless and supertraceless matrices
$T_A$, which, together with $\one_{2N}$
form a possible basis $\{S_{\alpha}\}$ of generators for $SU(N|N)$.
The two in the first line are the direct product of the $\tau_a$'s, generators of $SU(N)$,
with $\one_2$ and the Pauli matrix $\sigma_3$ respectively, and the last two
lines are the $\tau_a$'s and $\one_N$ in direct product with the Pauli matrices $\sigma_1$
and $\sigma_2$ respectively. Relating this basis with the matrices introduced in
section \ref{sub:1.2.1}, we have:
$B_a=\{\tau^{(1)}_{\alpha},\tau^{(2)}_{\alpha}\}$ and $F_a=\{s^{(1)}_{\alpha},s^{(2)}_{\alpha},s^{(1)}_0,s^{(2)}_0\}$.

In terms of the basis defined in eq.(\ref{app:basis}), we can rewrite the first relation of
eq.(\ref{commrules}) as:
\bea
\ [\tau^{(i)}_{\alpha},\tau^{(i)}_{\beta}]&=&i\ f_{\alpha\beta}^{\phantom{\alpha\beta}\gamma}\ \tau^{(1)}_{\gamma}\\
\ [\tau^{(1)}_{\alpha},\tau^{(2)}_{\beta}]&=&i\ f_{\alpha\beta}^{\phantom{\alpha\beta}\gamma}\ \tau^{(2)}_{\gamma}
\eea
The second can be rewritten:
\bea
\ [\tau^{(1)}_{\alpha},s^{(i)}_{\beta}]&=&i\ f_{\alpha\beta}^{\phantom{\alpha\beta}\gamma}\
s^{(i)}_{\gamma}\\
\ [\tau^{(2)}_{\alpha},s^{(1,2)}_{\beta}]&=&\pm i\
d_{\alpha\beta}^{\phantom{\alpha\beta}\gamma}\ s^{(2,1)}_{\gamma}\pm i\ \delta_{\alpha\beta}\ s^{(2,1)}_0\\
\ [\tau^{(i)}_{\alpha},s^{(j)}_0]&=&0
\eea
Finally the third:
\bea
\{s^{(1)}_{\alpha},s^{(2)}_{\beta}\}&=&f_{\alpha\beta}^{\phantom{\alpha\beta}\gamma}\
\tau^{(2)}_{\gamma}\\
\{s^{(i)}_{\alpha},s^{(i)}_{\beta}\}&=&d_{\alpha\beta}^{\phantom{\alpha\beta}\gamma}\
\tau^{(1)}_{\gamma}+ \delta_{\alpha\beta}\ \one_{2N}\\
\{s^{(i)}_{\alpha},s^{(j)}_0\}&=&2\ \delta_{ij}\ \tau^{(1)}_{\alpha}\\
\{s^{(i)}_0,s^{(j)}_0\}&=&2\delta_{ij}\one_{2N}
\eea
In all the equations above, $f$'s and $d$'s are indeed the antisymmetric
and symmetric structure constants of $SU(N)$.

\chapter{Completeness relation for $SU(N|N)$}\label{app:compl}

In order to derive the completeness relation for the $T_A$ generators of
$SU(N|N)$, let us first write a generic non constrained supermatrix,
expanded on the generators of $U(N|N)$, in this form:
\be\label{app:exp}
X=X_A T^A +X_0 \one+X_3\sigma_3
\ee
Recalling now that:
\be
X_{A}=2\ \mbox{str} T_{A}X
\ee
and:
\bea
X_0=\frac{1}{2N}\mbox{str}\sigma_3 X\\
X_3=\frac{1}{2N}\mbox{str}\one X
\eea
Then we can rewrite \eq{app:exp} as:
\be
X=2\ T^A \mbox{str} T_{A} X +\frac{1}{2N}\left(\one\ 
\mbox{str}\sigma_3 X+\sigma_3\ \mbox{str}\one X\right)
\ee
Writing both sides in components:
\be
\delta^l_{\ k}\delta^i_{\ m} X^k_{\ i}=2(T_A)^l_{\ m} (\sigma_3)^i_{\ j} (T^A)^j_{\ k}
X^k_{\ i}+\frac{1}{2N}\delta^l_{\ m}(\sigma_3)^i_{\ j} (\sigma_3)^j_{\ k}
X^k_{\ i}+\frac{1}{2N}(\sigma_3)^l_{\ m} (\sigma_3)^i_{\ j} \delta^j_{\ k} X^k_{\ i}
\ee
Considering the previous equation must be valid for any $ X^k_{\ i}$, and
rearranging the $SU(N|N)$ generators (without $\one$), on the LHS, we get:
\be
(T_A)^l_{\ m} (\sigma_3)^i_{\ j} (T^A)^j_{\ k}=\frac{1}{2}\delta^l_{\
k}\delta^i_{\ m} - \frac{1}{4N}\left[\delta^l_{\ m}(\sigma_3)^i_{\ j} (\sigma_3)^j_{\ k}+(\sigma_3)^l_{\ m} (\sigma_3)^i_{\ j} \delta^j_{\ k}\right]
\ee
If now we multiply both sides by $(\sigma_3)^n_{\ i}$, considering that:
\be
(\sigma_3)^n_{\ i}(\sigma_3)^i_{\ j}=\delta^n_{\ j}
\ee
and the fact that the elements of the generators are all ordinary commuting
(bosonic) numbers we get the completeness relation of eq.(\ref{complrel}).

\chapter{Four point equations at tree level}\label{app:4eq}

The four point tree level vertices whose equations were not explicitly
written in section \ref{subsec:2.1.5}, can be found here:
\bea
S^{AABB}_{\,\mu\,\nu\,\rho\,\sigma}(p,q,r,s)=-\int_{\Lam}^{\infty}\dfrac{d\Lam_1}{\Lam_1}\left\{\sh^{AABB}_{\,\mu\,\nu\,\rho\,\alpha}(p,q,r,s)\dDelta^{BB}_s\sh^{BB}_{\,\alpha\,\sigma}(s)+\sh^{AABB}_{\,\mu\,\nu\,\alpha\,\sigma}(p,q,r,s)\dDelta^{BB}_r\sh^{BB}_{\,\alpha\,\rho}(r)\right.\nonumber\\[5pt]
+\sh^{AABB}_{\,\mu\,\alpha\,\rho\,\sigma}(p,q,r,s)\dDelta^{AA}_q\sh^{AA}_{\,\alpha\,\nu}(q)+\sh^{AABB}_{\,\alpha\,\nu\,\rho\,\sigma}(p,q,r,s)\dDelta^{AA}_p\sh^{AA}_{\,\alpha\,\mu}(p)\nonumber\\[5pt]
-S^{AAA}_{\,\mu\,\nu\,\alpha}(p,q,r+s)\dDelta^{AA}_{p+q}S^{BBA}_{\,\rho\,\sigma\,\alpha}(r,s,p+q)+\sh^{AAA}_{\,\mu\,\nu\,\alpha}(p,q,r+s)\dDelta^{AA}_{p+q}S^{BBA}_{\,\rho\,\sigma\,\alpha}(r,s,p+q)\nonumber\\[5pt]
+S^{AAA}_{\,\mu\,\nu\,\alpha}(p,q,r+s)\dDelta^{AA}_{p+q}\sh^{BBA}_{\,\rho\,\sigma\,\alpha}(r,s,p+q)-S^{BBA}_{\,\alpha\,\sigma\,\mu}(q+r,s,p)\dDelta^{BB}_{p+s}S^{BBA}_{\,\rho\,\alpha\,\nu}(r,p+s,q)\nonumber\\[5pt]
+\sh^{BBA}_{\,\alpha\,\sigma\,\mu}(q+r,s,p)\dDelta^{BB}_{p+s}S^{BBA}_{\,\rho\,\alpha\,\nu}(r,p+s,q)+S^{BBA}_{\,\alpha\,\sigma\,\mu}(q+r,s,p)\dDelta^{BB}_{p+s}\sh^{BBA}_{\,\rho\,\alpha\,\nu}(r,p+s,q)\nonumber\\[5pt]
-\sh^{AABD\sigma}_{\,\mu\,\nu\,\rho}(p,q,r,s)\dDelta^{DD}_s\sh^{BD\sigma}_{\,\sigma}(s)-\sh^{BAAD\sigma}_{\,\sigma\,\mu\,\nu}(s,p,q,r)\dDelta^{DD}_r\sh^{BD\sigma}_{\,\rho}(r)\nonumber\\[5pt]
+S^{BAD\sigma}_{\,\sigma\,\mu}(s,p,q+r)\dDelta^{DD}_{p+s}S^{ABD\sigma}_{\,\nu\,\rho}(q,r,p+s)-\sh^{BAD\sigma}_{\,\sigma\,\mu}(s,p,q+r)\dDelta^{DD}_{p+s}S^{ABD\sigma}_{\,\nu\,\rho}(q,r,p+s)\nonumber\\[5pt]
-S^{BAD\sigma}_{\,\sigma\,\mu}(s,p,q+r)\dDelta^{DD}_{p+s}\sh^{ABD\sigma}_{\,\nu\,\rho}(q,r,p+s)+\sh^{BBA}_{\,\alpha\,\sigma\,\mu}(r+q,s,p)\dDelta^{B,AB}_{\,\rho}(r;q,p+s)\sh^{AA}_{\,\alpha\,\nu}(q)\nonumber\\[5pt]
+\sh^{BBA}_{\,\rho\,\alpha\,\nu}(r,s+p,q)\dDelta^{B,BA}_{\,\sigma}(s;q+r,p)\sh^{AA}_{\,\alpha\,\mu}(p)+\sh^{AAA}_{\,\mu\,\nu\,\alpha}(p,q,r+s)\dDelta^{B,BA}_{\,\sigma}(s;r,p+q)\sh^{BB}_{\,\alpha\,\rho}(r)\nonumber\\[5pt]
+\sh^{AAA}_{\,\mu\,\nu\,\alpha}(p,q,r+s)\dDelta^{B,AB}_{\,\rho}(r;p+q,s)\sh^{BB}_{\,\alpha\,\sigma}(s)+\sh^{BBA}_{\,\rho\,\alpha\,\nu}(r,p+s,q)\dDelta^{A,BB}_{\,\mu}(p;s,q+r)\sh^{BB}_{\,\alpha\,\sigma}(s)\nonumber\\[5pt]
+\sh^{BBA}_{\,\rho\,\sigma\,\alpha}(r,s,p+q)\dDelta^{A,AA}_{\,\mu}(p;r+s,q)\sh^{AA}_{\,\alpha\,\nu}(q)+\sh^{BBA}_{\,\rho\,\sigma\,\alpha}(r,s,p+q)\dDelta^{A,AA}_{\,\nu}(q;p,r+s)\sh^{AA}_{\,\alpha\,\mu}(p)\nonumber\\[5pt]
+\sh^{BBA}_{\,\alpha\,\sigma\,\mu}(q+r,s,p)\dDelta^{A,BB}_{\,\nu}(q;s+p,r)\sh^{BB}_{\,\alpha\,\rho}(r)+\sh^{AA}_{\,\mu\,\alpha}(p)\dDelta^{AB,AB}_{\,\nu\,\rho}(q,r;p,s)\sh^{BB}_{\,\alpha\,\sigma}(s)\nonumber\\[5pt]
+\sh^{AA}_{\,\nu\,\alpha}(q)\dDelta^{BB,AA}_{\,\rho\,\sigma}(r,s;q,p)\sh^{AA}_{\,\alpha\,\mu}(p)+\sh^{BB}_{\,\rho\,\alpha}(r)\dDelta^{BA,BA}_{\,\sigma\,\mu}(s,p;r,q)\sh^{AA}_{\,\alpha\,\nu}(q)\nonumber\\[5pt]
+\sh^{BB}_{\,\rho\,\alpha}(r)\dDelta^{AA,BB}_{\,\mu\,\nu}(p,q;s,r)\sh^{BB}_{\,\alpha\,\sigma}(s)+\sh^{BB}_{\,\sigma\,\alpha}(s)\dDelta^{A,B,BA}_{\,\mu\,\,\rho}(p;r;s,q)\sh^{AA}_{\,\alpha\,\nu}(q)\nonumber\\[5pt]
+\sh^{AA}_{\,\mu\,\alpha}(p)\dDelta^{A,B,AB}_{\,\nu,\,\,\sigma}(q;s;p,r)\sh^{BB}_{\,\alpha\,\rho}(r)-\sh^{ABD\sigma}_{\,\nu\,\rho}(q,r,p+s)\dDelta^{A,DD}_{\,\mu}(p;s,q+r)\sh^{BD\sigma}_{\,\sigma}(s)\nonumber\\[5pt]
\left.-\sh^{BAD\sigma}_{\,\sigma\,\mu}(s,p,q+r)\dDelta^{A,DD}_{\,\nu}(q;p+s,r)\sh^{BD\sigma}_{\,\rho}(r)-\sh^{BD\sigma}_{\,\rho}(r)\dDelta^{AA,DD}_{\,\mu\,\nu}(p,q;s,r)\sh^{BD\sigma}_{\,\sigma}(s)\right\}+I.C.\nonumber\\\label{app:aabb}
\eea
\bea
S^{AACC}_{\,\mu\,\nu}(p,q,r,s)=-\int_{\Lam}^{\infty}\dfrac{d\Lam_1}{\Lam_1}\left\{\sh^{AACC}_{\,\alpha\,\nu}(p,q,r,s)\dDelta^{AA}_p\sh^{AA}_{\,\alpha\,\mu}(p)+\sh^{AACC}_{\,\mu\,\alpha}(p,q,r,s)\dDelta^{AA}_q\sh^{AA}_{\,\alpha\,\nu}(q)\right.\nonumber\\[5pt]
-S^{AAA}_{\,\mu\,\nu\,\alpha}(p,q,r+s)\dDelta^{AA}_{p+q}S^{ACC}_{\,\alpha}(p+q,r,s)+\sh^{AAA}_{\,\mu\,\nu\,\alpha}(p,q,r+s)\dDelta^{AA}_{p+q}S^{ACC}_{\,\alpha}(p+q,r,s)\nonumber\\[5pt]
+S^{AAA}_{\,\mu\,\nu\,\alpha}(p,q,r+s)\dDelta^{AA}_{p+q}\sh^{ACC}_{\,\alpha}(p+q,r,s)+\sh^{AA}_{\,\nu\,\alpha}(q)\dDelta^{A,AA}_{\,\mu}(p;r+s,q)\sh^{ACC}_{\,\alpha}(p+q,r,s)\nonumber\\[5pt]
+\sh^{AA}_{\,\mu\,\alpha}(p)\dDelta^{A,AA}_{\,\nu}(q;p,r+s)\sh^{ACC}_{\,\alpha}(p+q,r,s)+\sh^{AACC}_{\,\mu\,\nu}(p,q,r,s)\dDelta^{CC}_s\sh^{CC}(s)\nonumber\\[5pt]
+\sh^{AACC}_{\,\mu\,\nu}(p,q,r,s)\dDelta^{CC}_r\sh^{CC}(r)-S^{ACC}_{\,\mu}(p,q+r,s)\dDelta^{CC}_{p+s}S^{ACC}_{\,\nu}(q,r,p+s)\nonumber\\[5pt]
+\sh^{ACC}_{\,\mu}(p,q+r,s)\dDelta^{CC}_{p+s}S^{ACC}_{\,\nu}(q,r,p+s)+S^{ACC}_{\,\mu}(p,q+r,s)\dDelta^{CC}_{p+s}\sh^{ACC}_{\,\nu}(q,r,p+s)\nonumber\\[5pt]
+\sh^{ACC}_{\,\nu}(q,r,p+s)\dDelta^{A,CC}_{\,\mu}(p;s,q+r)\sh^{CC}(s)+\sh^{ACC}_{\,\mu}(p,q+r,s)\dDelta^{A,CC}_{\,\nu}(q;p+s,r)\sh^{CC}(r)\nonumber\\[5pt]
\left.+\sh^{AA}_{\,\mu\,\alpha}(p)\dDelta^{CC,AA}(r,s;q,p)\sh^{AA}_{\,\nu\,\alpha}(q)+\sh^{CC}(s)\dDelta^{AA,CC}_{\,\mu\,\nu}(p,q;s,r)\sh^{CC}(r)\right\}+I.C.\nonumber\\\label{app:aacc}
\eea
\bea
S^{AADD}_{\,\mu\,\nu}(p,q,r,s)=-\int_{\Lam}^{\infty}\dfrac{d\Lam_1}{\Lam_1}\left\{\sh^{AADD}_{\,\mu\,\alpha}(p,q,r,s)\dDelta^{AA}_q\sh^{AA}_{\,\al\,\nu}(q)+\sh^{AADD}_{\,\al\,\nu}(p,q,r,s)\dDelta^{AA}_p\sh^{AA}_{\,\al\,\mu}(p)\right.\nonumber\\[5pt]
+\sh^{BAAD\si}_{\,\al\,\mu\,\nu}(s,p,q,r)\dDelta^{BB}_s\sh^{BD\si}_{\,\al}(s)+\sh^{AABD\si}_{\,\mu\,\nu\,\al}(p,q,r,s)\dDelta^{BB}_r\sh^{BD\si}_{\,\al}(r)\nonumber\\[5pt]
-S^{AAA}_{\,\mu\,\nu\,\al}(p,q,r+s)\dDelta^{AA}_{p+q}S^{ADD}_{\,\al}(p+q,r,s)+\sh^{AAA}_{\,\mu\,\nu\,\al}(p,q,r+s)\dDelta^{AA}_{p+q}S^{ADD}_{\,\al}(p+q,r,s)\nonumber\\[5pt]
+S^{AAA}_{\,\mu\,\nu\,\al}(p,q,r+s)\dDelta^{AA}_{p+q}\sh^{ADD}_{\,\al}(p+q,r,s)+S^{ABD\si}_{\,\mu\,\al}(p,q+r,s)\dDelta^{BB}_{q+r}S^{BAD\si}_{\,\al\,\nu}(p+s,q,r)\nonumber\\[5pt]
-\sh^{ABD\si}_{\,\mu\,\al}(p,q+r,s)\dDelta^{BB}_{q+r}S^{BAD\si}_{\,\al\,\nu}(p+s,q,r)-S^{ABD\si}_{\,\mu\,\al}(p,q+r,s)\dDelta^{BB}_{q+r}\sh^{BAD\si}_{\,\al\,\nu}(p+s,q,r)\nonumber\\[5pt]
+\sh^{BD\si}_{\,\al}(s)\dDelta^{A,BB}_{\,\mu}(p;s,q+r)\sh^{BAD\si}_{\,\al\,\nu}(p+s,q,r)+\sh^{ADD}_{\,\al}(p+q,r,s)\dDelta^{A,AA}_{\,\mu}(p;r+s,q)\sh^{AA}_{\,\al\,\nu}(q)\nonumber\\[5pt]
+\sh^{AA}_{\,\mu\,\al}(p)\dDelta^{A,AA}_{\,\nu}(q;p,r+s)\sh^{ADD}_{\,\al}(p+q,r,s)+\sh^{ABD\si}_{\,\mu\,\al}(p,q+r,s)\dDelta^{A,BB}_{\,\nu}(q;p+s,r)\sh^{BD\si}_{\,\al}(r)\nonumber\\[5pt]
-\sh^{BD\si}_{\,\al}(s)\dDelta^{AA,BB}_{\,\mu\,\nu}(p,q;s,r)\sh^{BD\si}_{\,\al}(r)+\sh^{AADD}_{\,\mu\,\nu}(p,q,r,s)\dDelta^{DD}_s\sh^{DD}(s)\nonumber\\[5pt]
\sh^{AADD}_{\,\mu\,\nu}(p,q,r,s)\dDelta^{DD}_r\sh^{DD}(r)-S^{ADD}_{\,\mu}(p,q+r,s)\dDelta^{DD}_{q+r}S^{ADD}_{\,\nu}(q,r,p+s)\nonumber\\[5pt]
+\sh^{ADD}_{\,\mu}(p,q+r,s)\dDelta^{DD}_{q+r}S^{ADD}_{\,\nu}(q,r,p+s)+S^{ADD}_{\,\mu}(p,q+r,s)\dDelta^{DD}_{q+r}\sh^{ADD}_{\,\nu}(q,r,p+s)\nonumber\\[5pt]
+\sh^{DD}(s)\dDelta^{A,DD}_{\,\mu}(p;s,q+r)\sh^{ADD}_{\,\nu}(q,r,p+s)+\sh^{ADD}_{\,\mu}(p,q+r,s)\dDelta^{A,DD}_{\,\nu}(q;p+s,r)\sh^{DD}(r)\nonumber\\[5pt]
+\sh^{AA}_{\,\nu\,\al}(q)\dDelta^{\sigma D,AB}(q;p,r+s)\sh^{ABD\si}_{\,\mu\,\al}(p,q+r,s)-\sh^{AAA}_{\,\mu\,\nu\,\al}(p,q,r+s)\dDelta^{D\sigma,AB}(r;p+q,s)\sh^{BD\si}_{\,\al}(s)\nonumber\\[5pt]
-\sh^{BAD\si}_{\,\al\,\nu}(p+s,q,r)\dDelta^{D\sigma,BA}(s;q+r,p)\sh^{AA}_{\,\al\,\mu}(p)+\sh^{BD\si}_{\,\al}(r)\dDelta^{D\sigma,BA}(s;r,p+q)\sh^{AAA}_{\,\mu\,\nu\,\al}(p,q,r+s)\nonumber\\[5pt]
+\sh^{AA}_{\,\mu\,\al}(p)\dDelta^{A,D\sigma,AB}_{\,\nu}(q;s;p,r)\sh^{BD\si}_{\,\al}(s)+\sh^{AA}_{\,\nu\,\alpha}(q)\dDelta^{A,D\sigma,BA}_{\,\mu}(p;r;s,q)\sh^{BD\sigma}_{\,\alpha}(s)\nonumber\\[5pt]
+\sh^{AA}_{\,\mu\,\alpha}(p)\dDelta^{AD\sigma,AB}_{\,\nu}(q,r;p,s)\sh^{BD\sigma}_{\,\alpha}(s)+\sh^{BD\si}_{\,\al}(r)\dDelta^{D\sigma
A,BA}_{\phantom{D\sigma}\,\mu}(s,p;r,q)\sh^{AA}_{\,\al\,\nu}(q)\nonumber\\[5pt]
\left.+\sh^{AA}_{\,\mu\,\alpha}(p)\dDelta^{D\sigma D\sigma,AA}(r,s;q,p)\sh^{AA}_{\,\nu\,\alpha}(q)+\sh^{DD}(s)\dDelta^{AA,DD}_{\,\mu\,\nu}(p,q;s,r)\sh^{DD}(r)\right\}+I.C.\nonumber\\[5pt]\label{saadd}
\eea
The integration constants, not specified in most of the
previous equations, are the bare action
vertices. As mentioned in section \ref{subsec:2.1.5}, although the request on the
seed action vertices to keep UV finiteness, ensures there are not classical
divergencies, we must at least prove this is possible for a particular
$\sh$. This check is done in section \ref{subsec:2.1.5}. 

\chapter{Integrated wines}\label{App:intwine}

First of all, once we have defined the zero point wine $\dDelta^{ff}_{RS}(p)$, we
can define its integrated form. We can define
\be
\dDelta^{ff}_{RS}(p)=-\Lam\de_{\Lam}\Delta^{ff}_{RS}(p)
\ee
$\Delta^{ff}_{RS}(p)$ now being the {\it integrated wine}. Let us now consider the two point equation for the generalised
field multiplet $f$:
\be\label{ff}
\Lam\de_{\Lam}S^{ff}_{MN}(p)=S^{ff}_{ML}(p)\dDelta^{ff}_{LS}(p)S^{ff}_{SN}(p)
\ee
It turns out that:
\be\label{app:ush}
S^{ff}_{RS}(k)\Delta^{ff}_{ST}(k)=\delta_{RT}-B^{ff}_{RT}(k)\label{app:inverse}
\ee
This relation will be crucial throughout the entire calculation that will
follow.
In the case of the $A$ field, in particular we have:
\be\label{inversea}
B^{AA}_{\al\beta}(k)=\frac{k_{\al}k_{\beta}}{k^2}
\ee
In the $C$ sector $B^{CC}(k)=0$ and finally in the $BD\sigma$ one we have:
\be\label{inversebds2}
B^{FF}_{RT}(k)=k'_R k_T
\ee
where $k$ and $k'$ are the generalised momenta of Eq.(\ref{genmom1}) and
(\ref{genmom2}). 
There is another relation which holds in this sector if we place the two
point vertex with the integrated wine the other way round:
\be
\Delta^{FF}_{RS}(k)S^{FF}_{ST}(k)=\delta_{RT}-A^{FF}_{RT}(k)\label{app:Art}
\ee
where:
\be\label{inversebds1}
A^{FF}_{RT}(k)=(-k)_R(-k')_T
\ee
but it can be seen that there is a relation between them which will
allow us to make use of only one of the two, namely:
\be\label{a-brel}
A^{FF}_{SR}(-k)=B^{FF}_{RS}(k)
\ee
For the $C$ sector the derivation of these relations is much easier and it is similar to the scalar field case in the $\phi^4$ theory considered in section \ref{sec:1.4}. Since the two point vertex is here invertible, and there is no remainder ($B^{CC}(k)=0$), it is easy to present the full derivation of the previous equations for the present case.

In the $C$-sector, we remind from \eq{ker:c} that the zero-point wine has the following
form:
\be
\dDelta^{CC}(p)=\frac{1}{\Lam^4x}\left(\frac{2 x^2
\tilde{c}_p}{x+2\lambda\tilde{c}_p}\right)^{\prime}
\ee
where $x=p^2/\Lambda^2$ in this case. Now, recalling that in this notation:
\be\label{simplebutimportant}
\de_x=-\frac{1}{2x}\Lambda\de_{\Lambda},
\ee
we can write the following identity:
\be\label{id}
\dDelta^{CC}(p)=-\Lam\de_{\Lam}\Delta^{CC}(p)
\ee
where:
\be\label{app:cc}
\Delta^{CC}(p)=\frac{1}{\Lam^4}\frac{\tilde{c}_p}{x+2\lambda\tilde{c}_p}
\ee
This is the integrated 0-point wine for the $C$ sector, which is
represented in fig.\ref{fig:0intcwine}. 
\begin{figure}[hh]
\begin{center}
\psfrag{=}{$=$}
\includegraphics[scale=.3]{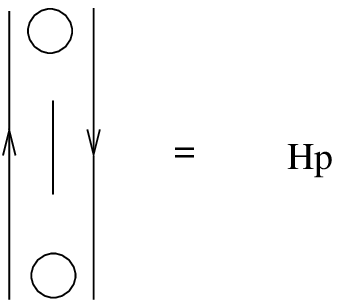}
\caption{Graphical representation of the integrated 0-point wine for the $C$ sector}\label{fig:0intcwine}
\end{center}
\end{figure}
Recalling now the equation for $S^{CC}$:
\be\label{scc}
\Lam\de_{\Lam} S^{CC}(p)=S^{CC}(p) \dDelta^{CC}(p) S^{CC}(p)
\ee
and since $S^{CC}(p)$ is invertible, we can rewrite it as:
\be\label{scc1}
-\Lam\de_{\Lam}(S^{CC}(p))^{-1}=\dDelta^{CC}(p)=-\Lam\de_{\Lam}
 \Delta^{CC}(p)
\ee
Since at $\Lam\to\infty$ we have $(S^{CC}(p))^{-1}- \Delta^{CC}(p)\to 0$ (choosing the integration constant here and later, so that the 
 `effective propagator' vanishes as $p\to\infty$), we see we must have:
\be\label{inversec}
(S^{CC}(p))^{-1}= \Delta^{CC}(p)
\ee
which we can indeed see explicitly from (\ref{hatcc}) and (\ref{app:cc}).

We represent the integrated wine as in \fig{fig:0wines}, but with a line down its spine,
 and thus \eq{inversec} is represented 
 diagrammatically as in fig.\ref{fig:id}. 
\begin{figure}[h]
\begin{center}
\includegraphics[scale=.3]{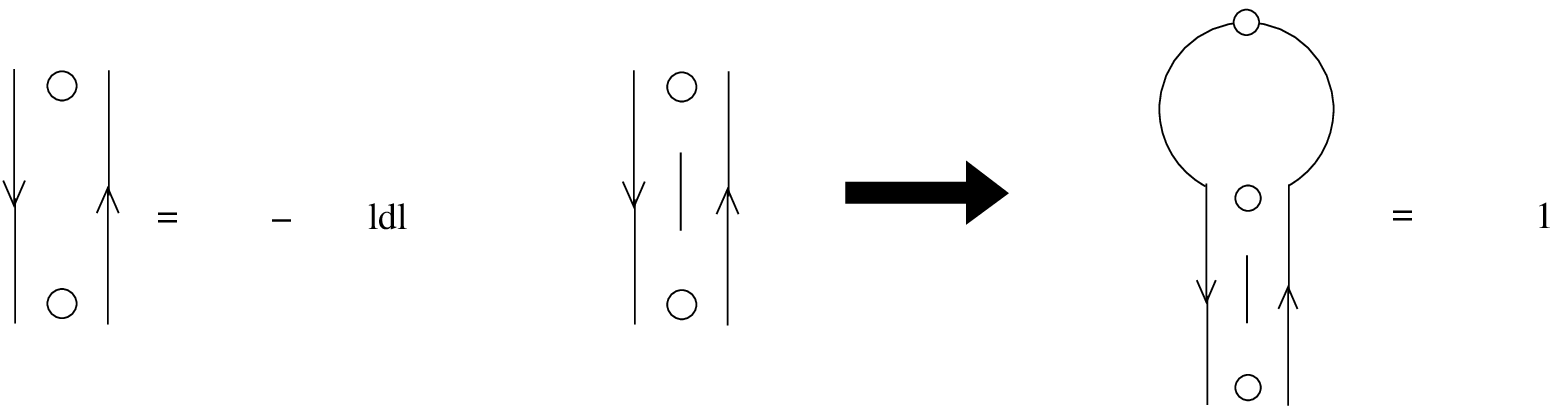}
\caption{The integrated wine in the case of the bosonic component of the scalar field is the inverse of the two-point vertex}\label{fig:id}
\end{center}
\end{figure}
Similarly to \eq{app:cc}, one can find all the other integrated
kernels. They are listed below:
\bea
\Delta^{AA}(p)&=&\frac{1}{2p^2}c_p\label{app:aa}\\
\Delta^{BB}(p)&=&\frac{1}{2\Lam^2}\ \frac{\tilde{c}c}{x\tilde{c}+2c}\label{app:bb}\\
\Delta^{DD}(p)&=&\frac{1}{\Lam^4}\ \frac{\tilde{c}^2}{x\tilde{c}+2c}\label{app:dd}
\eea
where $c$ and $\tilde{c}$ are intended as functions of $x=p^2/\Lam^2$.
Let us spend few words about the previous equations.  
The two $A$'s integrated kernel, in \eq{app:aa}, despite its similarity to
a regularised Feynman propagator, has no gauge fixing. Indeed this
 `effective propagator' is the inverse of the classical $AA$ kinetic
 term only in the transverse space, as one can see from
\eqs{app:ush}{inversea}. Since in practice $\Delta^{AA}$ will be connected to an $A$ point on some 
 other vertex, the remainder term above
 will simply generate gauge transformations 
 via \eq{bosonwi}. This observation proves crucial in the `magic' 
 of the calculation.

The $B$ and $D$ integrated kernels are described in \eqs{app:bb}{app:dd}.  Note that despite the classical $D$ kinetic term being that of a massless
 (Goldstone) field, the $D$ effective propagator like that of $C$ and $B$
 (but unlike $A$) has no massless pole. Of course this is nothing but the
 Higgs mechanism, arising here from
 the $B$ and $D$ two-point vertices being intimately related via
 \eqs{bb}{dd} (the $BD\sigma$ vertex being non zero). Similarly to the above reasoning, the pair of effective
 propagators \eqs{app:bb}{app:dd}, would form the inverse of the {\sl matrix} of
 these fermionic two-point vertices (see \eq{sFF}), if this matrix was invertible. It is
 not, for the same reason that these flows are necessarily entangled: $B$ and
 $D\sigma$ rotate into each other under the broken supergauge transformations
 \eq{fermionwi}.

\chapter{Special momenta}\label{App:specmom}

In the present appendix is contained a derivation of Eqs.(\ref{tre}) and (\ref{quattro}), plus some comments about the behaviour of the
vertices at some particular momenta. The symmetry which governs the theory
is gauge symmetry and also in this case, will give us a hint on how to derive
the expressions mentioned above. Let us consider first the three point vertex
$S^{Aff}_{\mu RS}$ at momenta $(0,k,-k)$. We can imagine to consider it
evaluated at momenta $(\epsilon,k-\epsilon,-k)$ and applying the Ward
identity with momentum $\epsilon_{\mu}$, we get:
\be
\epsilon_{\mu}S^{Aff}_{\mu
RS}(\epsilon,k-\epsilon,-k)=S^{ff}_{RS}(k)-S^{ff}_{RS}(k-\epsilon)
\ee
Expanding in $\epsilon$ both sides of the equation and taking order linear
in $\epsilon$, we get precisely the first equality of eq.(\ref{tre}). For
the second one, we can repeat the procedure the same way.

If we now want to have an espression for the four point vertex
$S^{AAff}_{\mu\nu RS}$ evaluated at momenta $(0,0,k,-k)$, we can as well
consider it instead at momenta $(\epsilon,-\epsilon,k,-k)$ and applying
twice the Ward identity, with $\epsilon_{\mu}$ first and then
$\epsilon_{\nu}$, we get:
\be
\epsilon_{\mu}\epsilon_{\nu}S^{AAff}_{\mu\nu
RS}(\epsilon,-\epsilon,k,-k)=\epsilon_{\mu}\de_{\mu}^k
S^{ff}_{RS}(k)-S^{ff}_{RS}(k)+S^{ff}_{RS}(k-\epsilon)
\ee
In this case, if we expand the equation in $\epsilon$ and take the order
$\epsilon^2$, what we get is eq.(\ref{quattro}).

Other vertices at particular momenta can be evaluated in similar ways.

\newpage
\chapter{Bare action vertices (finite part)} \label{App:bare}  

This is a list of finite part of the 4-point and 3-point vertices of the bare
action, when $\Lambda_0$ is sent to infinity that must be added to the
equations in section \ref{subsec:2.1.5}. They represent the finite part of the
integration constants, necessary in order to have gauge invariant tree level vertices.

\begin{center}
\noindent {\bf Four-point:}
\end{center}
\begin{eqnarray}
\left.\hat{S}^{AAAA}_{\,\mu\,\nu\,\rho\,\sigma}(p,q,r,s)\right |_{\Lambda_0\to\infty} &=&2\delta_{\mu\nu}\delta_{\rho\sigma} - 4\delta_{\mu\rho}\delta_{\nu\sigma} + 2\delta_{\mu\sigma}\delta_{\nu\rho}\\[5pt]
\left.\hat{S}^{AABB}_{\,\mu\,\nu\,\rho\,\sigma}(p,q,r,s)\right |_{\Lambda_0\to\infty} &=&2\delta_{\mu\nu}\delta_{\rho\sigma} - 4\delta_{\mu\rho}\delta_{\nu\sigma} + 2\delta_{\mu\sigma}\delta_{\nu\rho} - 4\tilde{c}^{\prime}_0\delta_{\mu\nu}\delta_{\rho\sigma}\\[5pt]
\left.\hat{S}^{AADD}_{\,\mu\,\nu}(p,q,r,s)\right |_{\Lambda_0\to\infty}
&=&\tilde{c}^{\prime}_0\left(p_{\nu}r_{\mu} + q_{\mu}s_{\nu}
- r_{\mu}r_{\nu} + 2r_{\mu}s_{\nu}
- s_{\mu}s_{\nu}\right.\nonumber\\[3pt]
&&\left. - \delta_{\mu\nu}p^2 - 2\delta_{\mu\nu}p\cdot s +\delta_{\mu\nu}r\cdot s - \delta_{\mu\nu}s^2\right)\\[5pt]
\left.\hat{S}^{AACC}_{\,\mu\,\nu}(p,q,r,s)\right |_{\Lambda_0\to\infty}
&=&\tilde{c}^{\prime}_0\left(p_{\nu}r_{\mu} + q_{\mu}s_{\nu}
- r_{\mu}r_{\nu} + 2r_{\mu}s_{\nu}
- s_{\mu}s_{\nu}\right.\nonumber\\[3pt]
&&\left. - \delta_{\mu\nu}p^2 - 2\delta_{\mu\nu}p\cdot s + \delta_{\mu\nu}r\cdot s - \delta_{\mu\nu}s^2\right)\\[5pt]
\left.\hat{S}^{AABDS}_{\,\mu\,\nu\,\rho}(p,q,r,s)\right |_{\Lambda_0\to\infty} &=&- 2\tilde{c}^{\prime}_0\left(\delta_{\mu\nu}s_{\rho} +\delta_{\mu\rho}p_{\nu}-\delta_{\mu\rho}r_{\nu} +\delta_{\mu\rho}s_{\nu}\right)\\[5pt]
\left.\hat{S}^{BAADS}_{\,\mu\,\nu\,\rho}(p,q,r,s)\right |_{\Lambda_0\to\infty}
&=&2\tilde{c}^{\prime}_0\left(\delta_{\mu\rho}p_{\nu} -
\delta_{\mu\rho}r_{\nu} - \delta_{\mu\rho}s_{\nu} -
\delta_{\nu\rho}s_{\mu}\right)
\end{eqnarray}

\begin{center}
\noindent {\bf Three-point:}
\end{center}
\begin{eqnarray}
\left.\hat{S}^{BADS}_{\,\mu\,\nu}(p,q,r)\right |_{\Lambda_0\to\infty} &=&2\tilde{c}^{\prime}_0\left(p_{\nu}r_{\mu} - r_{\mu}r_{\nu} - \delta_{\mu\nu}p^2\right)\\[5pt]
\left.\hat{S}^{ABDS}_{\,\mu\,\nu}(p,q,r)\right |_{\Lambda_0\to\infty} &=&- 2\tilde{c}^{\prime}_0\left(q_{\mu}r_{\nu} -r_{\mu}r_{\nu} -\delta_{\mu\nu}q^2\right)\\[5pt]
\left.\hat{S}^{AAA}_{\,\mu\,\nu\,\rho}(p.q.r)\right |_{\Lambda_0\to\infty} &=&2\delta_{\mu\nu}p_{\rho} - 2\delta_{\mu\nu}q_{\rho} - 2\delta_{\mu\rho}p_{\nu} + 2\delta_{\mu\rho}r_{\nu} + 2\delta_{\nu\rho}q_{\mu}\nonumber\\[3pt]
&&  - 2\delta_{\nu\rho}r_{\mu}\\[5pt]
\left.\hat{S}^{BBA}_{\,\mu\,\nu\,\rho}(p,q,r)\right |_{\Lambda_0\to\infty}
&=&2\delta_{\mu\nu}p_{\rho} - 2\delta_{\mu\nu}q_{\rho} -
2\delta_{\mu\rho}p_{\nu} + 2\delta_{\mu\rho}r_{\nu} +
2\delta_{\nu\rho}q_{\mu}\nonumber\\[3pt]
&&  - 2\delta_{\nu\rho}r_{\mu}- 4\tilde{c}^{\prime}_0(\delta_{\mu\nu}p_{\rho} -\delta_{\mu\nu}q_{\rho})\\[5pt]
\left.\hat{S}^{ADD}_{\,\mu}(p,q,r)\right |_{\Lambda_0\to\infty} &=&- \tilde{c}^{\prime}_0(q_{\mu}q^2 -q_{\mu}q\cdot r +r_{\mu}q\cdot r -r_{\mu}r^2)\\[5pt]
\left.\hat{S}^{ACC}_{\mu}(p,q,r)\right |_{\Lambda_0\to\infty} &=&- \tilde{c}^{\prime}_0(q_{\mu}q^2 -q_{\mu}q\cdot r +r_{\mu}q\cdot r-r_{\mu}r^2)
\end{eqnarray}
\newpage


\end{document}